\documentclass[journal]{IEEEtran}
\usepackage{epsfig,amsmath,amssymb,epsf,cite,scalefnt,subfig,array} 
\usepackage{cases,color}
\usepackage{graphicx,pifont,epstopdf,grffile}
\usepackage{multirow}

   \def\bd{{\mathbf{d}}}

\def\bz{{\mathbf{z}}}

\def\bA{{\mathbf{A}}}  \def\bC{{\mathbf{C}}} \def\bD{{\mathbf{D}}} 
   \def\bI{{\mathbf{I}}}

 \def\ibb{{\pmb{b}}} \def\ibc{{\pmb{c}}} \def\ibd{{\pmb{d}}} \def\ibe{{\pmb{e}}}
    
 \def\ibl{{\pmb{l}}}   
    
\def\ibu{{\pmb{u}}} \def\ibv{{\pmb{v}}} \def\ibw{{\pmb{w}}} \def\ibx{{\pmb{x}}} \def\iby{{\pmb{y}}}
\def\ibz{{\pmb{z}}}

   \def\ibI{{\pmb{I}}}

\renewcommand{\baselinestretch}{1.0}

     \def\d4{\!\!\!\!}

\newcommand{\bef}{\begin{figure}}
\newcommand{\eef}{\end{figure}}
\newcommand{\beq}{\begin{eqnarray}}
\newcommand{\eeq}{\end{eqnarray}}

\newcommand{\qed}{\nobreak \ifvmode \relax \else
\ifdim\lastskip<1.5em \hskip-\lastskip \hskip1.5em plus0em
minus0.5em \fi \nobreak \vrule height0.5em width0.5em
depth0.25em\fi}

\begin{document}

\title{ADMM-MCP Framework for Sparse Recovery with Global Convergence}

\author{Hao~Wang,
        Zhanglei~Shi,
        Chi-Sing~Leung,~\IEEEmembership{Member,~IEEE,}
        Hing Cheung So,~\IEEEmembership{Fellow,~IEEE,}

\thanks{Hao~Wang, Zhanglei~Shi, Chi-Sing~Leung, and Hing Cheung So are with the Department of Electronic Engineering, City University of Hong Kong, Hong Kong, China.}}

\markboth{IEEE TRANSACTIONS ON SIGNAL PROCESSING,~Vol.~, No.~, September~2018}%
{Shell \MakeLowercase{\textit{et al.}}: ADMM-MCP Framework for Sparse Recovery with Global Convergence}

\maketitle

\begin{abstract}
In compressed sensing, the $l_0$-norm minimization of sparse signal reconstruction is NP-hard. Recent work shows that compared with the best convex relaxation ($l_1$-norm), nonconvex penalties can better approximate the $l_0$-norm and can reconstruct the signal based on fewer observations. In this paper, the original problem is relaxed by using minimax concave penalty (MCP). Then alternating direction method of multipliers (ADMM) and modified iterative hard thresholding method are used to solve the problem. Under certain reasonable assumptions, the global convergence of the proposed method is proved. The parameter setting is also discussed.
Finally, through simulations and comparisons with several state-of-the-art algorithms, the effectiveness of proposed method is confirmed.
\end{abstract}


\IEEEpeerreviewmaketitle

\section{Introduction}\label{section1}
In compressed sensing~\cite{donoho2006compressed,blumensath2009iterative}, a fundamental problem is to recover an unknown sparse signal $\ibx \in \mathbb{R}^N$ given an observation vector $\ibb\in \mathbb{R}^M$. Under the noiseless environment, the relationship between them is given by
\begin{equation} \label{eq:x&b}
\ibb = \bA \ibx,
\end{equation}
where $\bA \in \mathbb{R}^{M \times N}\,\,(M<N)$ is a measurement matrix with full rank $M$.
Apparently, there is an infinite number of solutions for \eqref{eq:x&b} if there is no restriction on $\ibx$. Given that $\ibx$ is sparse, the compressed sensing problem is to solve:
\begin{subequations} \label{eq:SR}
\begin{align}
&\min \|\ibx\|_0, \\
&\mbox{s.t. } \bA \ibx=\ibb,
\end{align}
\end{subequations}
where the "true" sparsity measure is the $l_0$-norm term (the number of non-zero elements) in \eqref{eq:SR}. In fact, observation noise in $\ibb$ is inevitable.
If we assume that the noise follows Gaussian distribution.
Then, \eqref{eq:SR} can be modified to:
\begin{subequations}
\label{eq:QC}
\begin{align}
&\min \|\ibx\|_0, \\
&\mbox{s.t. } || \ibb - \bA \ibx ||_2^2 \leq M\sigma^2,
\end{align}
\end{subequations}
or a quadratic programming problem given by
\begin{subequations}
\label{eq:QP}
\begin{align}
&\min || \ibb - \bA \ibx ||_2^2, \\
&\mbox{s.t. } \|\ibx\|_0 \leq \tau,
\end{align}
\end{subequations}
where $\sigma > 0$ is the standard deviation of observation noise and $\tau > 0$ is the sparsity level.
Based on Lagrangian duality, there is a one-to-one relationship between $\sigma$ and $\tau$.
The selection of these two functions depends on which parameter is known in advance.
If we have the statistical properties of noise, we select function (\ref{eq:QC}) for signal recovery. Otherwise, if we know the sparsity, (\ref{eq:QP}) is used.
Unfortunately, due to the $\|\ibx\|_0$ term, both \eqref{eq:QC} and \eqref{eq:QP} are NP-hard \cite{natarajan1995sparse}. Thus, to solve these problems, an alternative function of $\|\ibx\|_0$ must be introduced. The function should be continuous and result in an unbiased sparse estimate $\hat{\ibx}$ \cite{fan2001variable}.

Commonly, $l_1$-norm is used to replace the $l_0$-norm term.
Because $l_1$-norm is continuous and convex, the estimate of it is the sparsest solution under certain conditions~\cite{donoho2006most}.  Substituting the sparse measurement function by $l_1$-norm term, (\ref{eq:QC}) is transformed to the famous basis pursuit denoising (BPDN) problem and (\ref{eq:QP}) is the well-known LASSO problem \cite{tibshirani1996}. In the past decades,  a large collection of numerical algorithms are proposed for solving them, including BPDN-interior~\cite{chen2001atomic}, PDCO~\cite{saunders2002pdco}, Homotopy~\cite{osborne2000new}, LARS~\cite{efron2004least}, OMP~\cite{tropp2007signal}, STOMP~\cite{donoho2006bsparse}. Besides, some elegant implementation packages, for example $l_1$ magic~\cite{candes2007l1} and SPGL1 \cite{van2008probing,spgl12007}, are also available.

Although $l_1$-norm relaxation approaches are the best convex approximation of the original problem and have satisfactory performance under certain conditions, they have two major drawbacks. First, the conditions needed by $l_1$-norm approaches may still be too strict in many cases. Second, comparing with the nonconvex relaxation approaches (e.g., $l_p$-norm with $0<p<1$), $l_1$-norm approaches normally need more observations to reconstruct the sparse signal of interest.

$l_p$-norm is another family of compressed sensing methods, which can obtain unbiased sparse solution under weaker conditions.
However, comparing with the $l_1$-norm methods, due to the nonconvex and non-differentiable $l_p$-norm function, these approaches normally have worse convergence and higher computational resource consumption.
Therefore, many practitioners consider introducing other auxiliary functions to replace the $l_p$-norm term.
For instance, a method called focal underdetermined system solver (FOCUSS) is proposed in \cite{engan2000regularized,gorodnitsky1997sparse}, and a smoothly clipped absolute deviation (SCAD) penalty function is introduced in \cite{fan2001variable}. Besides, a minimax concave penalty (MCP) function is used in~\cite{zhang2010nearly}.

In \cite{blumensath2009iterative,blumensath2010normalized}, the iterative hard thresholding (IHT) method is discussed which can be used to solve the problem given in \eqref{eq:QP}.
To ensure the sparsity, an element-wise hard thresholding operator is used in this algorithm. IHT consists of the iteration
\begin{equation} \label{eq:IHT}
\ibx^{k+1} =H_{\tau}(\ibx^k+\bA^\mathrm{T}(\ibb-\bA\ibx^k)),
\end{equation}
where $H_{\tau}(\ibz)$ is a non-linear operator which sets all but the largest $\tau$ elements (in magnitude) of $\ibz$ to zero.
IHT has near-optimal error guarantees, and it is efficient when handling large-scale problems.
The simple structure and efficiency make this algorithm prevalent recently.
While IHT can only find locally optimal solutions, and its initialization is of first importance. In practice, this method is usually initialized by the solutions of $l_1$-norm relevant methods.
To overcome this defect and further improve its performance, several modified approaches are devised, including the TST method \cite{maleki2010approximate}, AMP and its variants \cite{maleki2010approximate,metzler2016denoising}. Although comparing with IHT, the convergence of these methods are largely improved, they are still sensitive to the initial value of the estimated signal. None of them theoretically guarantees to converge to a globally optimal solution.

In this paper, we further modify the IHT method, combine it with MCP and ADMM framework. Finally, we propose a new approach with theoretically global convergence and better performance in practice.
The main contributions of this paper are given as follows:

(i). We prove the global convergence of the proposed method. In other words, there is no need for us to carefully select its starting value.

(ii). The parameter selection for our method is analyzed and two effective parameter setting methods are proposed.

(iii). Comparing with several sparse signal reconstruction methods, the performance of our proposed method is superior.

The rest of this paper is organized as follows. Section~\ref{section2} introduces the background of the ADMM framework, several $l_0$-norm approximate functions and basic one-dimensional optimization problem with MCP. Section~\ref{section3} presents the development of the proposed algorithm for solving \eqref{eq:QP}. In Section~\ref{section4}, the convergence of the proposed algorithm is analyzed. The parameter setting of our proposed method is discussed in Section~\ref{section5}. Simulation results and numerical comparisons are given in Section~\ref{section6}. Finally, conclusions are drawn in Section~\ref{section7}.

\section{Background}\label{section2}

\subsection{Notation}
In this paper, we use a lower-case or upper-case letter to represent a scalar while vectors and matrices are denoted by bold lower-case and upper-case letters, respectively. The transpose operator is denoted as $(\centerdot)^ \mathrm{T}$. And $\ibI$ denotes an identity matrix with appropriate dimension. Other mathematical symbols are defined in their first appearance.

\subsection{ADMM}
The ADMM framework is an iterative approach for solving optimization problems by breaking them into several subproblems \cite{boyd2011distributed}. This algorithm is normally used to handle problems in the following standard form:
\begin{subequations}
\beq
& & \min\limits_{\ibz,\ibv}:~\psi\left(\ibz\right) + g\left(\ibv\right) \\
& & s.t. ~~~~\bC \ibz + \bD \ibv = \iby,
\eeq
\label{eq:admm_obj}
\end{subequations}
with variables $\ibz\in \mathbb{R}^n$ and $\ibv\in \mathbb{R}^m$, where $\iby\in \mathbb{R}^p$, $\bC \in \mathbb{R}^{p\times n}$ and $\bD \in \mathbb{R}^{p\times m}$. To solve this problem, first we need to construct an augmented Lagrangian function
\beq
&L \left(\ibz, \ibv, \ibl \right) = & \psi\left(\ibz\right) + g\left(\ibv\right) + \ibl^\mathrm{T} \left( \bC \ibz + \bD \ibv  - \iby \right) \nonumber\\
& &+ \frac{\rho}{2} \left\|\bC \ibz + \bD \ibv  - \iby \right\|_2^2,
\label{eq:admm_augmented}
\eeq
where $\ibl\in \mathbb{R}^p$ is the Lagrange multiplier, and $\rho>0$ is a trade-off parameter.
The general ADMM scheme is given by
\begin{subequations}
\label{eq:admm_scheme}
\beq
\ibv^{k+1} &=& \arg\min\limits_{\ibv} L(\ibz^{k}, \ibv, \ibl^k),\\
\ibz^{k+1} &=& \arg\min\limits_{\ibz} L(\ibz, \ibv^{k+1}, \ibl^k),\\
\ibl^{k+1} &=& \ibl^k + \rho \left( \bC \ibz^{k+1} + \bD \ibv^{k+1} - \iby \right).
\eeq
\end{subequations}

\subsection{Nonconvex $l_0-norm$ approximate functions}
Problems with $l_0$-norm are NP-hard. Hence a lot of proximate functions are proposed in the past two decades to replace the $l_0$-norm term. Several typical examples are shown as follows.

1. Exponential type function (ETF):
\begin{equation}\label{eq:ETF2}
P_{\lambda, \gamma}(u_i)=\frac{\lambda(1-e^{-\gamma|u_i|})}{1-e^{-\gamma}},
\end{equation}
where $\gamma>0$ and $\lambda>0$.

2. Logarithmic type function (LTF):
\begin{equation}\label{eq:LTF2}
P_{\lambda, \gamma}(u_i)=\frac{\lambda \log (\gamma|u_i|+1)}{\log (\gamma+1)},
\end{equation}
where $\gamma>0$ and $\lambda>0$.

3. Smoothly clipped absolute deviation (SCAD) function:
\beq
P_{\lambda,\gamma}(u_i) = \left\{ \begin{array}{lcl}
\lambda |u_i|, && if\,\,\,\, |u_i| \leq \lambda, \\
-\frac{u_i^2-2\gamma\lambda|u_i|+\lambda^2}{2(\gamma-1)}, && if\,\,\,\, \lambda<|u_i| \leq \gamma\lambda, \\
\frac{1}{2}(\gamma+1)\lambda^2, && if\,\,\,\, |u_i| > \gamma\lambda,
\end{array}\right.
\label{SCAD}
\eeq
where $\gamma>2$ and $\lambda>0$.

4. Minimax concave penalty (MCP) function:
\beq
P_{\lambda,\gamma}(u_i) = \left\{ \begin{array}{lcl}
\lambda |u_i|-\frac{u_i^2}{2\gamma}, & if\,\,\,\, |u_i| \leq \gamma\lambda, \\
\frac{1}{2}\gamma\lambda^2, & if\,\,\,\, |u_i| > \gamma\lambda,
\end{array}\right.
\label{MCP}
\eeq
where $\gamma>1$ and $\lambda>0$.

The definitions of above approximate functions are all in an element-wise manner. For any vector $\ibu=[u_1, u_2, \dots, u_n]^\mathrm{T}$, their definitions are
\begin{equation}\label{eq:penalty}
P_{\lambda, \gamma}(\ibu)=\sum_{i=1}^n P_{\lambda, \gamma}(u_i).
\end{equation}
Comparing these four approximate functions, although the MCP is a piecewise function, it empirically has better performance and needs fewer computations in practice. Hence we use MCP function to approximate $l_0$-norm in this paper.

\begin{figure}[htb]
\centering
\includegraphics[width=3.4in]{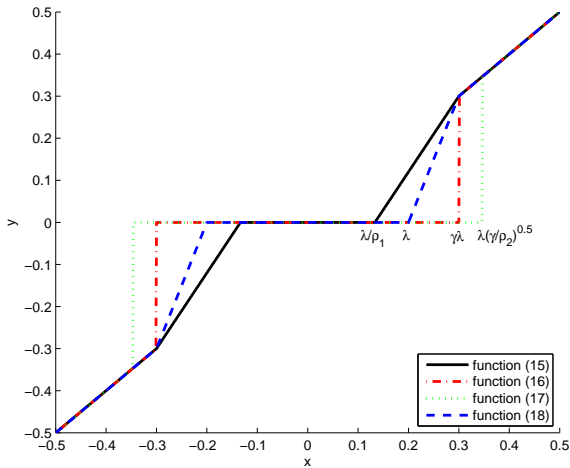}
\caption{Function shapes in \eqref{mcp_ui}-\eqref{mcp_ui_3}.}
\label{fig:mcp_u}
\end{figure}

Then, we discuss a basic one-dimensional optimization problem which occurs in the ADMM iterations in Section~\ref{section3},
\begin{equation}\label{eq:L_i}
\arg\min_{u_i} L_i(u_i)=P_{\lambda, \gamma}(u_i)+\frac{\rho}{2}(s_i-u_i)^2,
\end{equation}
where $s_i$ can be treated as a constant and $P_{\lambda, \gamma}(u_i)$ denotes the MCP in~\eqref{MCP}. Thus, $L_i(u_i)$ is a continuous piecewise function. If $\rho>\frac{1}{\gamma}$, its first order derivative is non-decreasing. Further we know that \eqref{eq:L_i} is convex and has a unique global optimal solution which depends on the value of $s_i$,
\beq \label{mcp_ui}
u_i = \left\{ \begin{array}{lcl}
s_i, && if\,\,\,\, |s_i| > \gamma\lambda, \\
sign(s_i)\frac{|s_i|-\frac{\lambda}{\rho}}{1-\frac{1}{\gamma\rho}}, && if\,\,\,\, \frac{\lambda}{\rho}<|s_i|\leq \gamma\lambda, \\
0, && if\,\,\,\, |s_i| \leq \frac{\lambda}{\rho}.
\end{array}\right.
\eeq
If $\rho=\frac{1}{\gamma}$, the analytical solution of \eqref{eq:L_i} is
\beq \label{mcp_ui_1}
u_i = \left\{ \begin{array}{lcl}
s_i, && if\,\,\,\, |s_i| > \gamma\lambda, \\
0, && if\,\,\,\, |s_i| \leq \gamma\lambda.
\end{array}\right.
\eeq
If $\rho<\frac{1}{\gamma}$, we can deduce
\beq \label{mcp_ui_2}
u_i = \left\{ \begin{array}{lcl}
s_i, && if\,\,\,\, |s_i| > \sqrt{\frac{\gamma}{\rho}}\lambda, \\
0, && if\,\,\,\, |s_i| \leq \sqrt{\frac{\gamma}{\rho}}\lambda.
\end{array}\right.
\eeq
However, in practice, for all these three cases, we can use a unified approximate solution
\beq \label{mcp_ui_3}
u_i = \left\{ \begin{array}{lcl}
s_i, && if\,\,\,\, |s_i| > \gamma\lambda, \\
sign(s_i)\frac{|s_i|-\lambda}{1-\frac{1}{\gamma}}, && if\,\,\,\, \lambda<|s_i|\leq \gamma\lambda, \\
0, && if\,\,\,\, |s_i| \leq \lambda.
\end{array}\right.
\eeq
The shapes of \eqref{mcp_ui}-\eqref{mcp_ui_3} are shown in Fig.\ref{fig:mcp_u} where $\rho_1>\frac{1}{\gamma}$ and $\rho_2<\frac{1}{\gamma}$. The blue line denotes our unified approximate solution.
It retains all variables whose magnitude are greater than the threshold $\gamma\lambda$, compresses the value between $\lambda$ and $\gamma\lambda$, and sets variables smaller than $\lambda$ to 0.
It is an efficient approximation to function \eqref{mcp_ui}-\eqref{mcp_ui_2}.
Because compared with \eqref{mcp_ui}, it just modifies its compressed range $[\lambda/\rho_1,\lambda\gamma]$ to $[\lambda,\lambda\gamma]$. Comparing with \eqref{mcp_ui_1}, it smoothes its hard threshold. And comparing with \eqref{mcp_ui_2}, it changes its threshold and smoothes it. Since $\lambda$ is a tunable hyper-parameter, if we use an appropriate $\lambda$, the influence of the threshold changing can be avoided.

\section{Development of Proposed Algorithm} \label{section3}
In this section, a novel approach with global convergence is devised to solve the problem in \eqref{eq:QP}. 

First, to ensure the problem \eqref{eq:QP} can be solved by ADMM framework, we introduce an indicator function
\beq \label{indicator}
i_{\ibc(\tau)}(\ibu) = \left\{\begin{array}{ll}
0, &if\,\,\,\,\ibu\in \ibc(\tau), \\
+\infty, &otherwise, \\
\end{array}\right.
\eeq
where the set $\ibc(\tau)=\{\ibu:\|\ibu\|_0\leq \tau\}$ ($\tau\leq M$).
After that, the problem in \eqref{eq:QP} can be rewritten as
\begin{subequations}\label{admm_prob1}
\beq
&\min\limits_{\ibx,\ibu} &\|\ibb-\bA\ibx\|_2^2 + i_{\ibc(\tau)}(\ibu), \\
& s.t. &\ibx=\ibu.
\eeq
\end{subequations}
Thus the problem follows the standard form given in \eqref{eq:admm_obj}. From \eqref{admm_prob1}, we construct its augmented Lagrangian
\beq\label{eq:Lagr_meth1}
L(\ibx, \ibu, \ibw)&=&\|\ibb-\bA\ibx\|_2^2 + i_{\ibc(\tau)}(\ibu)+\ibw^\mathrm{T}(\ibx-\ibu) \nonumber\\
& &+ \frac{\rho}{2} \left\|\ibx - \ibu\right\|_2^2,
\eeq
where $\ibw \in \mathbb{R}^N$.
Then, according to the ADMM in \eqref{eq:admm_scheme}, the variables $\ibu, \ibx, \ibw$ are iteratively updated as:
\beq
\ibu^{k+1} &=& \arg\min\limits_{\ibu} L(\ibx^k, \ibu, \ibw^k) \nonumber \\
&=& \arg\min\limits_{\ibu} i_{\ibc(\tau)}(\ibu) + \frac{\rho}{2} \left\|\ibx^k-\ibu+\frac{\ibw^k}{\rho} \right\|_2^2\nonumber \\
&\approx& H_{\tau}\left(\ibx^k + \frac{\ibw^k}{\rho}\right),\label{meth1_u}\\
\ibx^{k+1} &=& \arg\min\limits_{\ibx} L(\ibx, \ibu^{k+1}, \ibw^k) \nonumber \\
&=& \arg\min\limits_{\ibx} \|\ibb-\bA\ibx\|_2^2+\frac{\rho}{2} \left\|\ibx-\ibu^{k+1}+\frac{\ibw^k}{\rho} \right\|_2^2\nonumber \\
&=&\left[2\bA^\mathrm{T}\bA+\rho\bI\right]^{-1} \left[2\bA^\mathrm{T}\ibb+\rho\ibu^{k+1}-\ibw^k\right], \label{meth1_x}\\
\ibw^{k+1} &=& \ibw^k + \rho \left(\ibx^{k+1}-\ibu^{k+1}\right). \label{meth1_w}
\eeq
In \eqref{meth1_u}, similar with \cite{blumensath2009iterative}, an element-wise hard thresholding operator is used to calculate its approximate solution. For this method, it is hard to prove its global convergence. Nevertheless, according to our preliminary simulation result, for compressed sensing problems, methods with nonconvex $l_0$-norm approximate functions normally have better performance than the hard thresholding operator based methods. Hence, we further modify the problem in \eqref{admm_prob1} as
\begin{subequations}\label{admm_prob2}
\beq
&\min\limits_{\ibx,\ibu} &\|\ibb-\bA\ibx\|_2^2 + P_{\lambda,\gamma}(\ibu), \\
& s.t. &\ibx=\ibu.
\eeq
\end{subequations}
Where the function $P_{\lambda,\gamma}(\ibu)$ denotes MCP in the following sections. (Actually, other nonconvex $l_0$-norm approximate functions in Section~\ref{section2} can also be used here.)
The augmented Lagrangian is given by
\beq\label{eq:Lagr_meth2}
L(\ibx, \ibu, \ibw)&=&\|\ibb-\bA\ibx\|_2^2 +P_{\lambda,\gamma}(\ibu) +\ibw^\mathrm{T}(\ibx-\ibu) \nonumber\\
& &+ \frac{\rho}{2} \left\|\ibx - \ibu\right\|_2^2.
\eeq
We still utilize ADMM framework to solve the problem. Comparing with the scheme in \eqref{meth1_u}-\eqref{meth1_w}, we just need to change the update procedure of $\ibu$ to
\beq\label{meth2_u}
\ibu^{k+1} &=& \arg\min\limits_{\ibu} L(\ibx^k, \ibu, \ibw^k) \nonumber \\
&=& \arg\min\limits_{\ibu} P_{\lambda,\gamma}(\ibu) + \frac{\rho}{2} \left\|\ibx^k-\ibu+\frac{\ibw^k}{\rho} \right\|_2^2.
\eeq
According to \eqref{mcp_ui}-\eqref{mcp_ui_2}, we can deduce that
$\ibu^{k+1}=[u_1^{k+1},\dots,u_N^{k+1}]^\mathrm{T}$, for any $i\in[1,\dots,N]$, if $\rho>\frac{1}{\gamma}$ then
\beq \label{meth2_ui_mcp1}
u_i^{k+1}=\left\{ \begin{array}{lcl}
\displaystyle \frac{S\left( x_i^k + \frac{w_i^k}{\rho},\frac{\lambda}{\rho} \right)}{1-1/ (\gamma\rho)}, & if |x_i^k + \frac{w_i^k}{\rho}| \leq \gamma\lambda, \\
\displaystyle x_i^k + \frac{w_i^k}{\rho}, & if |x_i^k + \frac{w_i^k}{\rho}| > \gamma\lambda,
\end{array}\right.
\eeq
where $S$ denotes the soft-threshold operator \cite{donoho1994ideal} given by
\beq
S(z,\eta)=sign(z)\max\{|z|-\eta,0\}, \nonumber
\eeq
if $\rho=\frac{1}{\gamma}$,
\beq \label{meth2_ui_mcp2}
u_i^{k+1}=\left\{ \begin{array}{lcl}
\displaystyle 0, & if |x_i^k + \frac{w_i^k}{\rho}| \leq \gamma\lambda, \\
\displaystyle x_i^k + \frac{w_i^k}{\rho}, & if |x_i^k + \frac{w_i^k}{\rho}| > \gamma\lambda,
\end{array}\right.
\eeq
if $\rho<\frac{1}{\gamma}$,
\beq \label{meth2_ui_mcp3}
u_i^{k+1}=\left\{ \begin{array}{lcl}
\displaystyle 0, & if |x_i^k + \frac{w_i^k}{\rho}| \leq \sqrt{\frac{\gamma}{\rho}}\lambda, \\
\displaystyle x_i^k + \frac{w_i^k}{\rho}, & if |x_i^k + \frac{w_i^k}{\rho}| > \sqrt{\frac{\gamma}{\rho}}\lambda.
\end{array}\right.
\eeq
According to \eqref{mcp_ui_3}, all these three cases have a unified approximate solution
\beq \label{meth2_ui_mcp}
u_i^{k+1}=\left\{ \begin{array}{lcl}
\displaystyle \frac{S\left( x_i^k + \frac{w_i^k}{\rho},\lambda \right)}{1-1/\gamma}, & if |x_i^k + \frac{w_i^k}{\rho}| \leq \gamma\lambda, \\
\displaystyle x_i^k + \frac{w_i^k}{\rho}, & if |x_i^k + \frac{w_i^k}{\rho}| > \gamma\lambda.
\end{array}\right.
\eeq

\section{Proof of Global Convergence} \label{section4}
In this section, the convergence of our proposed method is analyzed.
First, we discuss the case when $\rho>\frac{1}{\gamma}$ and the solution of $u_i$ is exact which is given by~\eqref{meth2_ui_mcp1}. The sketch of the proof is shown in Theorem 1.

\textbf{Theorem 1:} If the proposed method satisfies the following three conditions:

\textbf{C1} (Sufficient decrease condition)
For each iteration step $k$, $\exists \tau_1>0$ let
\beq
\small L(\ibx^{k+1}, \ibu^{k+1}, \ibw^{k+1})-L(\ibx^{k}, \ibu^{k}, \ibw^{k}) \leq -\tau_1\|\ibx^{k+1}-\ibx^{k}\|_2^2.
\eeq

\textbf{C2} (Boundness condition)
The sequences $\{\ibx^{k}, \ibu^{k}, \ibw^{k}\}$ generated by the proposed method are bounded and its Lagrangian $L(\ibx^{k}, \ibu^{k}, \ibw^{k})$ is lower bounded.

\textbf{C3} (Sub-gradient boundness condition) There exists
$\bd^{k+1}\in \partial L(\ibx^{k+1},\ibu^{k+1},\ibw^{k+1})$, and $\tau_2>0$ such that
\beq
\|\bd^{k+1}\|_2^2\leq \tau_2\|\ibx^{k+1}-\ibx^{k}\|_2^2.
\eeq

Then, based on \textbf{C1} to \textbf{C3}, we further deduce that the sequence $\{\ibx^{k}, \ibu^{k}, \ibw^{k}\}$ generated by the proposed method has at least a limit point $\{\ibx^{*}, \ibu^{*}, \ibw^{*}\}$ and any limit point is a stationary point. For the penalty function MCP in~\eqref{MCP} with $\rho>\frac{1}{\gamma}$, the proposed method has global convergence.

\textbf{Proof:} The theorem is similar as the Proposition 2 in \cite{wang2015global}. First, by \textbf{C1} and \textbf{C2}, we know that the sequence generated by the proposed algorithm is bounded and exists a convergent subsequence $\{\ibx^{k_s}, \ibu^{k_s}, \ibw^{k_s}\}$ $(s\in \mathbb{N})$. When $s\to +\infty$, we have $\{\ibx^{k_s}, \ibu^{k_s}, \ibw^{k_s}\}\to\{\ibx^{\ast}, \ibu^{\ast}, \ibw^{\ast}\}$.
Besides, from \textbf{C1} and \textbf{C2} we also see that $L(\ibx^{k},\ibu^{k},\ibw^{k})$ is monotonically non-increasing and lower bounded, i.e., for $k\to \infty$, we have $\|\ibx^{k+1}-\ibx^k\|_2^2\to0$, $\|\ibu^{k+1}-\ibu^k\|_2^2\to0$ and $\|\ibw^{k+1}-\ibw^k\|_2^2\to0$. Combining with \textbf{C3}, we can further deduce that $\|\ibd^{k+1}\|_2^2\to0$ for any $\ibd^{k+1}\in \partial L(\ibx^{k+1},\ibu^{k+1},\ibw^{k+1})$ as $k\to\infty$, i.e. $0 \in \partial L(\ibx^{*},\ibu^{*},\ibw^{*})$. In other words, any limit point $\{\ibx^{*}, \ibu^{*}, \ibw^{*}\}$ is a stationary point.

If $\rho>\frac{1}{\gamma}$ for MCP in~\eqref{MCP}, $L(\ibx^{k},\ibu^{k},\ibw^{k})$ is convex, it has a unique global optimal solution. Thus, the proposed method has global convergence.
$\blacksquare$

Then, we discuss the proof of three conditions \textbf{C1}, \textbf{C2} and \textbf{C3} for our proposed method.
%

\textbf{Proof of C1:}
The Lagrangian in \eqref{eq:Lagr_meth2} can be rewritten as
\beq\label{eq:Lagr_meth2_1}
L(\ibx, \ibu, \ibw) = \psi(\ibx)+P_{\lambda,\gamma}(\ibu) + \frac{\rho}{2} \left\|\ibx-\ibu+\frac{\ibw}{\rho} \right\|_2^2-\frac{\|\ibw\|_2^2}{2\rho},
\eeq
where $\psi(\ibx)=\|\ibb-\bA\ibx\|_2^2$.
From \eqref{eq:Lagr_meth2_1}, we see $\frac{\partial^2 L(\ibx, \ibu, \ibw)}{\partial \ibx^2}=2\bA^\mathrm{T}\bA+\rho\bI$ is positive definite. Hence the Lagrangian is strongly convex with respect to $\ibx$. Based on the definition of strongly convex function, we have
\beq \label{eq:1st}
L(\ibx^{k+1}, \ibu^{k+1}, \ibw^{k})-L(\ibx^{k}, \ibu^{k+1}, \ibw^{k}) \nonumber \\ \leq -\frac{a}{2}\|\ibx^{k+1}-\ibx^{k}\|_2^2,
\eeq
where $a>0$.

Equation \eqref{meth1_x} implies
\beq
\nabla \psi(\ibx^{k+1})+\ibw^{k} +\rho(\ibx^{k+1}-\ibu^{k+1})=0.
\eeq
Direct combination of it with \eqref{meth1_w} yields
\beq \label{eq:func1}
\nabla\psi(\ibx^{k+1})=-\ibw^{k+1},
\eeq
and
\beq \label{eq:func2}
\ibx^{k+1}-\ibu^{k+1} = \frac{1}{\rho}\left(\ibw^{k+1}-\ibw^k\right).
\eeq
Thus, based on \eqref{eq:func1} and \eqref{eq:func2} we obtain
\beq \label{eq:2nd}
&&L(\ibx^{k+1}, \ibu^{k+1}, \ibw^{k+1})-L(\ibx^{k+1}, \ibu^{k+1}, \ibw^{k}) \nonumber \\
&=& \left(\ibw^{k+1}-\ibw^{k}\right)^\mathrm{T}\left(\ibx^{k+1}-\ibu^{k+1}\right) \nonumber \\ &=&\frac{1}{\rho}\|\ibw^{k+1}-\ibw^{k}\|_2^2
=\frac{1}{\rho}\|-\nabla\psi(\ibx^{k+1})+\nabla\psi(\ibx^{k})\|_2^2 \nonumber \\
&\leq & \frac{l_{\psi}^2}{\rho}\|\ibx^{k+1}-\ibx^{k}\|_2^2,
\eeq
where $l_{\psi}$ is a Lipschitz constant of $\psi(\ibx)$, and the last inequality is due to the fact that $\psi(\ibx)$ has Lipschitz continuous gradient.

Finally, from \eqref{meth2_u} we see that
\beq \label{eq:3rd}
L(\ibx^{k}, \ibu^{k+1}, \ibw^{k})-L(\ibx^{k}, \ibu^{k}, \ibw^{k})\leq 0
\eeq
Combining \eqref{eq:1st}, \eqref{eq:2nd} and \eqref{eq:3rd}, results in
\beq \label{eq:suff_desc}
&&L(\ibx^{k+1}, \ibu^{k+1}, \ibw^{k+1})-L(\ibx^{k}, \ibu^{k}, \ibw^{k}) \nonumber \\
&=&L(\ibx^{k+1}, \ibu^{k+1}, \ibw^{k+1})-L(\ibx^{k+1}, \ibu^{k+1}, \ibw^{k}) \nonumber \\
&&+L(\ibx^{k+1}, \ibu^{k+1}, \ibw^{k})-L(\ibx^{k}, \ibu^{k+1}, \ibw^{k}) \nonumber \\
&&+L(\ibx^{k}, \ibu^{k+1}, \ibw^{k})-L(\ibx^{k}, \ibu^{k}, \ibw^{k}) \nonumber \\
&\leq& \left(\frac{l_{\psi}^2}{\rho}-\frac{a}{2}\right)\|\ibx^{k+1}-\ibx^{k}\|_2^2.
\eeq
Let $\tau_1=\frac{a}{2}-\frac{l_{\psi}^2}{\rho}$ in \textbf{C1}, when $\rho>\frac{2l_{\psi}^2}{a}$, \textbf{C1} is satisfied. $\blacksquare$


\textbf{Proof of C2:}
First, we prove that $L(\ibx^{k}, \ibu^{k}, \ibw^{k})$ is lower bounded for any $k$.
The proof is mainly based on the following lemma.

\textbf{Lemma 1} (Descent lemma) $\nabla \psi$ is $l_{\psi}$-Lipschitz continuous, for any two point $\ibx^k$ and $\ibu^k$,
\beq
\psi(\ibu^k)-\psi(\ibx^k)\leq\nabla\psi(\ibx^k)^\mathrm{T}(\ibu^k-\ibx^k) + \frac{l_{\psi}}{2}\|\ibu^k-\ibx^k\|_2^2.
\eeq

\textbf{Proof:} The proof of Lemma 1 is provided in Appendix~\ref{append:1}. $\blacksquare$

Then, according to Lemma 1, we deduce that
\beq\label{eq:Lagr_meth2_2}
L(\ibx^k, \ibu^k, \ibw^k)=\psi(\ibx^k)+P_{\lambda,\gamma}(\ibu^{k}) +{\ibw^k}^\mathrm{T}(\ibx^k-\ibu^k)\nonumber\\
+ \frac{\rho}{2} \left\|\ibx^k - \ibu^k\right\|_2^2, \nonumber\\
\geq \psi(\ibu^k)+P_{\lambda,\gamma}(\ibu^{k})
+\left(\frac{\rho}{2}-\frac{l_{\psi}}{2}\right)\|\ibu^k-\ibw^k\|_2^2.
\eeq

Obviously, if $\rho\geq l_{\psi}$, $\psi(\ibu^k)+P_{\lambda,\gamma}(\ibu)+\left(\frac{\rho}{2}- \frac{l_{\psi}}{2}\right)\|\ibu^k-\ibw^k\|_2^2 >-\infty$ for any $k$ and $L(\ibx^k, \ibu^k, \ibw^k)$ is lower bounded. According to the proof of \textbf{C1}, we know that $L(\ibx^k, \ibu^k, \ibw^k)$ is sufficient descent. Hence $L(\ibx^k, \ibu^k, \ibw^k)$ is upper bounded by $L(\ibx^0, \ibu^0, \ibw^0)$.

Next, we prove the sequence $\{\ibx^{k}, \ibu^{k}, \ibw^{k}\}$ is also bounded. From \eqref{eq:suff_desc}, we have
\beq \label{eq:wk1_wk}
&&\|\ibx^{k+1}-\ibx^k\|_2^2\nonumber \\ &&\leq\frac{1}{\tau_1}\left(L(\ibx^k, \ibu^k, \ibw^k)- L(\ibx^{k+1}, \ibu^{k+1}, \ibw^{k+1})\right), \nonumber
\eeq
then
\beq \label{eq:wk1_wk}
&& \sum_{k=0}^{l} \|\ibx^{k+1}-\ibx^k\|_2^2\nonumber \\ &&\leq\frac{1}{\tau_1}\left(L(\ibx^0, \ibu^0, \ibw^0)- L(\ibx^{l+1}, \ibu^{l+1}, \ibw^{l+1})\right) \nonumber \\
&&< \infty.
\eeq
If $l\to\infty$, we still have $\sum_{k=0}^{\infty}\|\ibx^{k+1}-\ibx^k\|_2^2 < \infty$. Thus $\ibx^{k}$ is bounded.

From \eqref{eq:2nd}, we know
\beq
\|\ibw^{k+1}-\ibw^{k}\|_2^2 \leq l_{\psi}^2 \|\ibx^{k+1}-\ibx^{k}\|_2^2. \nonumber
\eeq
Hence it is deduced that
\beq
\sum_{i=0}^{\infty} \|\ibw^{k+1}-\ibw^k\|_2^2<\infty.
\eeq

In addition, according to \eqref{eq:func2}, we have
\beq
&&\|\ibu^{k+1}-\ibu^k\|_2^2\nonumber \\
&=&\|\ibx^{k+1}-\ibx^k-\frac{1}{\rho} \left(\ibw^{k+1}-\ibw^k\right)+\frac{1}{\rho} \left(\ibw^{k-1}-\ibw^k\right)\|_2^2 \nonumber \\
&\leq& 2\|\ibx^{k+1}-\ibx^k\|_2^2+\frac{2}{\rho^2}\|\ibw^{k+1}-\ibw^k\|_2^2\nonumber\\ &&+\frac{2}{\rho^2}\|\ibw^{k}-\ibw^{k-1}\|_2^2.
\eeq
Thus
\beq
\sum_{i=1}^{\infty} \|\ibu^{k+1}-\ibu^k\|_2^2 < \infty.
\eeq
So the sequence $\{\ibx^{k}, \ibu^{k}, \ibw^{k}\}$ is bounded. $\blacksquare$


\textbf{Proof of C3}:
\beq
&&\left. \frac{\partial L}{\partial \ibx} \right|_{(\ibx^{k+1}, \ibu^{k+1}, \ibw^{k+1})} \nonumber \\
&=&\nabla \psi(\ibx^{k+1})+\rho\left(\ibx^{k+1}-\ibu^{k+1}\right)+\ibw^{k+1} \nonumber \\
&=&\ibw^{k+1}-\ibw^{k}, \label{eq:dldw}\\
&&\left. \frac{\partial L}{\partial \ibu} \right|_{(\ibx^{k+1}, \ibu^{k+1}, \ibw^{k+1})} \nonumber \\
&=&\partial P_{\lambda,\gamma}(\ibu^{k+1})-\rho\left(\ibx^{k+1}-\ibu^{k+1}\right)-\ibw^{k+1} \nonumber \\
&\owns&\rho\left(\ibx^{k}-\ibx^{k+1}\right)+\ibw^{k}-\ibw^{k+1}, \label{eq:dldu}\\
&&\left. \frac{\partial L}{\partial \ibw} \right|_{(\ibx^{k+1}, \ibu^{k+1}, \ibw^{k+1})} \nonumber \\
&=&\ibx^{k+1}-\ibu^{k+1}=\frac{1}{\rho}\left(\ibw^{k+1}-\ibw^{k}\right), \label{eq:dldupsi}
\eeq
where the second equality in \eqref{eq:dldu} is based on
\beq
0\in\partial P_{\lambda,\gamma}(\ibu^{k+1})-\ibw^{k}-\rho\left(\ibx^k-\ibu^{k+1}\right),
\eeq
which can be deduced from \eqref{meth2_u}.
Thus there exists
\beq
\bd^{k+1}&:=&
\left[\begin{array}{c}
\ibw^{k+1}-\ibw^{k} \\
\rho\left(\ibx^{k}-\ibx^{k+1}\right)+\ibw^{k}-\ibw^{k+1}\\
\frac{1}{\rho}\left(\ibw^{k+1}-\ibw^{k}\right)
\end{array}\right] \nonumber\\
&\in&\partial L\left(\ibx^{k+1}, \ibu^{k+1}, \ibw^{k+1}\right)
\eeq
Combining it with \eqref{eq:2nd}, we deduce that there exists $\tau_2>0$ such that
\beq
\|\bd^{k+1}\|_2^2 \leq \tau_2\|\ibx^{k+1}-\ibx^k\|_2^2.
\eeq $\blacksquare$

In summary, for MCP,  if the $\rho>\max\{\frac{1}{\gamma}, \frac{2l_{\psi}^2}{a}, l_{\psi}\}$, the above three conditions \textbf{C1}-\textbf{C3} are satisfied. 
Thus, we can say that the proposed method has global convergence.

Then we analyze a more general case when $u_i$ is generated by the approximate solution in \eqref{meth2_ui_mcp} and there is not a fixed relationship between $\rho$ and $\frac{1}{\gamma}$.
In this case, under some assumptions, we can still prove the global convergence of our proposed method.

\textbf{Theorem 2} Assume that $\rho>\max\{\frac{2l_{\psi}^2}{a}, l_{\psi}\}$ and the approximate solution $\ibu^{k+1}$ in \eqref{meth2_ui_mcp} is not worse than the solution in the last iteration $\ibu^{k}$, i.e.,
\beq
L(\ibx^{k}, \ibu^{k+1}, \ibw^{k})-L(\ibx^{k}, \ibu^{k}, \ibw^{k})\leq 0.
\eeq
Then the proposed method satisfies  \textbf{C1}, \textbf{C2} and \textbf{C3}. Further, if the Lagrangian in \eqref{eq:Lagr_meth2} is a Kurdyka-{\L}ojasiewicz (K{\L}) function, then the corresponding sequence $\{\ibx^k, \ibu^k, \ibw^k \}$ globally converges to a unique stationary point $\{\ibx^*, \ibu^*, \ibw^* \}$.

\textbf{Proof:}
\eqref{meth2_ui_mcp} is an approximate optimal solution to function \eqref{meth2_u}, hence
\beq
L(\ibx^{k}, \ibu^{k+1}, \ibw^{k})-L(\ibx^{k}, \ibu^{k}, \ibw^{k})\leq 0
\eeq
is a reasonable assumption. Combining it with $\rho>\max\{\frac{2l_{\psi}^2}{a}, l_{\psi}\}$, we prove that \textbf{C1}, \textbf{C2} and \textbf{C3} are satisfied. Thus, the sequence generated by the proposed algorithm can also converge to a stationary point.

Similar to the Proposition 2 in \cite{wang2015global}, we can claim the global convergence of the corresponding sequence $\{\ibx^k, \ibu^k, \ibw^k \}$ under the K{\L} assumption of its Lagrangian in~\eqref{eq:Lagr_meth2}. $\blacksquare$

Finally, we prove that the Lagrangian in~\eqref{eq:Lagr_meth2} is a K{\L} function. Before that, we need to point out several fundamental definitions.

For a function $f:\,\,\mathbb{R}^n\to\mathbb{R}$, if the domain of $f$ is not empty and it can never attain $-\infty$, then $f$ is proper. If
\beq
\liminf \limits_{\ibx\to\ibx_0}\,f(\ibx)\geq f(\ibx_0), \nonumber
\eeq
then $f$ is lower semi-continuous at a point $\ibx_0$. If at every point in its domain,
$f$ is lower semi-continuous, then $f$ is a lower semi-continuous function.

A subset $S$ of $\mathbb{R}^d$ is a real semi-algebraic set if there exists a finite number of real polynomial functions $l_{ij},\,\,h_{ij}: \mathbb{R}^d\to\mathbb{R}$ such that
\beq
S=\bigcup_{j=1}^{q_1}\bigcap_{i=1}^{q_2}\{\bz\in\mathbb{R}^d:\, l_{ij}(\bz)=0, \,h_{ij}(\bz)<0\}. \nonumber
\eeq
A function $g: \mathbb{R}^d\to(-\infty,\infty]$ is semi-algebraic if its graph
\beq
\{(\bz,t)\in\mathbb{R}^{d+1}:g(\bz)=t\}
\eeq
is a semi-algebraic set in $\mathbb{R}^{d+1}$.

The definition of K{\L} function is given by \cite{attouch2013convergence}. Here we use the following lemma to determine if a function is a K{\L} function.

\textbf{Lemma 2:} Let $f: \mathbb{R}^n\to(-\infty,\infty]$ be a proper and lower semi-continuous function. If $f$ is semi-algebraic, then it satisfies the K{\L} property at any point of its domain. In other words, $f$ is a K{\L} function.

\textbf{Proof:} The proof of Lemma 2 is given by \cite{bolte2006nonsmooth, bolte2007lojasiewicz}. $\blacksquare$

Apparently, the Lagrangian in~\eqref{eq:Lagr_meth2} is proper and continuous. Hence it is also a semi-continuous function. Besides, MCP function is obviously semi-algebraic. Thus the Lagrangian function in~\eqref{eq:Lagr_meth2} is also semi-algebraic. In summary, we see that the Lagrangian in~\eqref{eq:Lagr_meth2} is a K{\L} function. Further, the proposed method has global convergence.

\begin{figure}[htb]
\centering
\includegraphics[width=3.4in]{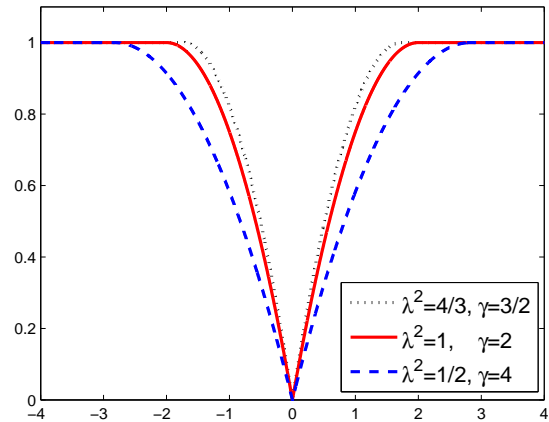}
\caption{MCP shapes under different parameter settings.}
\label{fig:mcpShape}
\end{figure}

\begin{figure*}[!ht]
\centering
\begin{tabular}{c@{\extracolsep{2mm}}c@{\extracolsep{2mm}}c@{\extracolsep{2mm}}c}
\mbox{\epsfig{figure=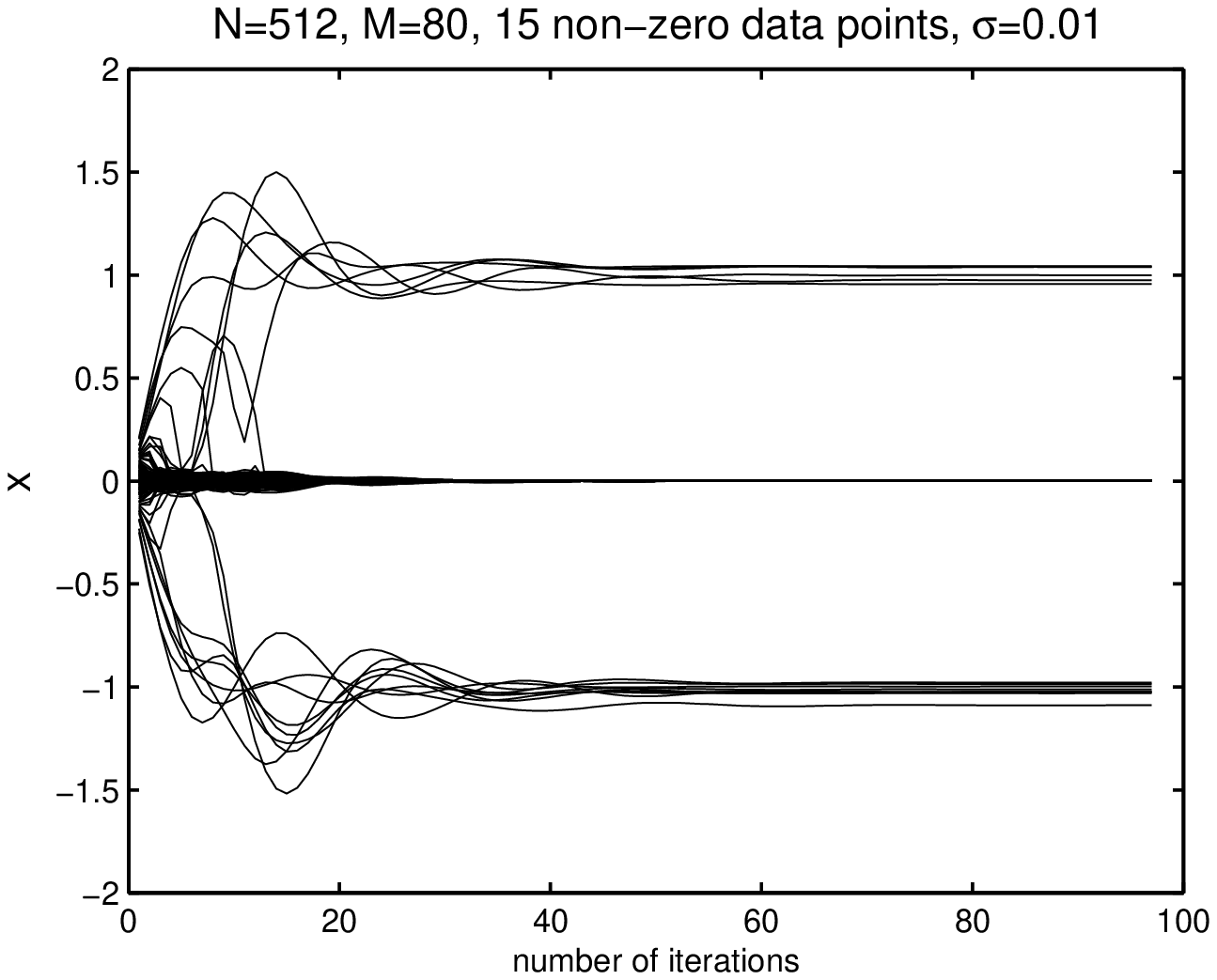,width=1.65in}} &
\mbox{\epsfig{figure=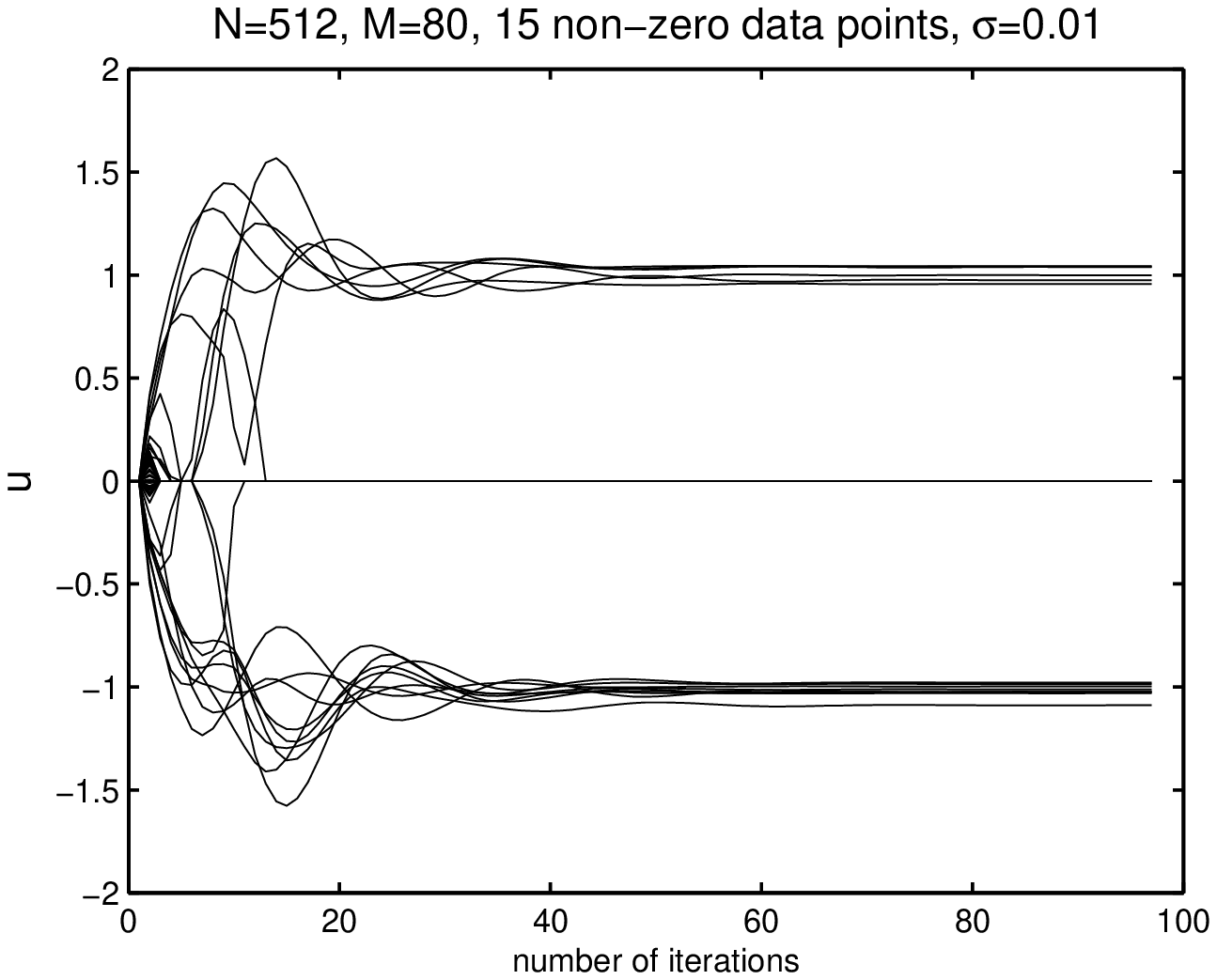,width=1.65in}} &
\mbox{\epsfig{figure=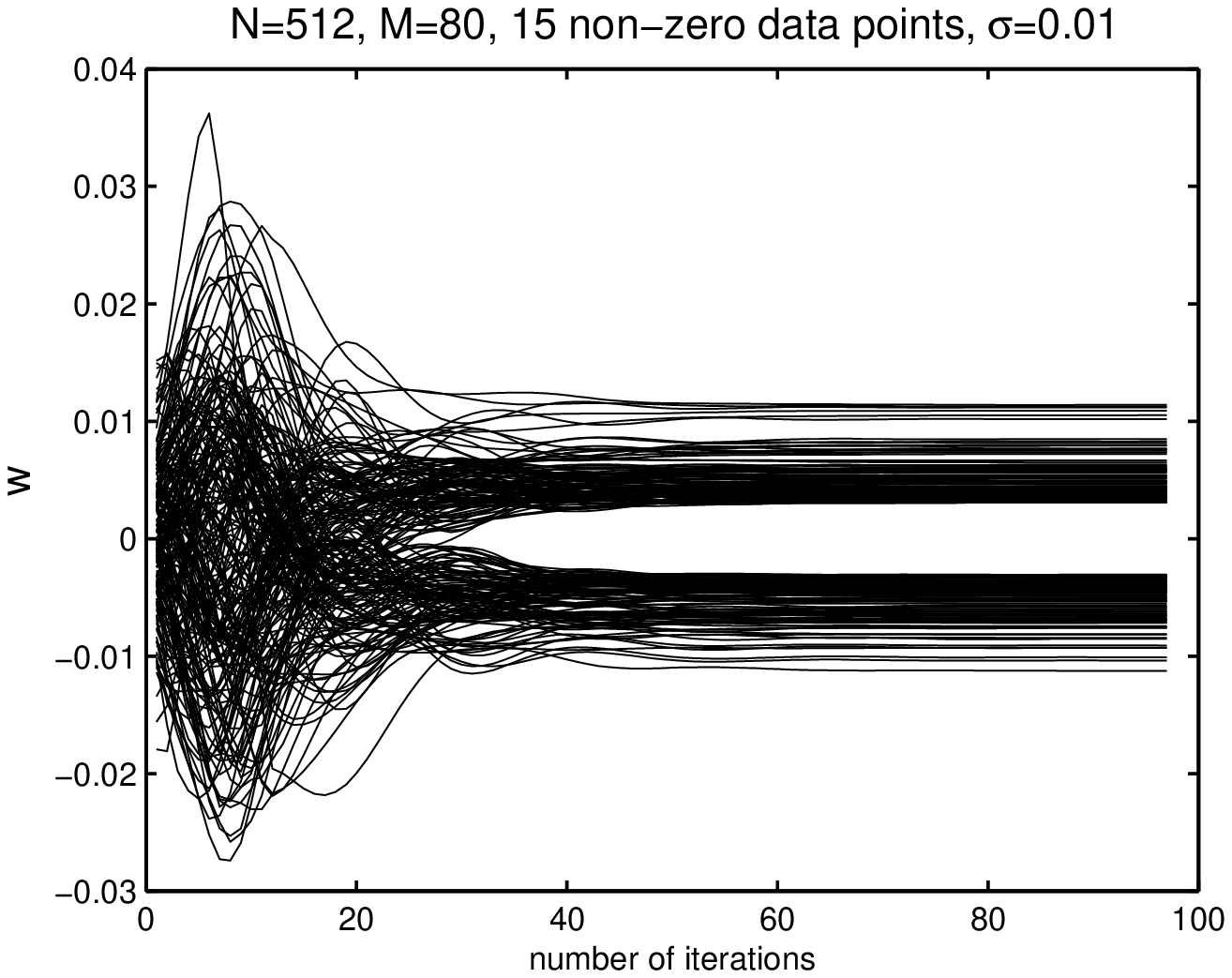,width=1.65in}} &
\mbox{\epsfig{figure=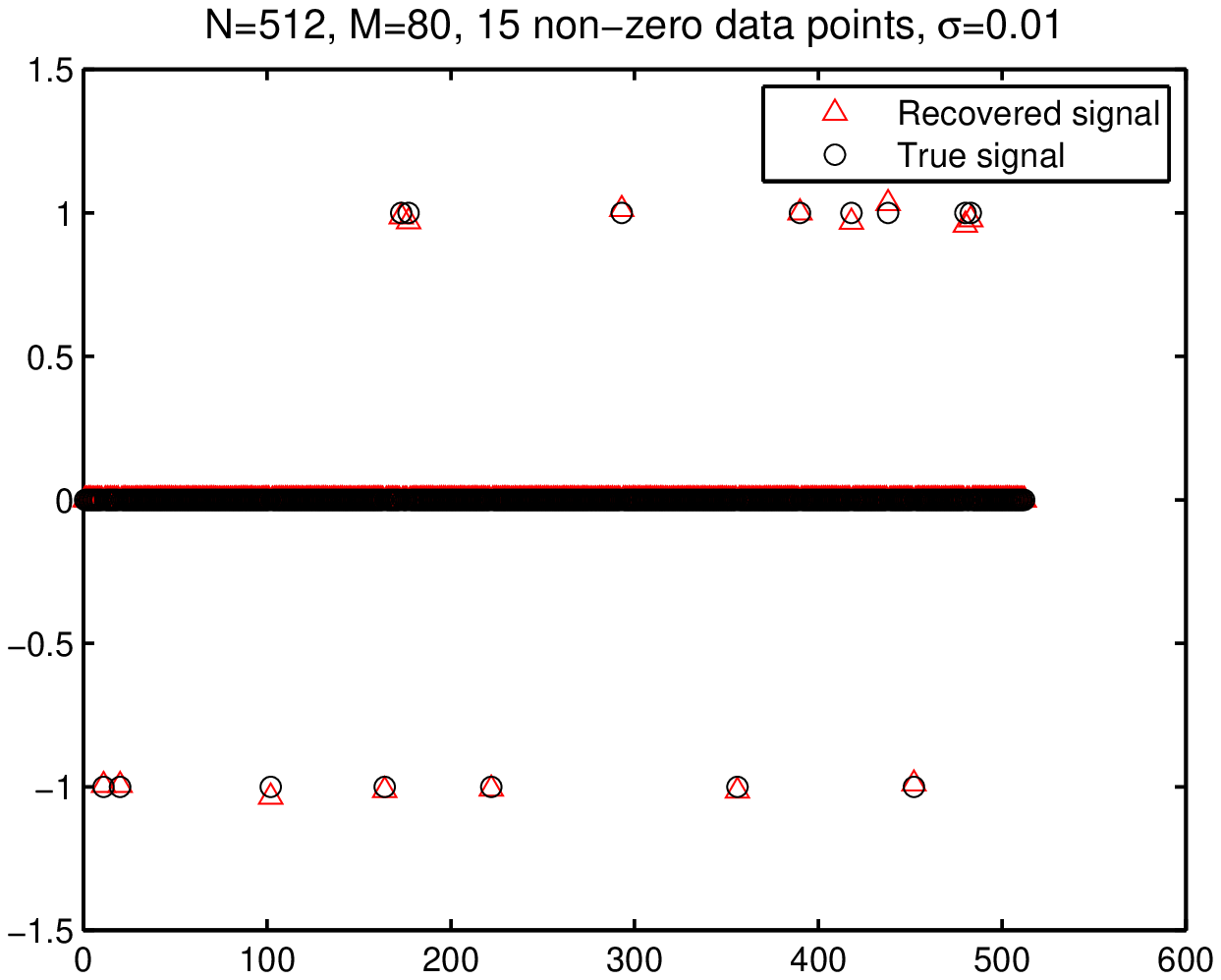,width=1.6in}} \\
\mbox{\epsfig{figure=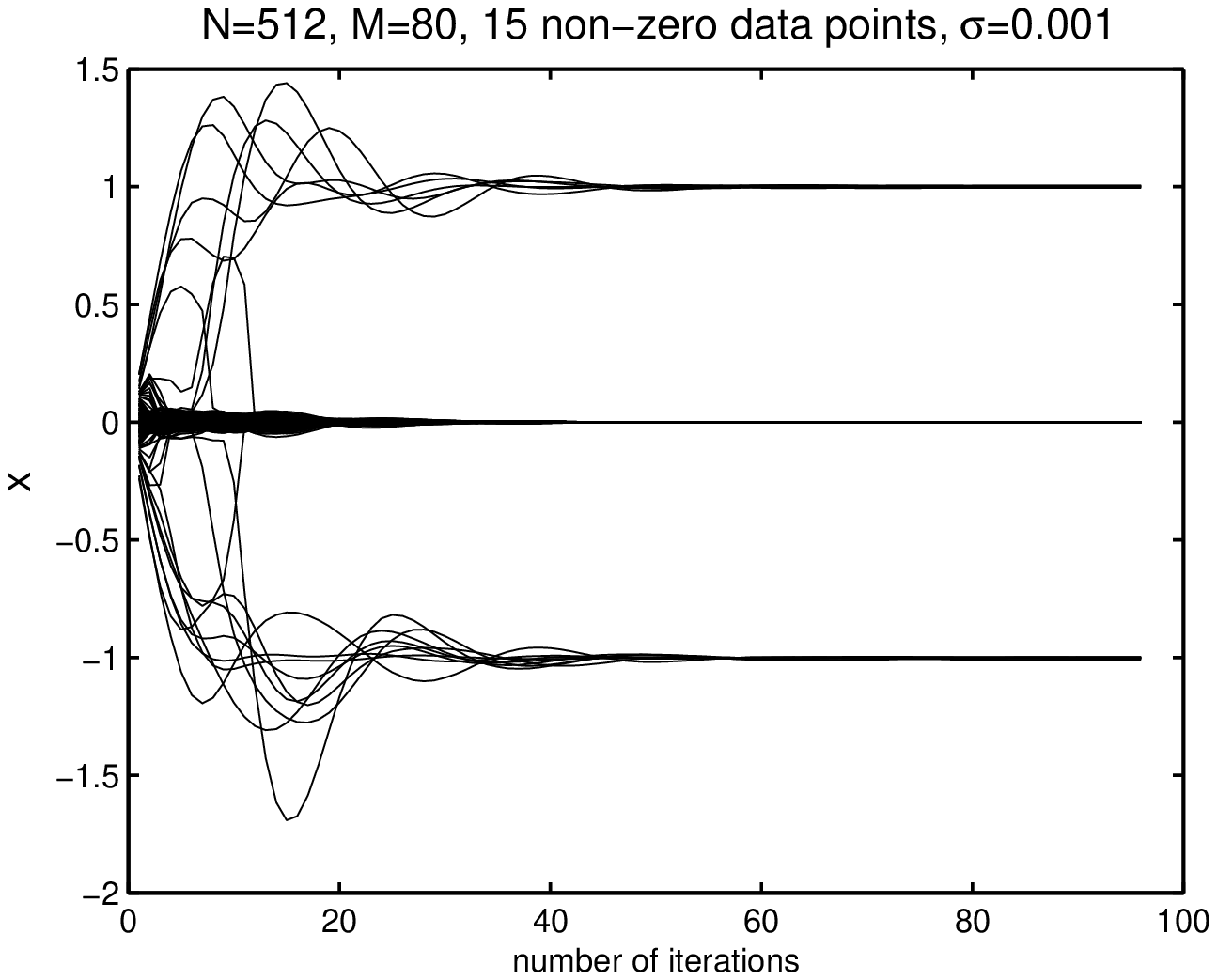,width=1.65in}} &
\mbox{\epsfig{figure=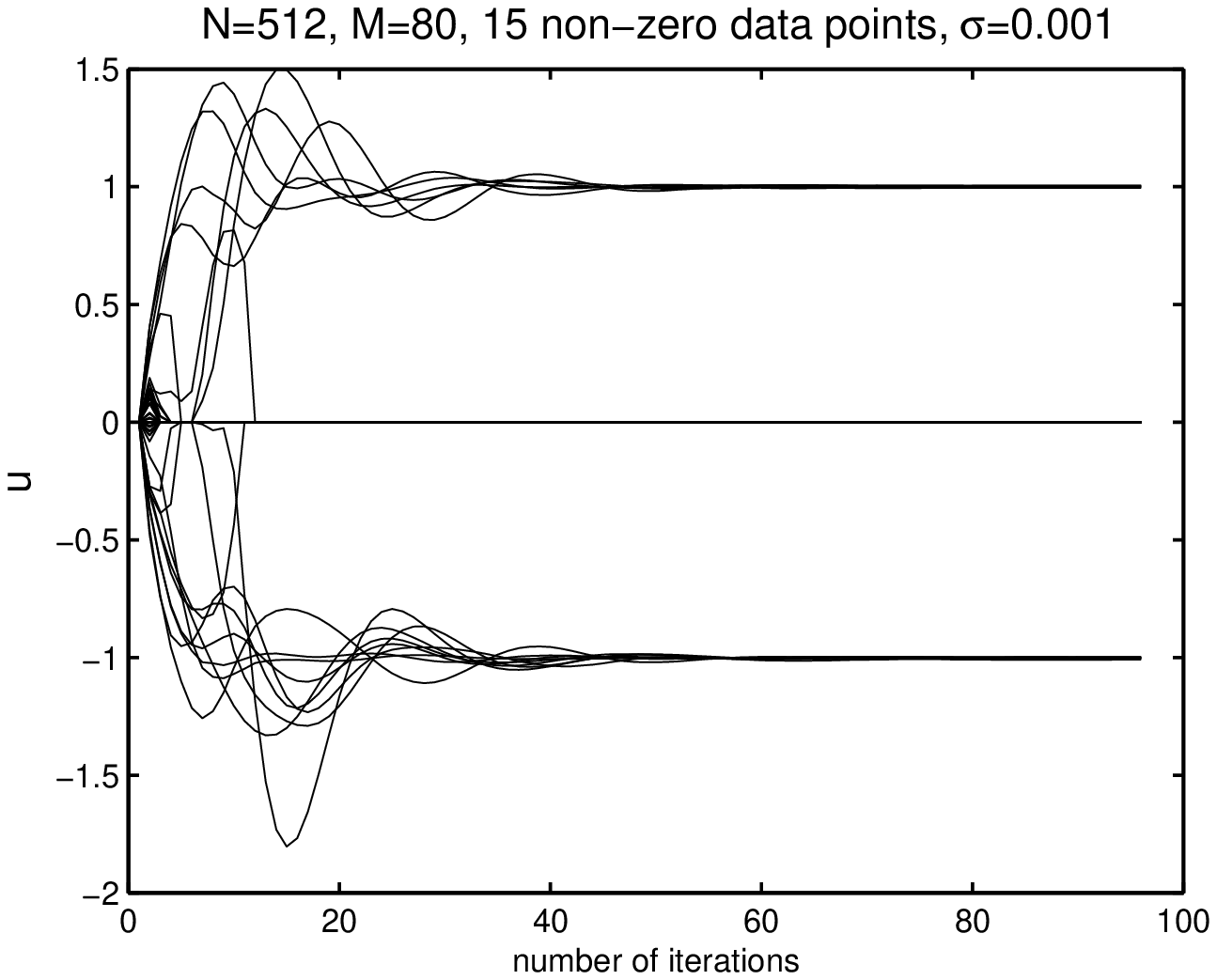,width=1.65in}} &
\mbox{\epsfig{figure=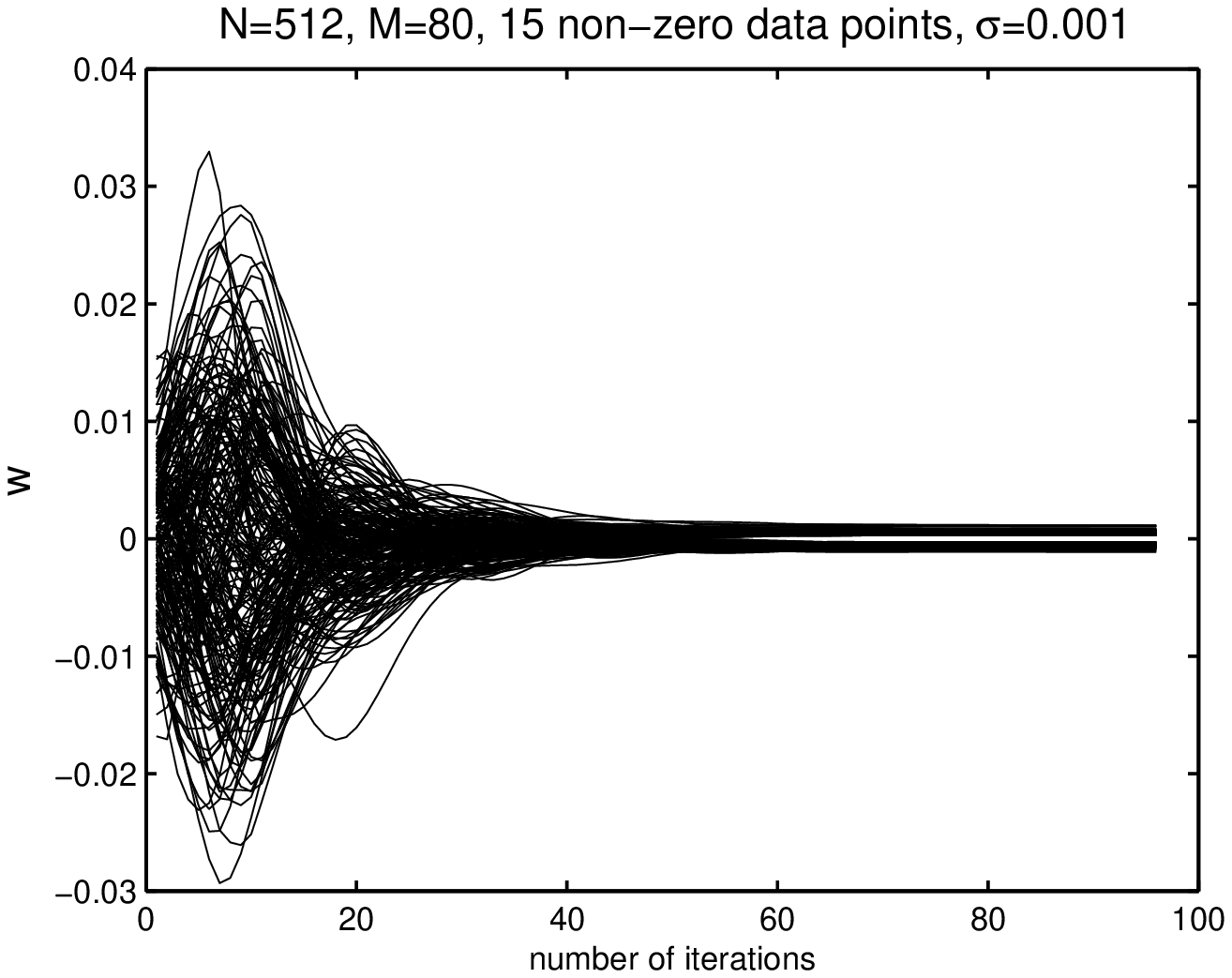,width=1.65in}} &
\mbox{\epsfig{figure=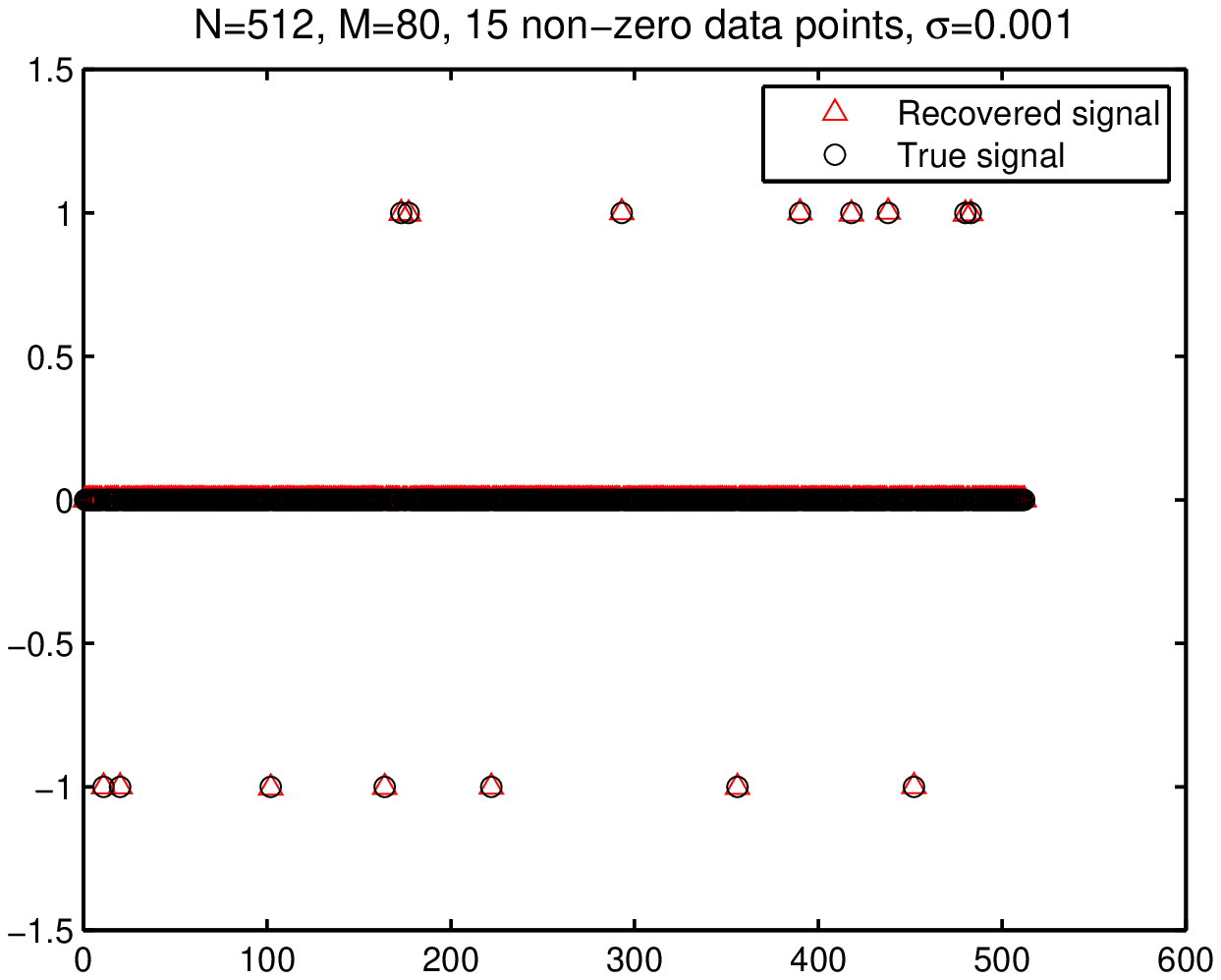,width=1.6in}} \\
\mbox{\epsfig{figure=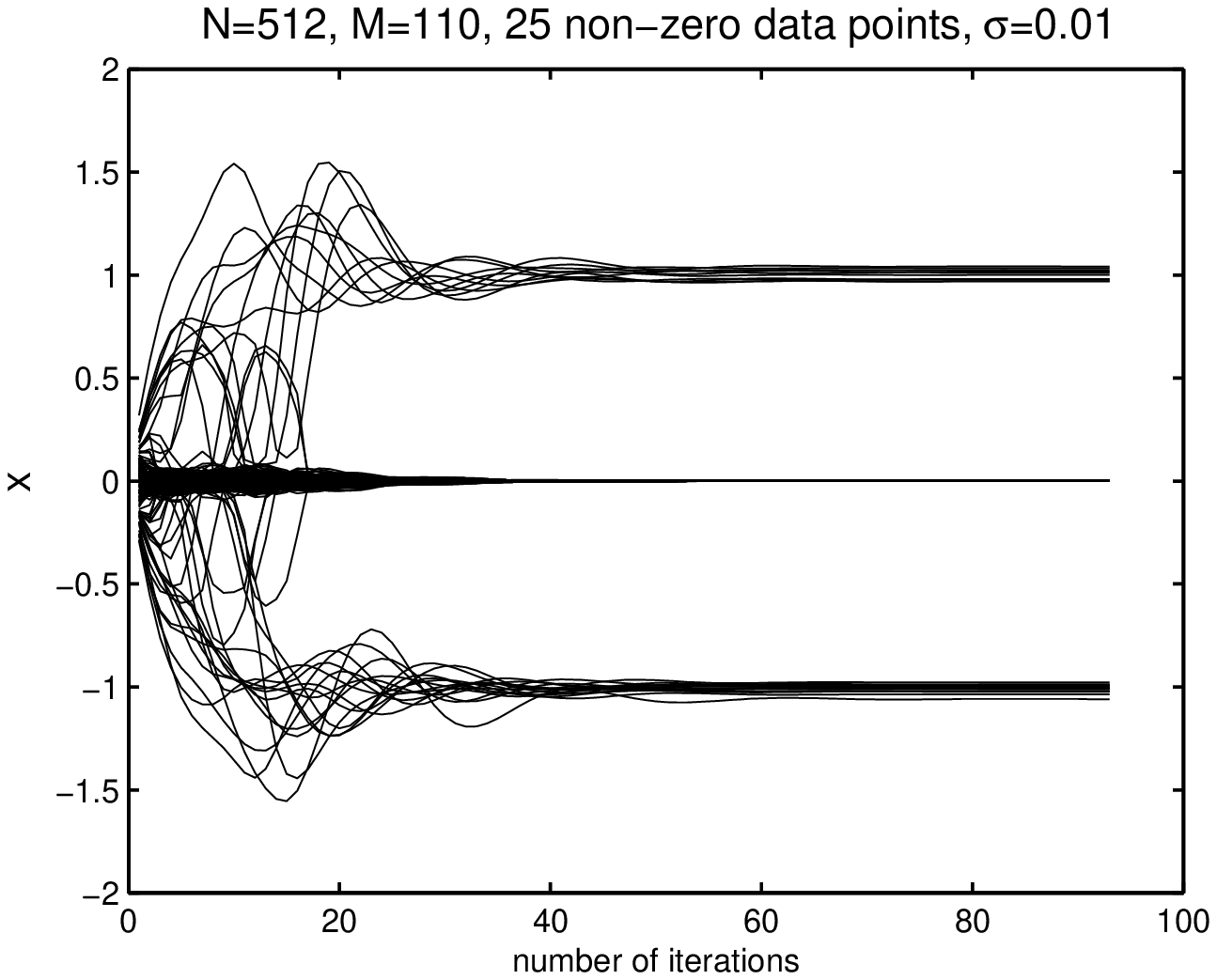,width=1.65in}} &
\mbox{\epsfig{figure=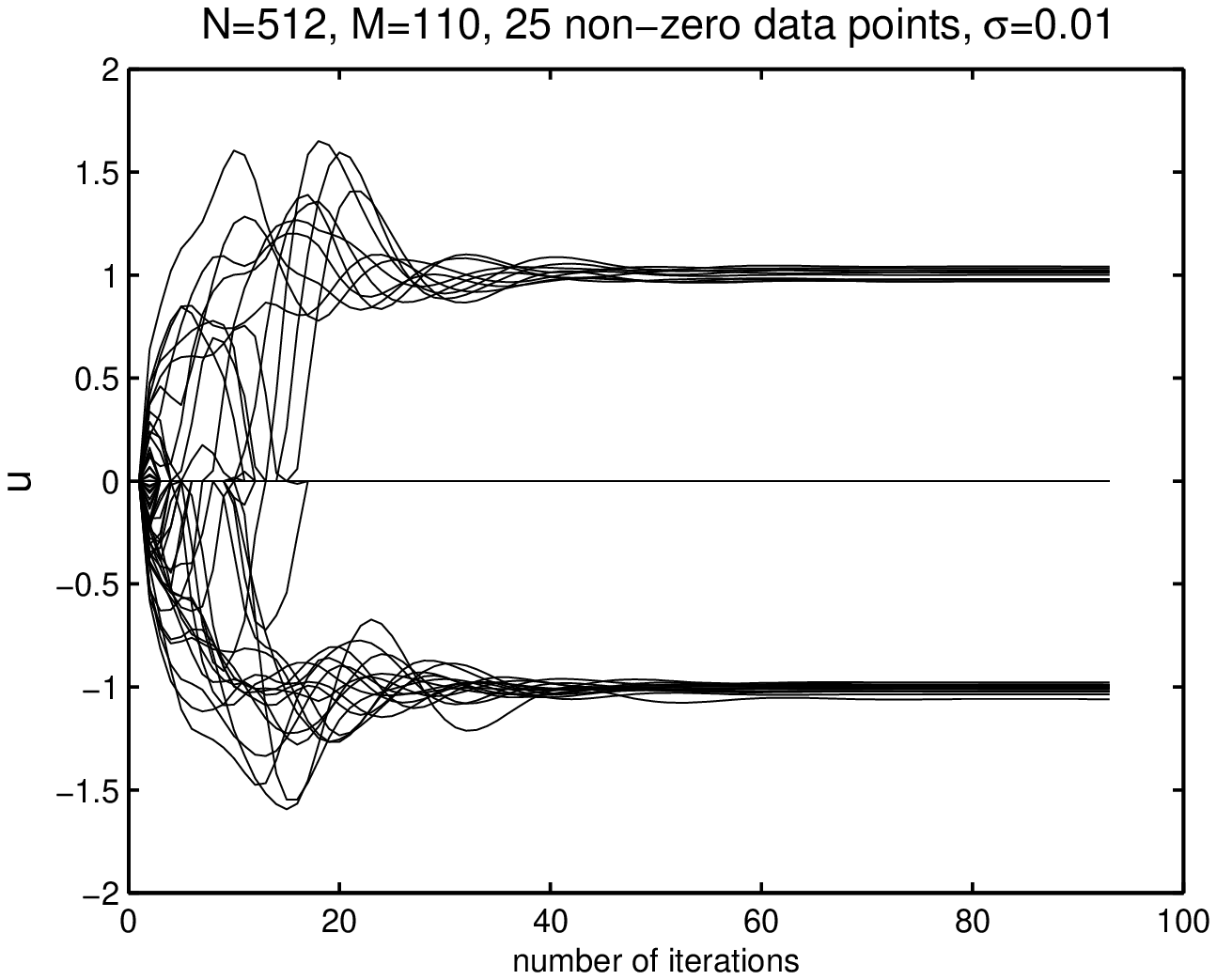,width=1.65in}} &
\mbox{\epsfig{figure=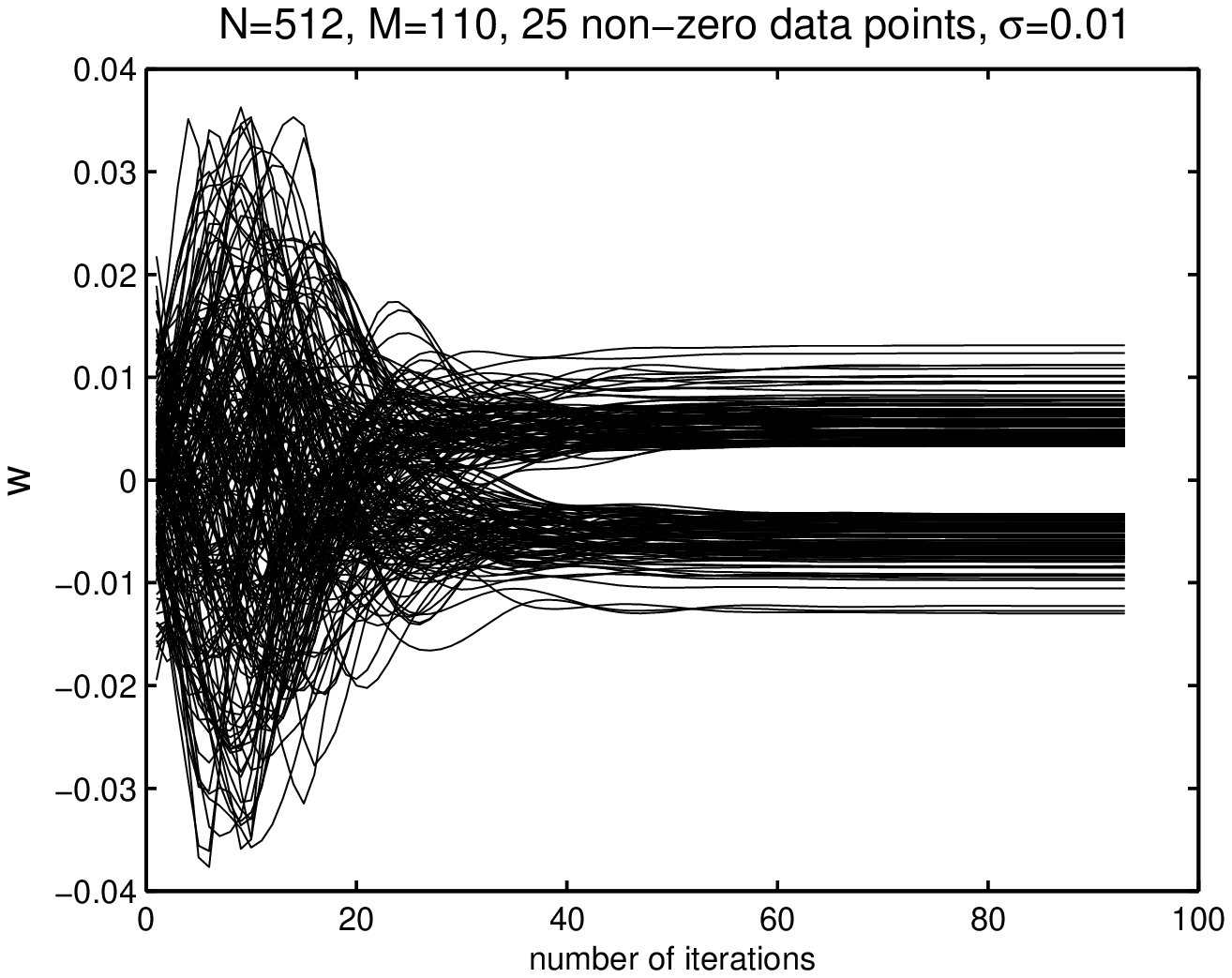,width=1.65in}} &
\mbox{\epsfig{figure=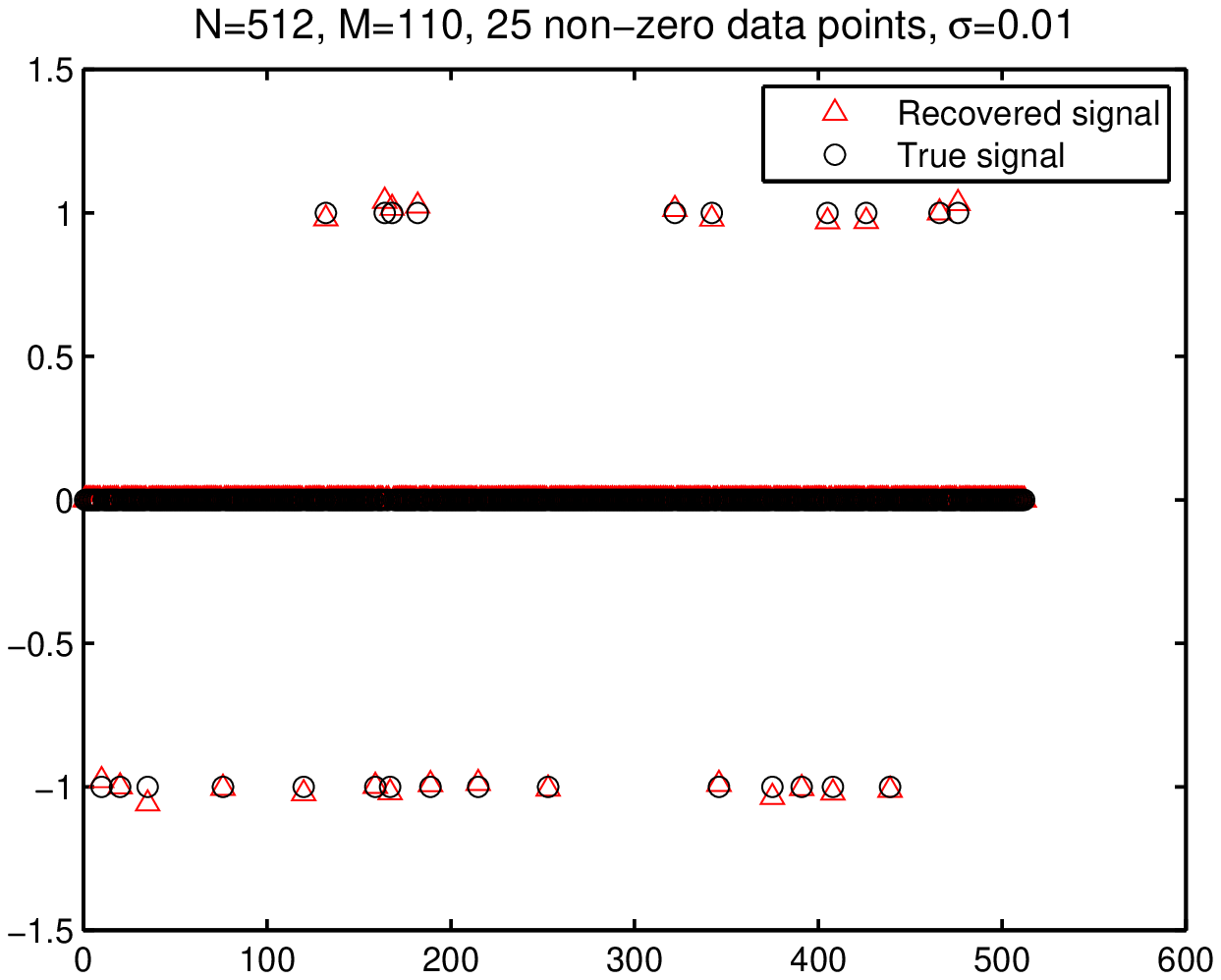,width=1.6in}} \\
\mbox{\epsfig{figure=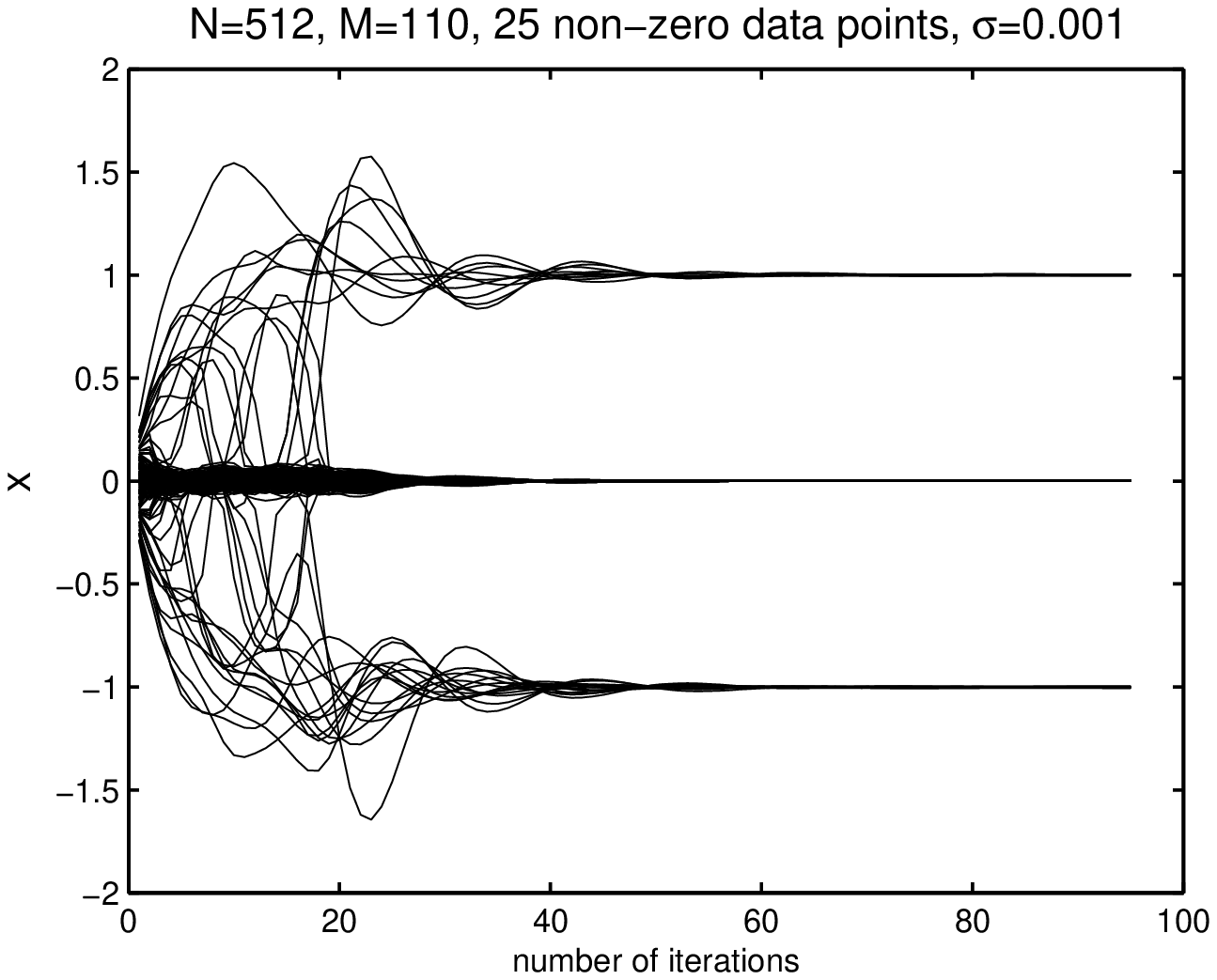,width=1.65in}} &
\mbox{\epsfig{figure=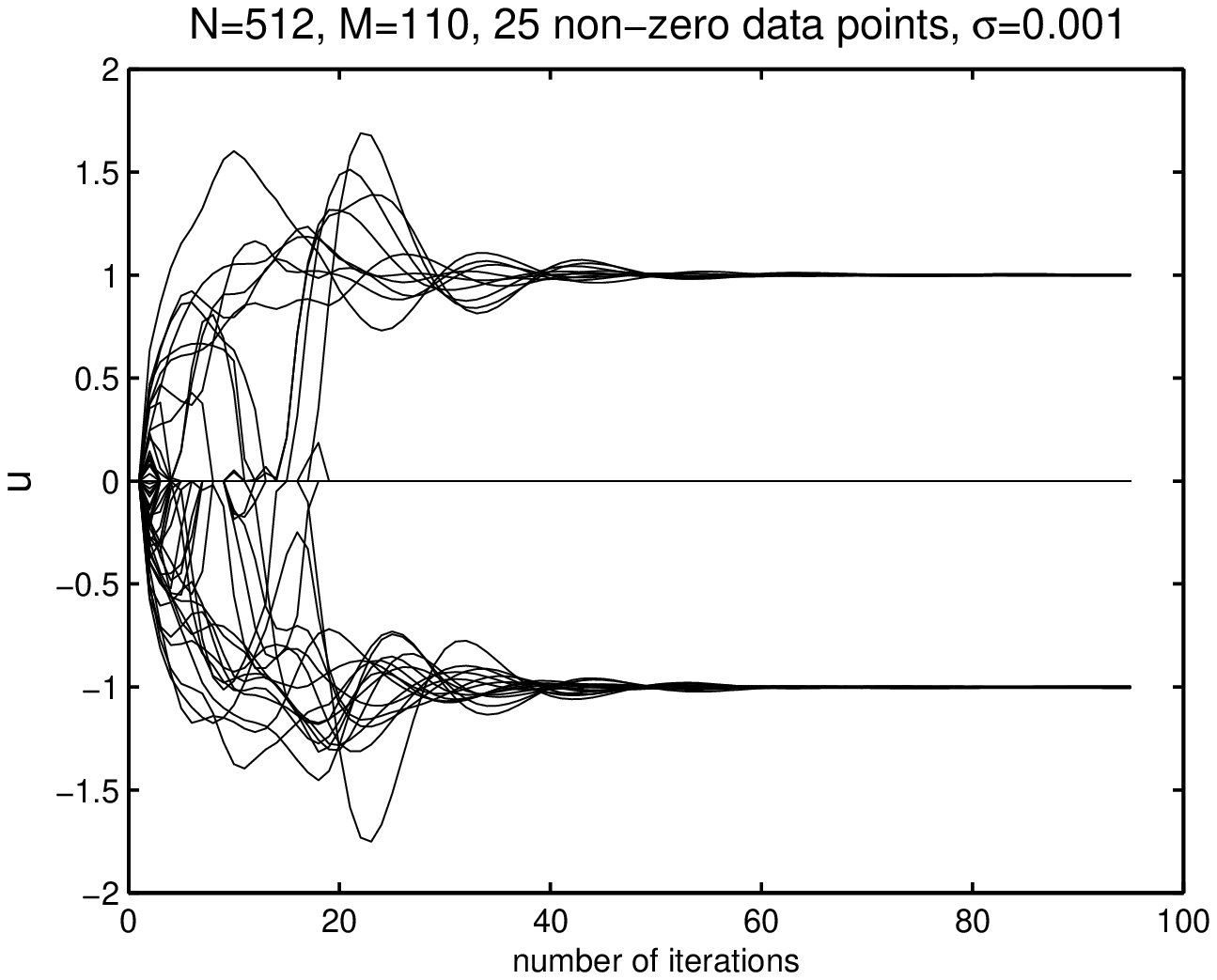,width=1.65in}} &
\mbox{\epsfig{figure=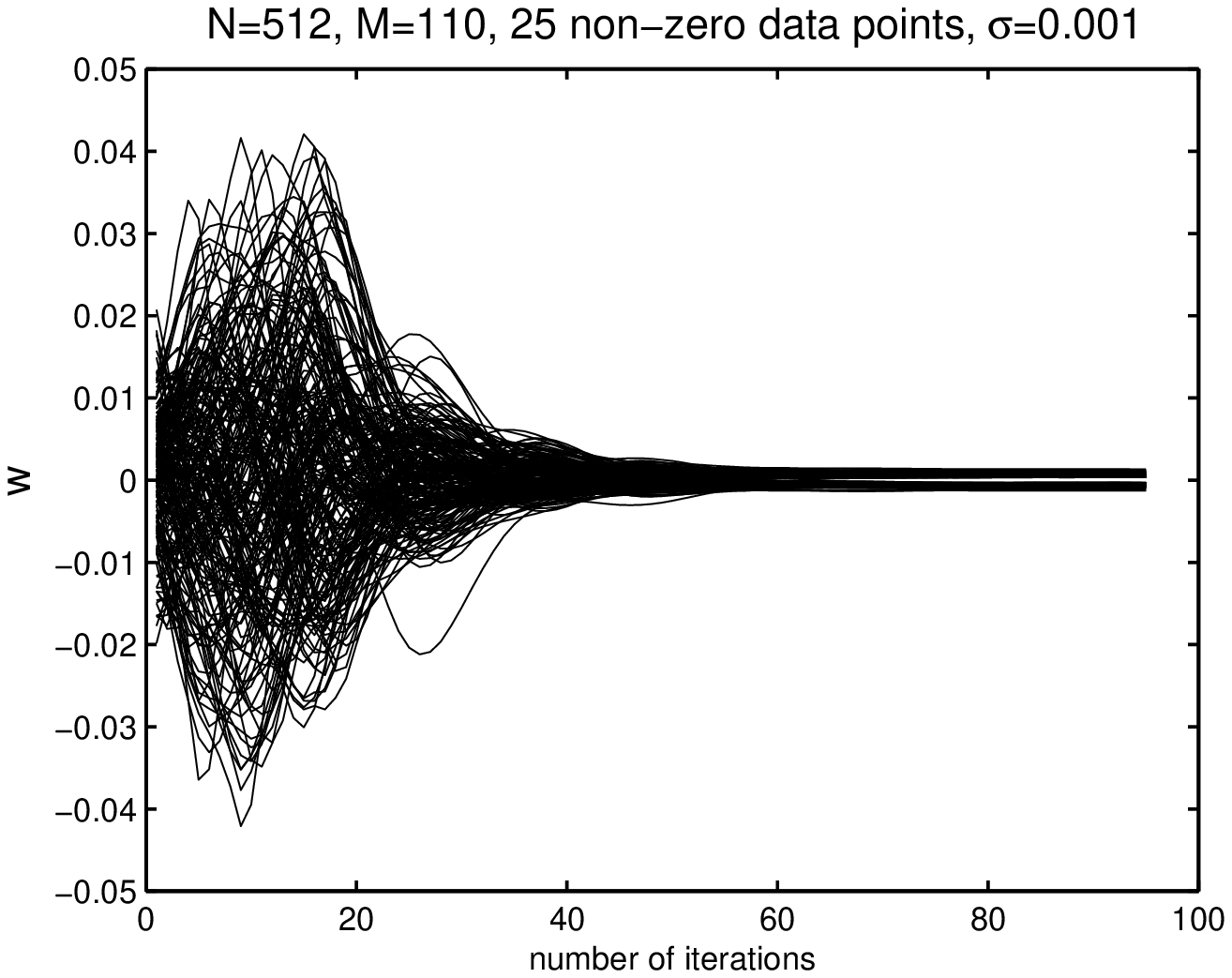,width=1.65in}} &
\mbox{\epsfig{figure=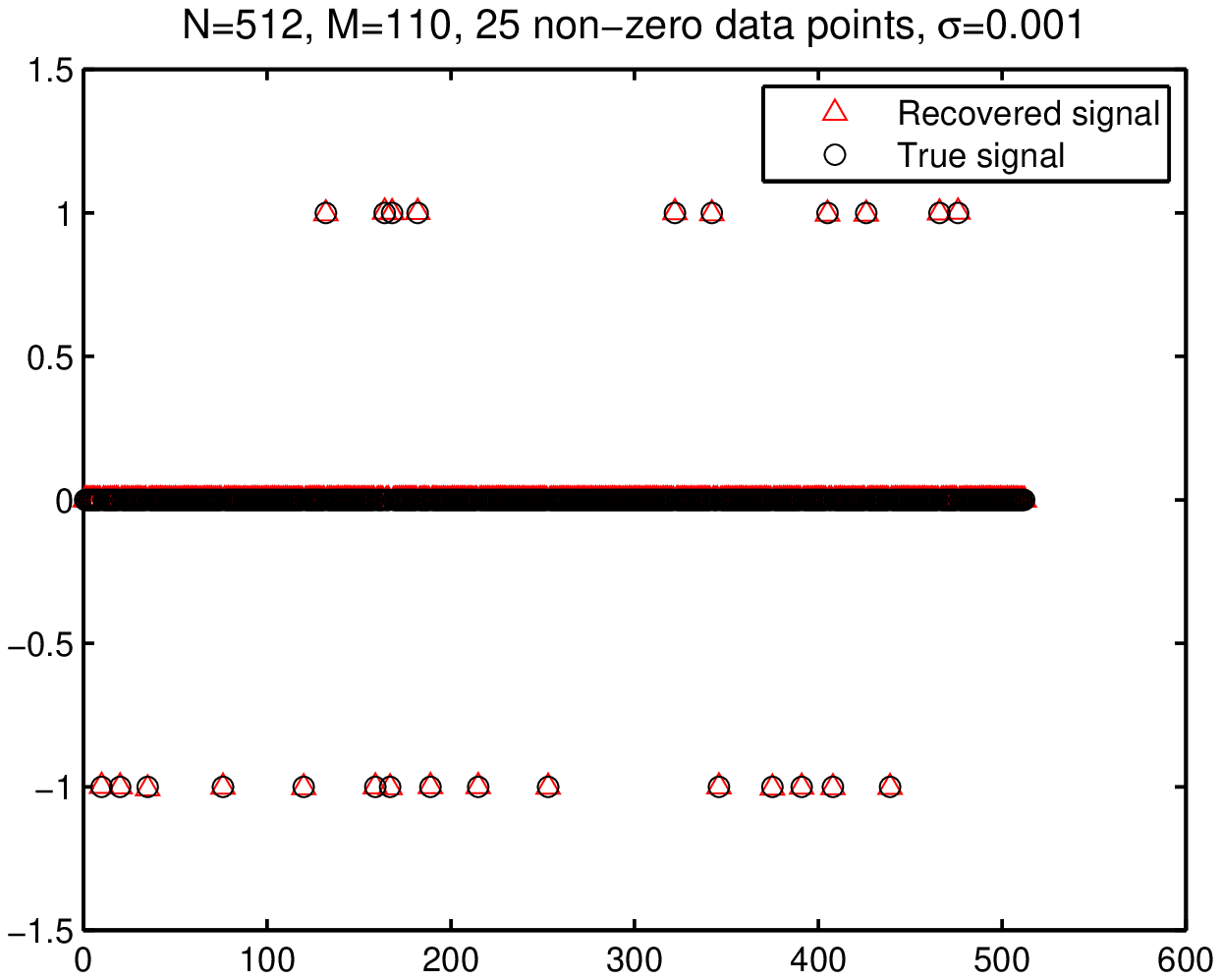,width=1.6in}} \\
\end{tabular}
\caption{Convergence of proposed method with $N=512$.}
\label{fig:convergence512}
\end{figure*}

\section{Parameter setting} \label{section5}
For our proposed method, there are three parameters which are respectively $\lambda$ and $\gamma$ in MCP function and the trade-off parameter $\rho$ in ADMM framework. All of them may influence the performance of the proposed method.

First, we illuminate the setting of parameters $\lambda$ and $\gamma$ for MCP. The shape of MCP is shown in Fig.~\ref{fig:mcpShape}. There are three different settings. For all of them, their thresholds $\frac{1}{2}\gamma\lambda^2$ are identical.
We can see that, for MCP function, the smaller $\gamma$ used, the steeper slope obtained. In other words, using smaller $\gamma$ can make the MCP function approximate $l_0$-norm better. Based on the definition of MCP in \eqref{MCP}, we have $\gamma>1$. Hence, we use $\gamma=\frac{3}{2}$ in our experiments.

Then, to find an appropriate value of $\lambda$, we propose two approaches. The first one is a trial and error method. For each trial, the applicable $\lambda$ is selected from a presupposed range $[10^{-2}, 10^{-1.9}, 10^{-1.8}, \dots, 10^{-0.1}]$. We implement the experiment with all different $\lambda$ in this range and select the $\lambda$ with the most sparse solution. If the most sparse solution is not unique, we choose the one with the smallest difference in the degree of sparsity between its neighbors.
Although this method can ensure the algorithm perform well, it should be carried out 20 times with different $\lambda$ for each trial. Hence, it consumes a large number of computing resources. To handle this issue, we propose the second method which can automatically regularize the parameter $\lambda$ during iteration. For each iteration, we let $\lambda=z^k_{\tau}/\gamma$, where $z^k_{\tau}$ is the $\tau$th largest element of $|\ibx^k+\ibw^k/\rho|$. Based on this scheme we can guarantee the largest $\tau$ elements of $\ibu$ are preserved and other elements will be compressed towards zero. It is similar with IHT and can be treated as a smooth hard threshold. Besides, the disadvantage of the IHT such as the result can only be calculated when the initial value close to the true signal is avoided. And, comparing with IHT, this method has similar or even better performance. 

For the trade-off parameter $\rho$, according to Section~\ref{section4}, for convergence, we must have $\rho>\max\{\frac{2l_{\psi}^2}{a}, l_{\psi}\}$. While, if $\rho$ is too large, the optimal solution of the proposed method will be far away from the true solution of problem \eqref{eq:QP}. Hence, the selection of $\rho$ is a trade-off problem, and in our experiments, we set $\rho=0.1$.

\begin{figure*}[!ht]
\centering
\begin{tabular}{c@{\extracolsep{2mm}}c@{\extracolsep{2mm}}c@{\extracolsep{2mm}}c}
\mbox{\epsfig{figure=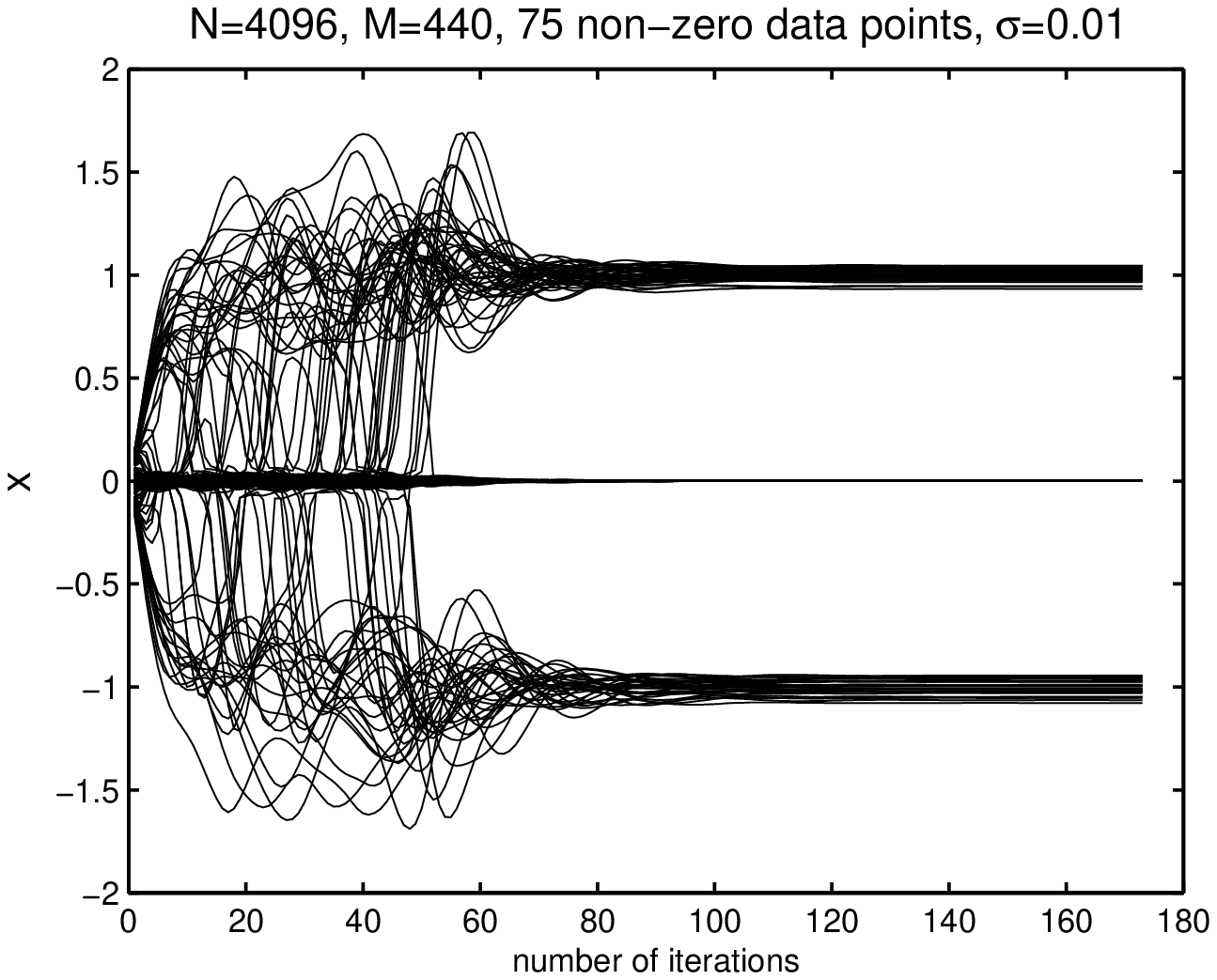,width=1.65in}} &
\mbox{\epsfig{figure=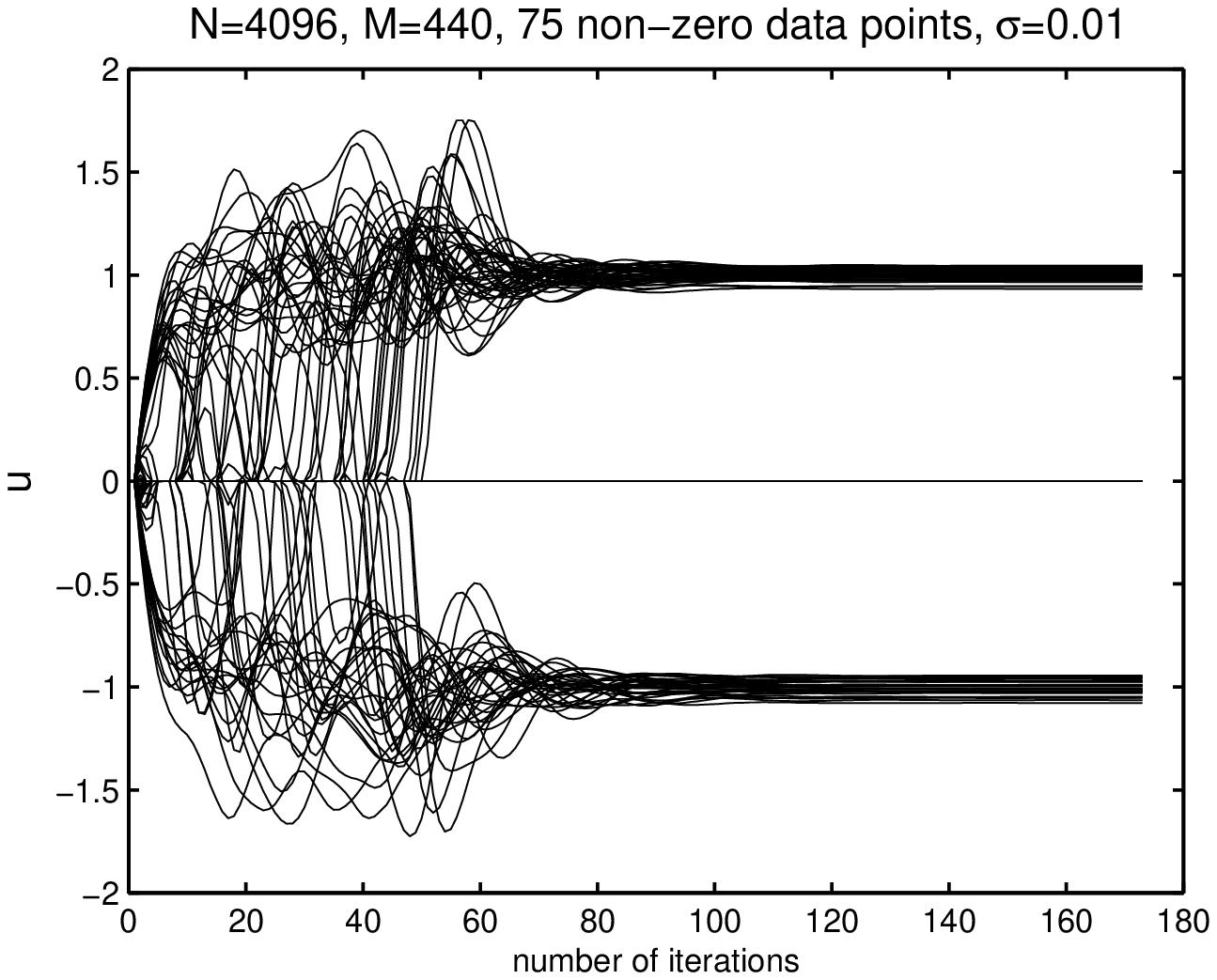,width=1.65in}} &
\mbox{\epsfig{figure=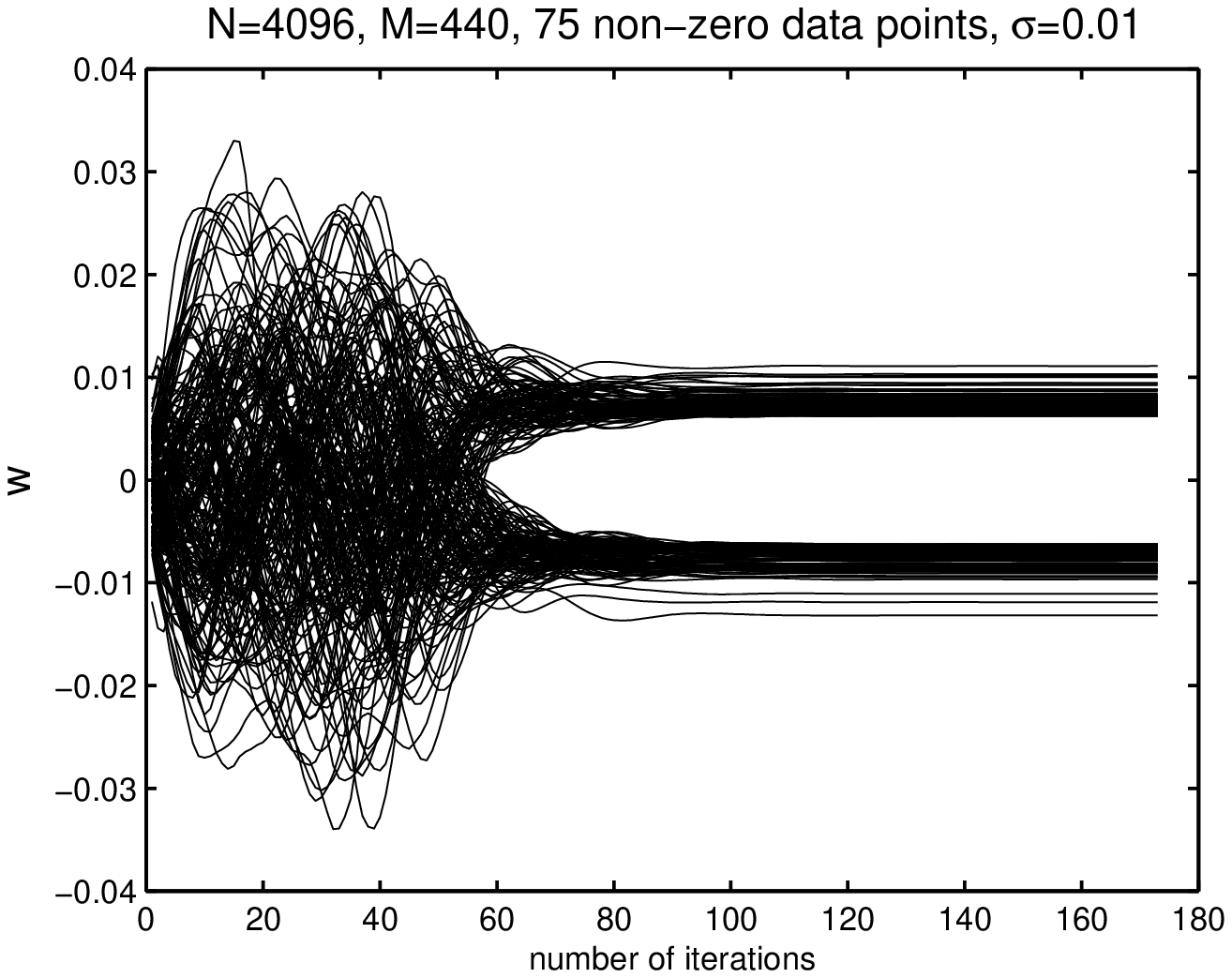,width=1.65in}} &
\mbox{\epsfig{figure=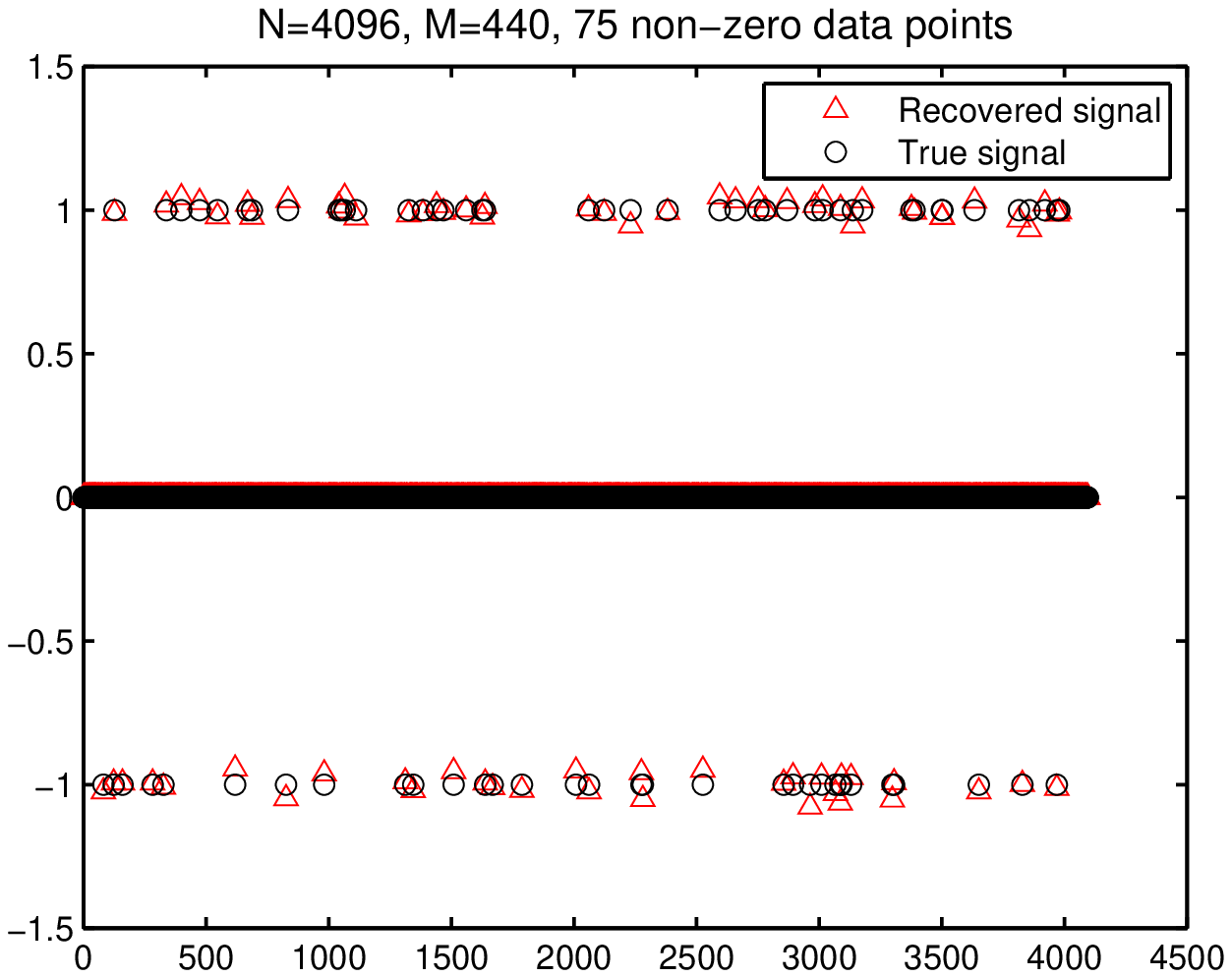,width=1.6in}}
\\
\mbox{\epsfig{figure=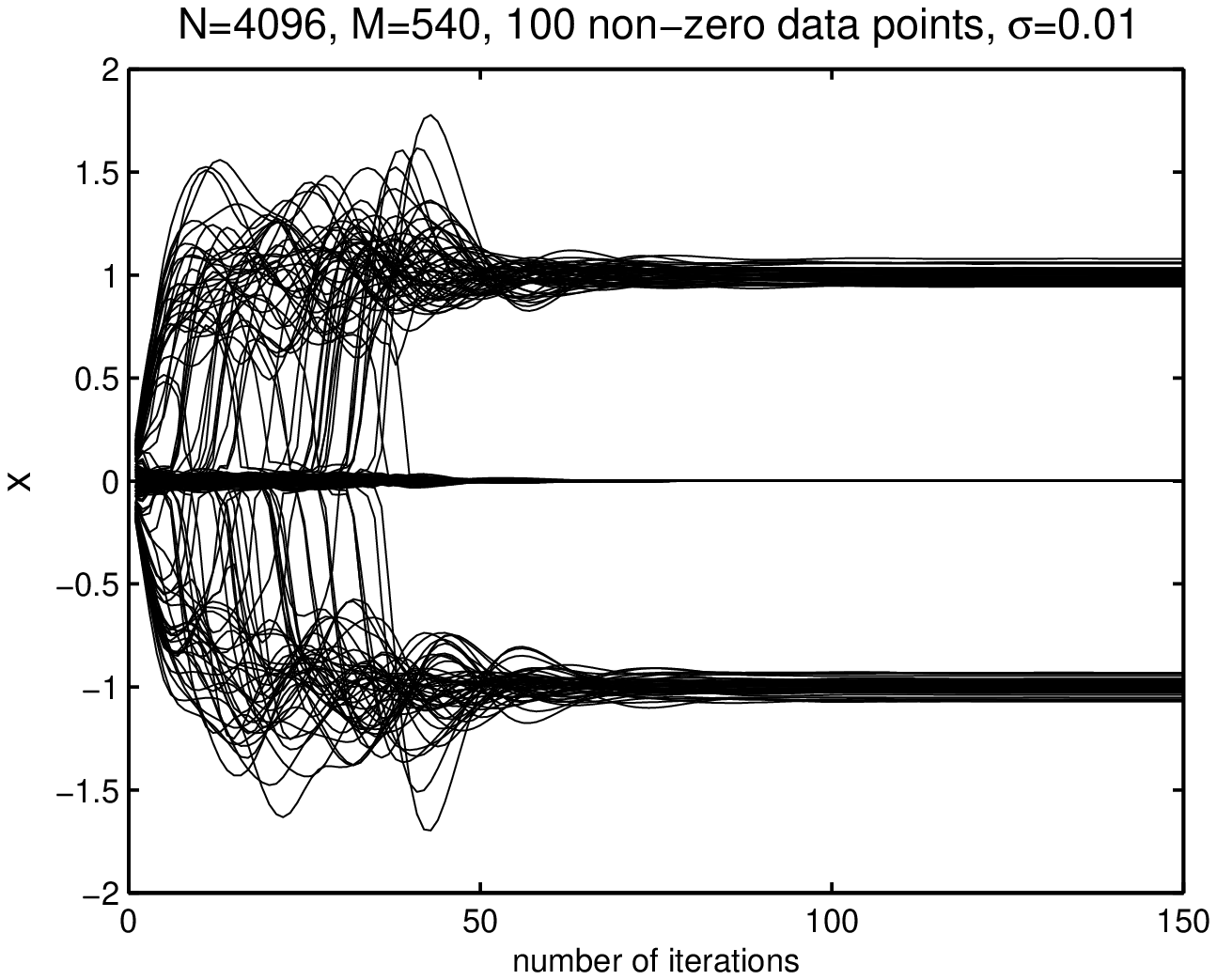,width=1.65in}} &
\mbox{\epsfig{figure=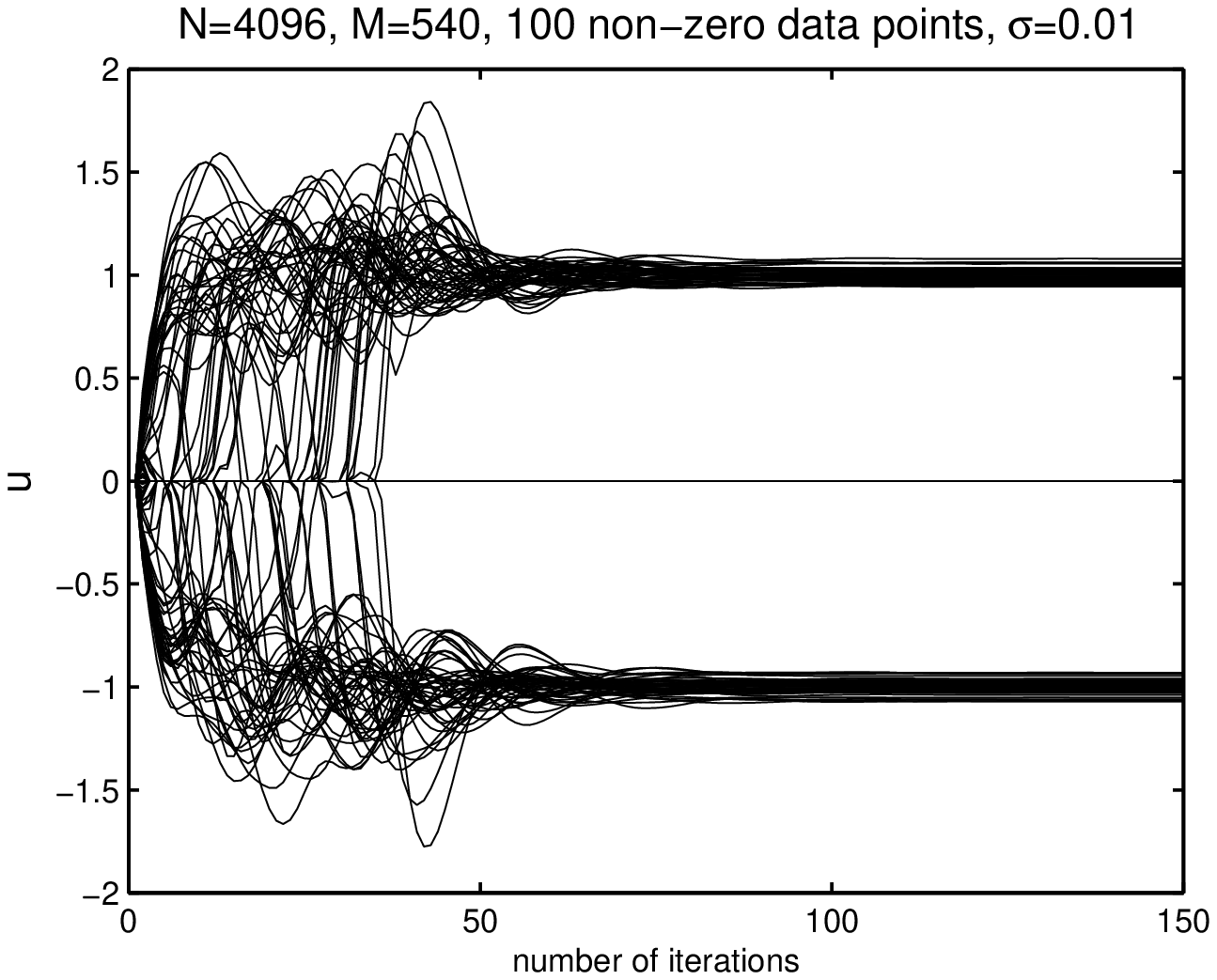,width=1.65in}} &
\mbox{\epsfig{figure=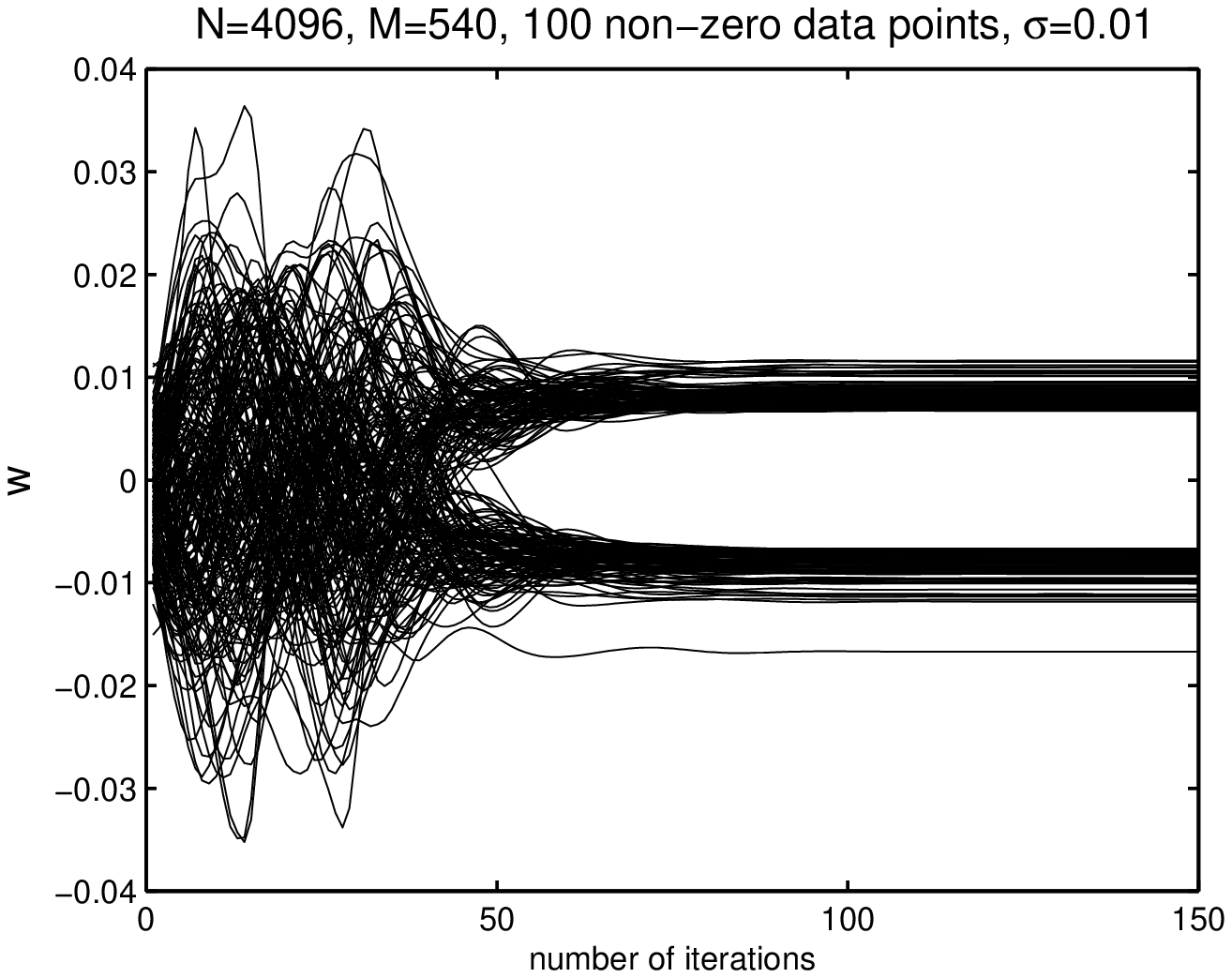,width=1.65in}} &
\mbox{\epsfig{figure=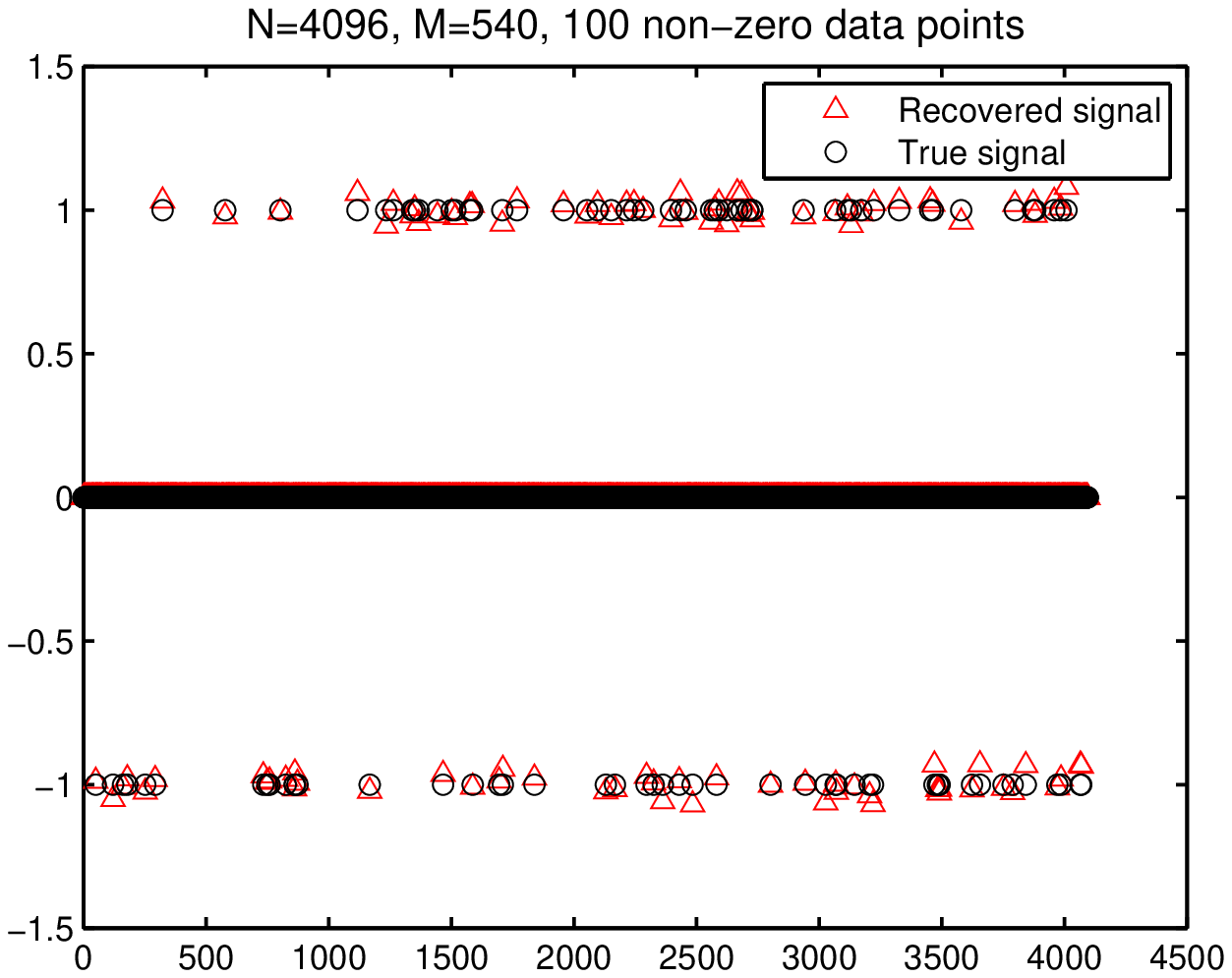,width=1.6in}} \\
\mbox{\epsfig{figure=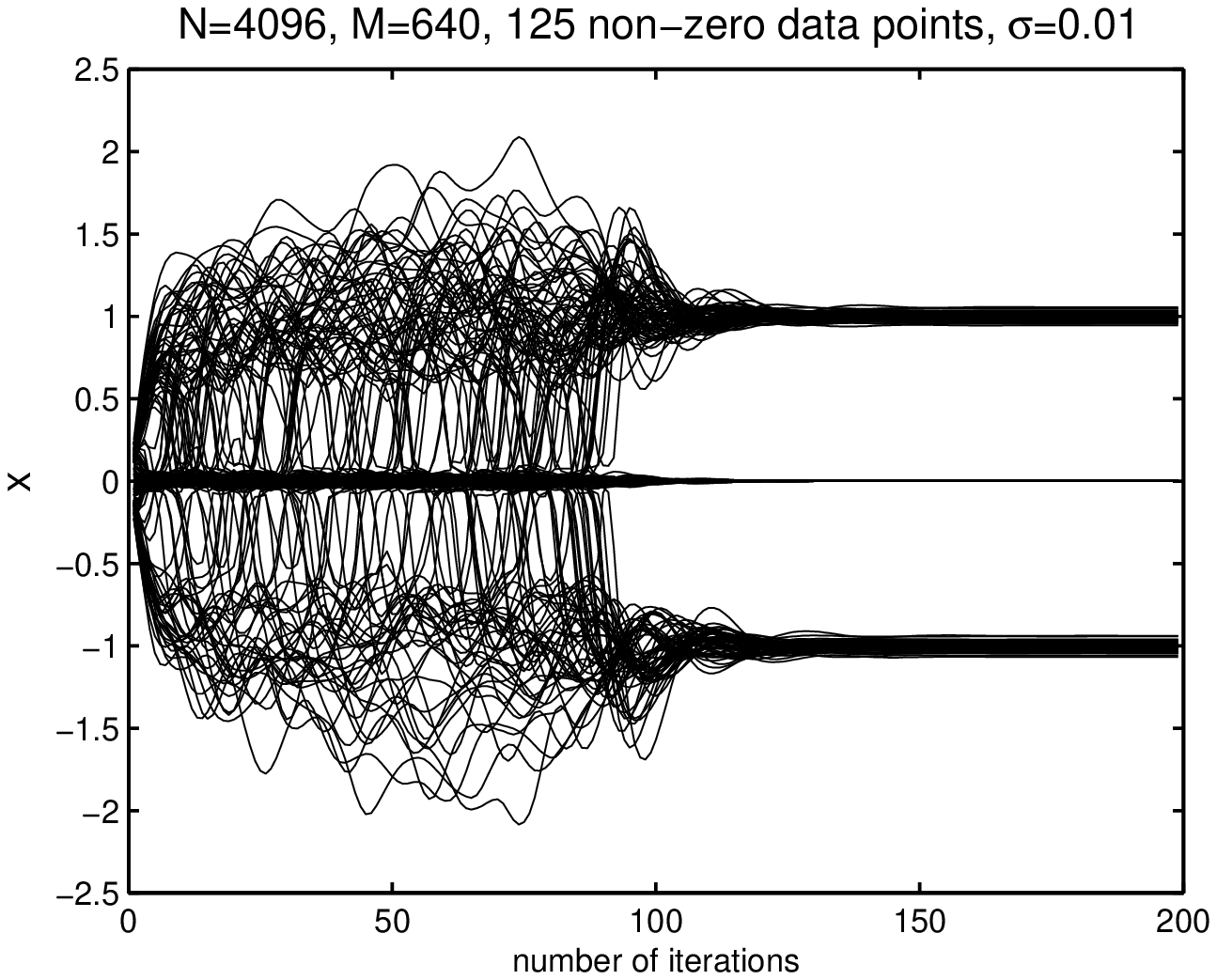,width=1.65in}} &
\mbox{\epsfig{figure=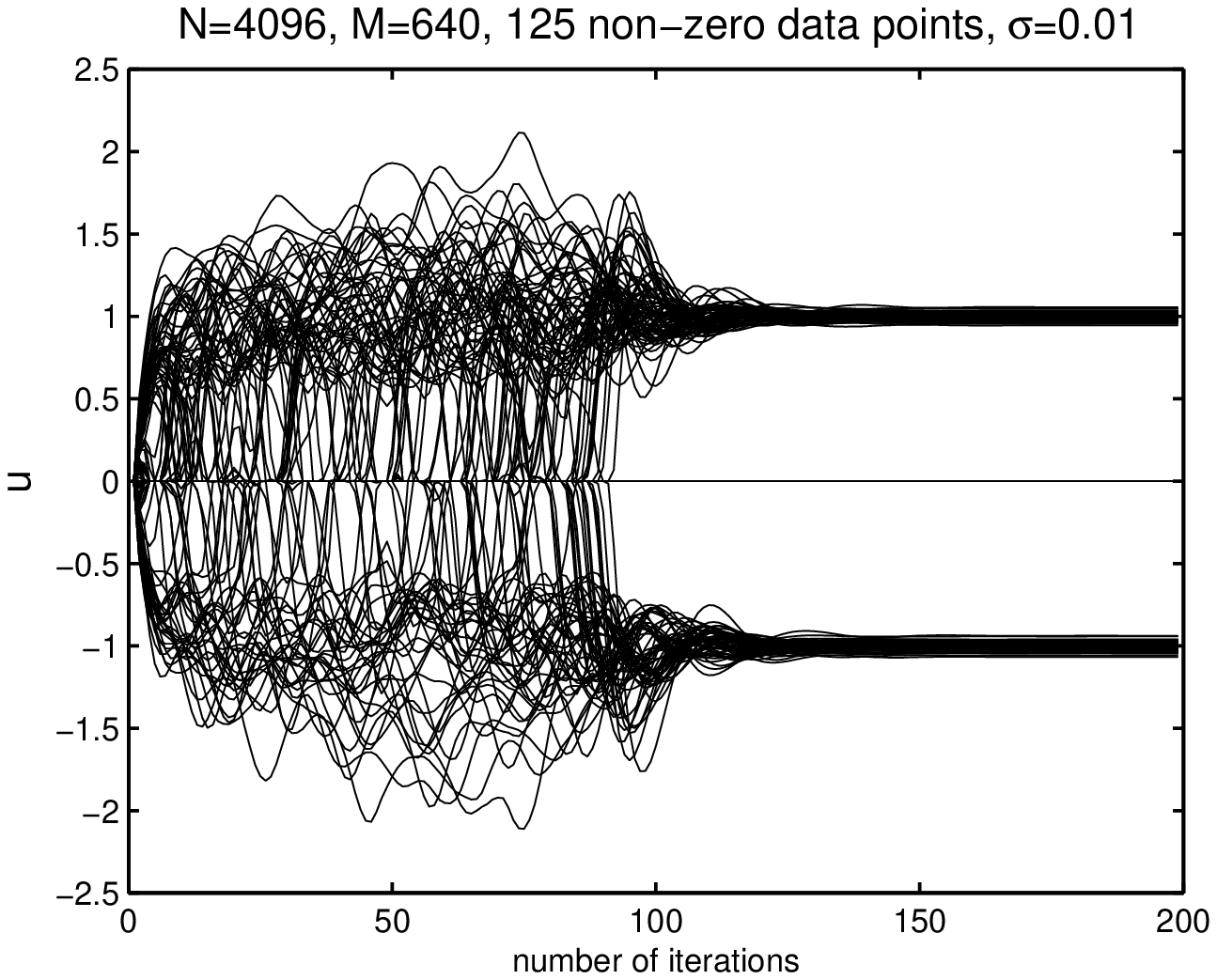,width=1.65in}} &
\mbox{\epsfig{figure=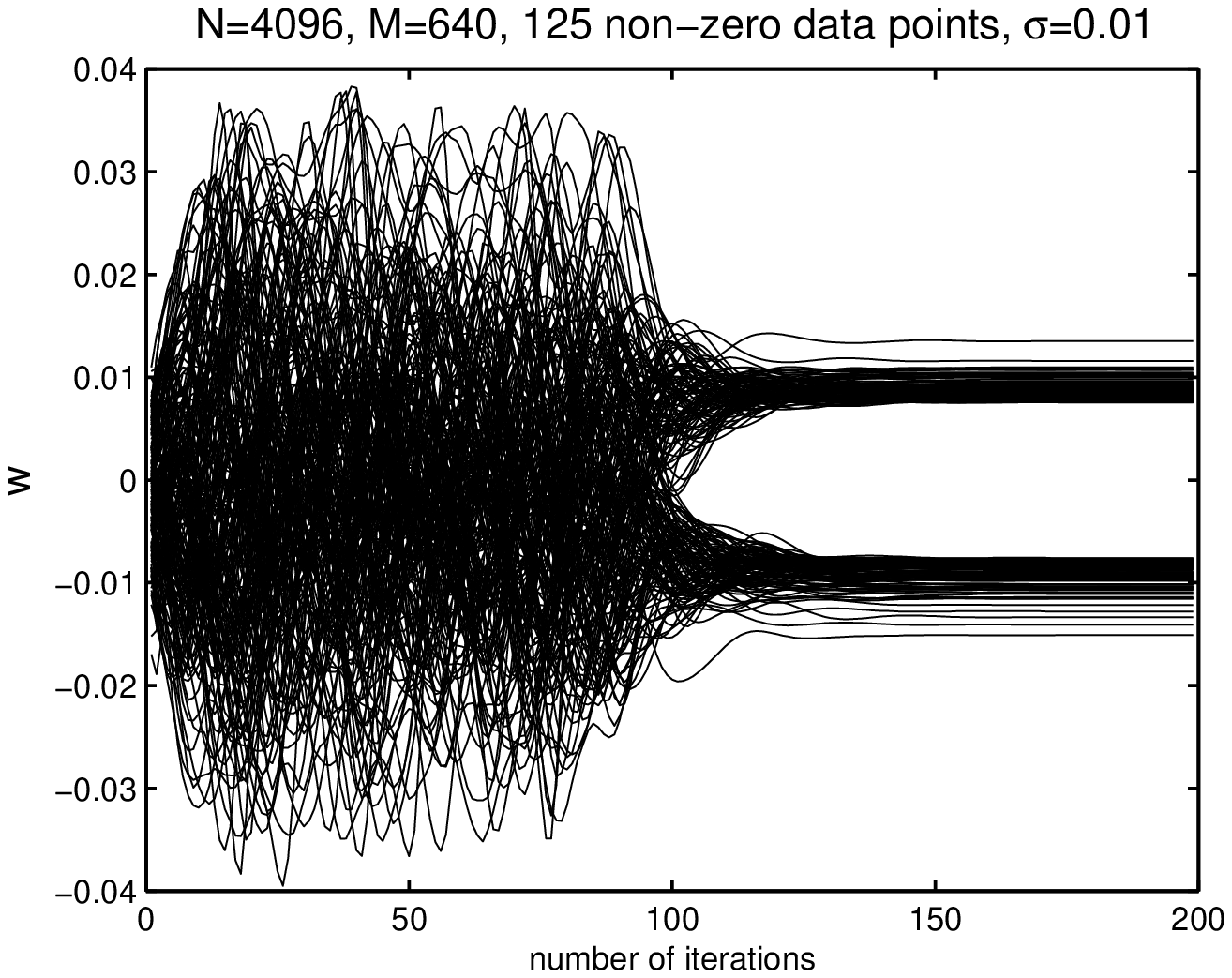,width=1.65in}} &
\mbox{\epsfig{figure=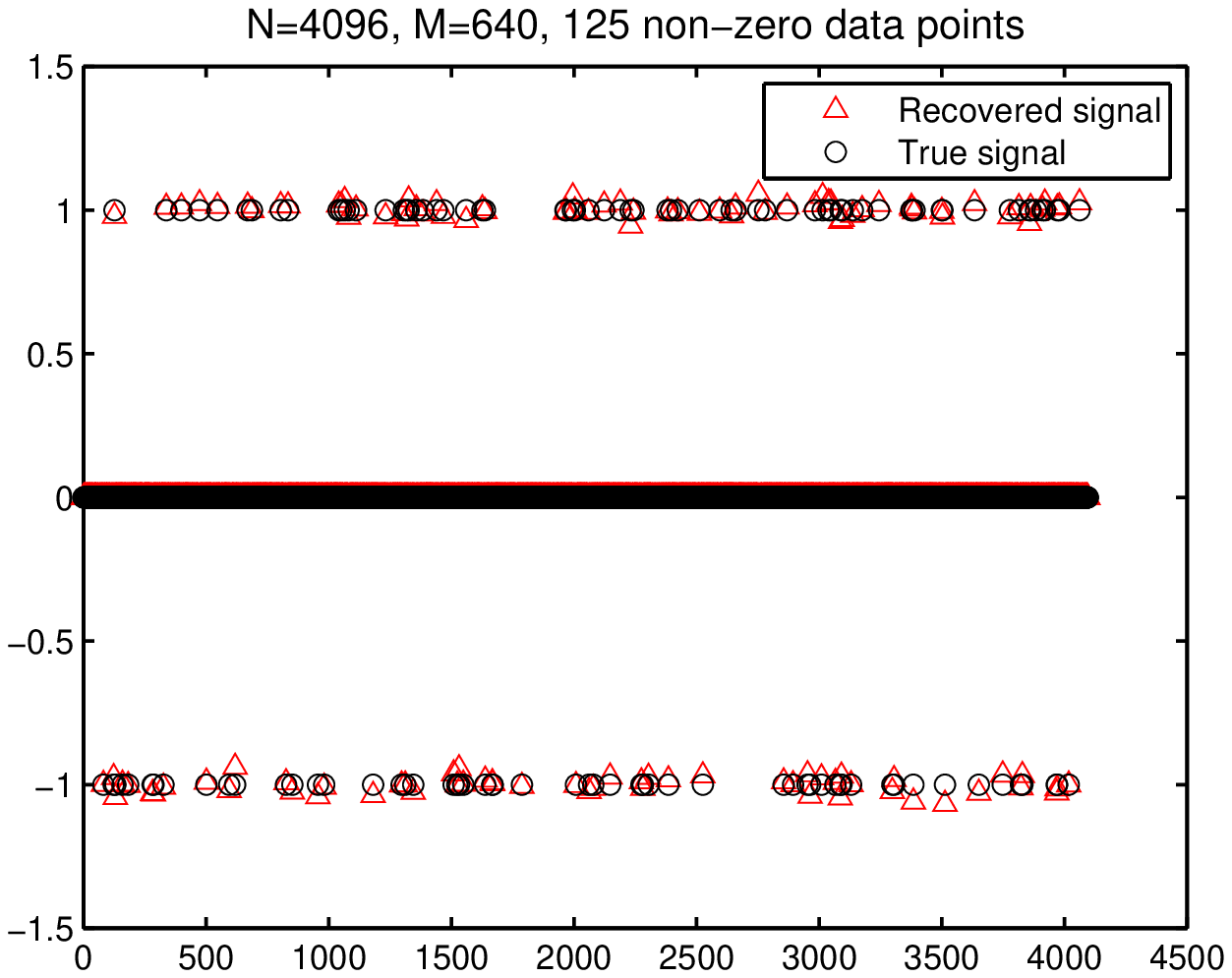,width=1.6in}} \\
\end{tabular}
\caption{Convergence of proposed method with $N=4096$.}
\label{fig:convergence4096}
\end{figure*}

\begin{figure*}[!ht]
\centering
\begin{tabular}{c@{\extracolsep{2mm}}c@{\extracolsep{2mm}}c}
\mbox{\epsfig{figure=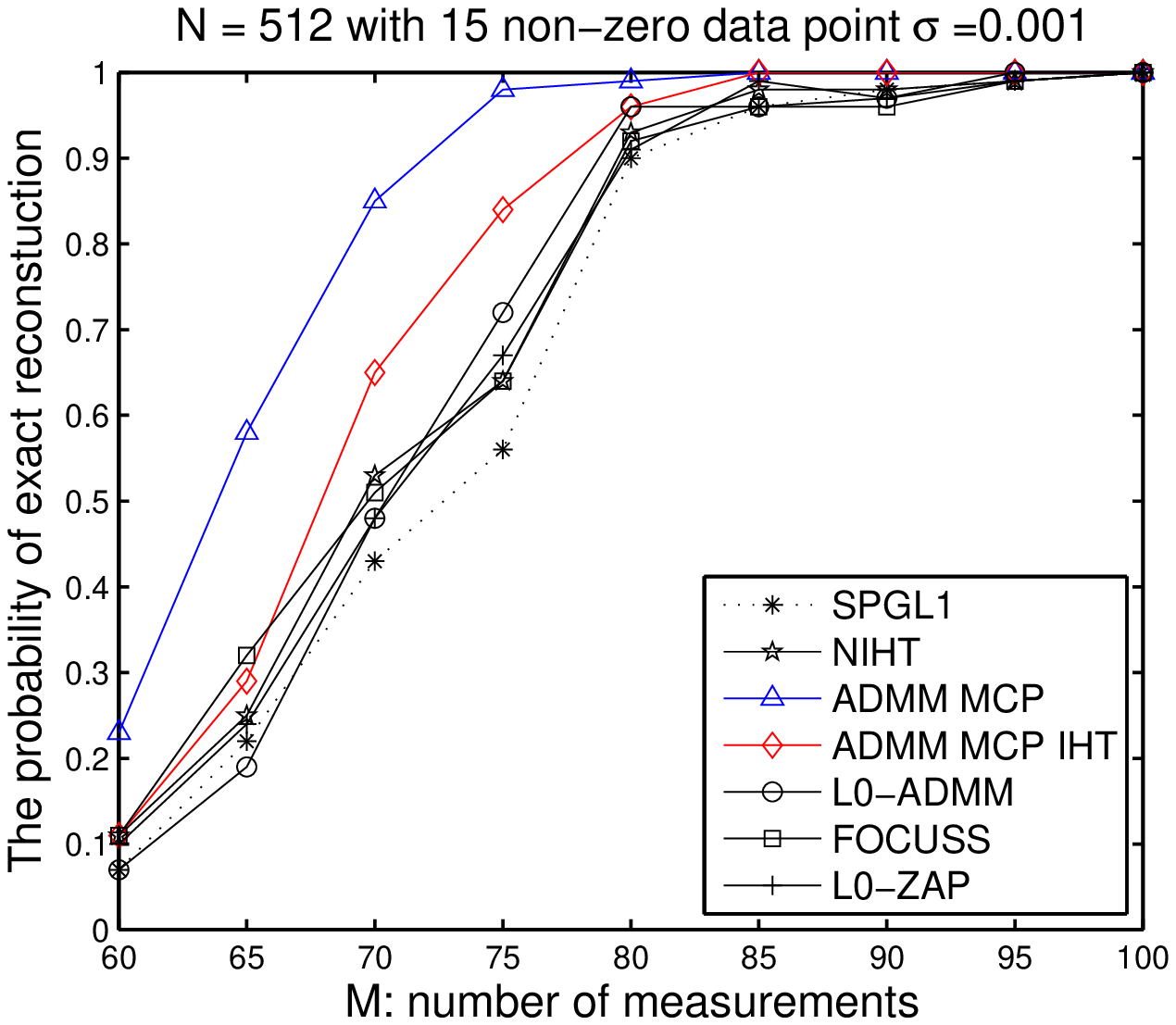,width=2.05in}} &
\mbox{\epsfig{figure=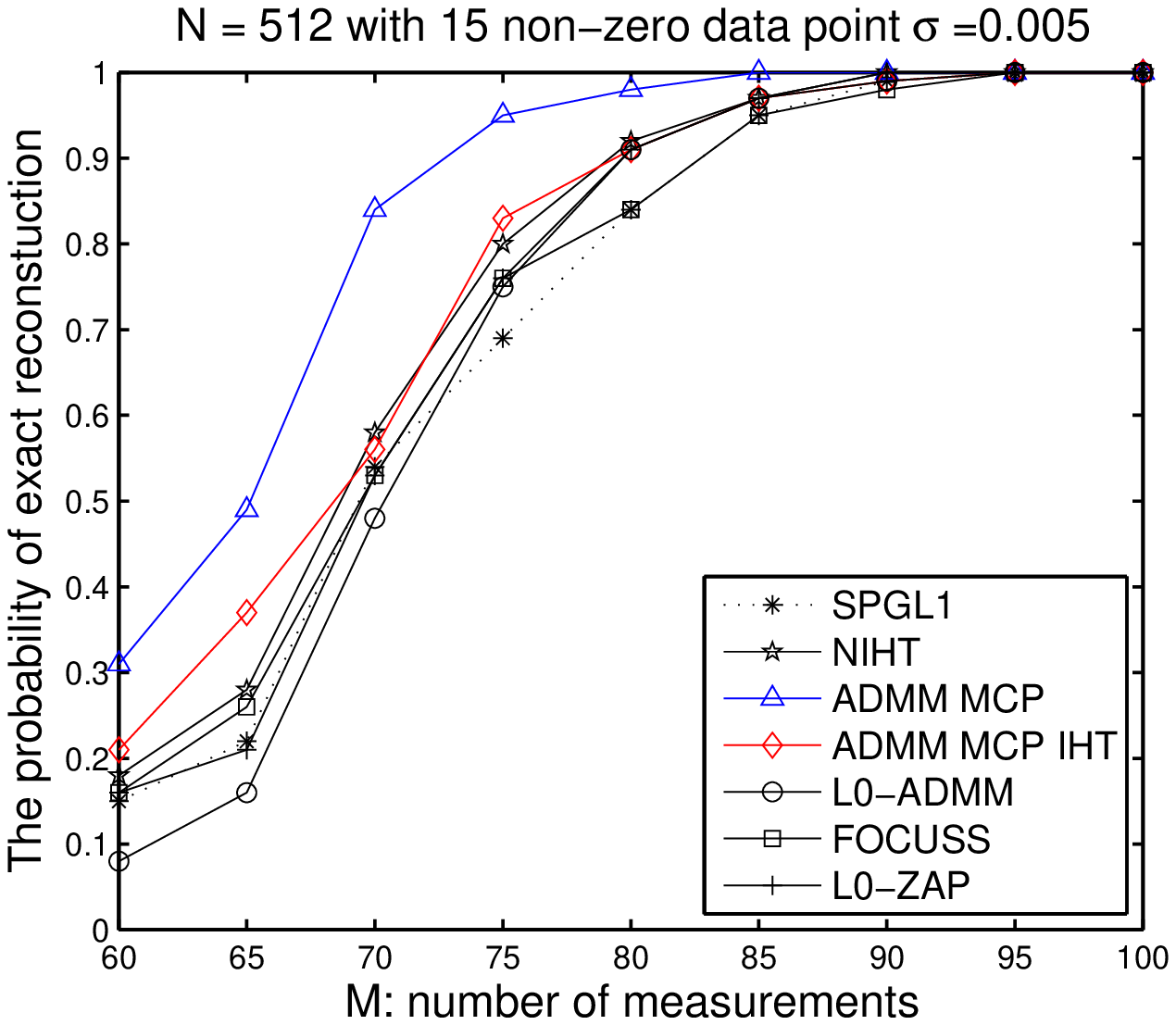,width=2.05in}} &
\mbox{\epsfig{figure=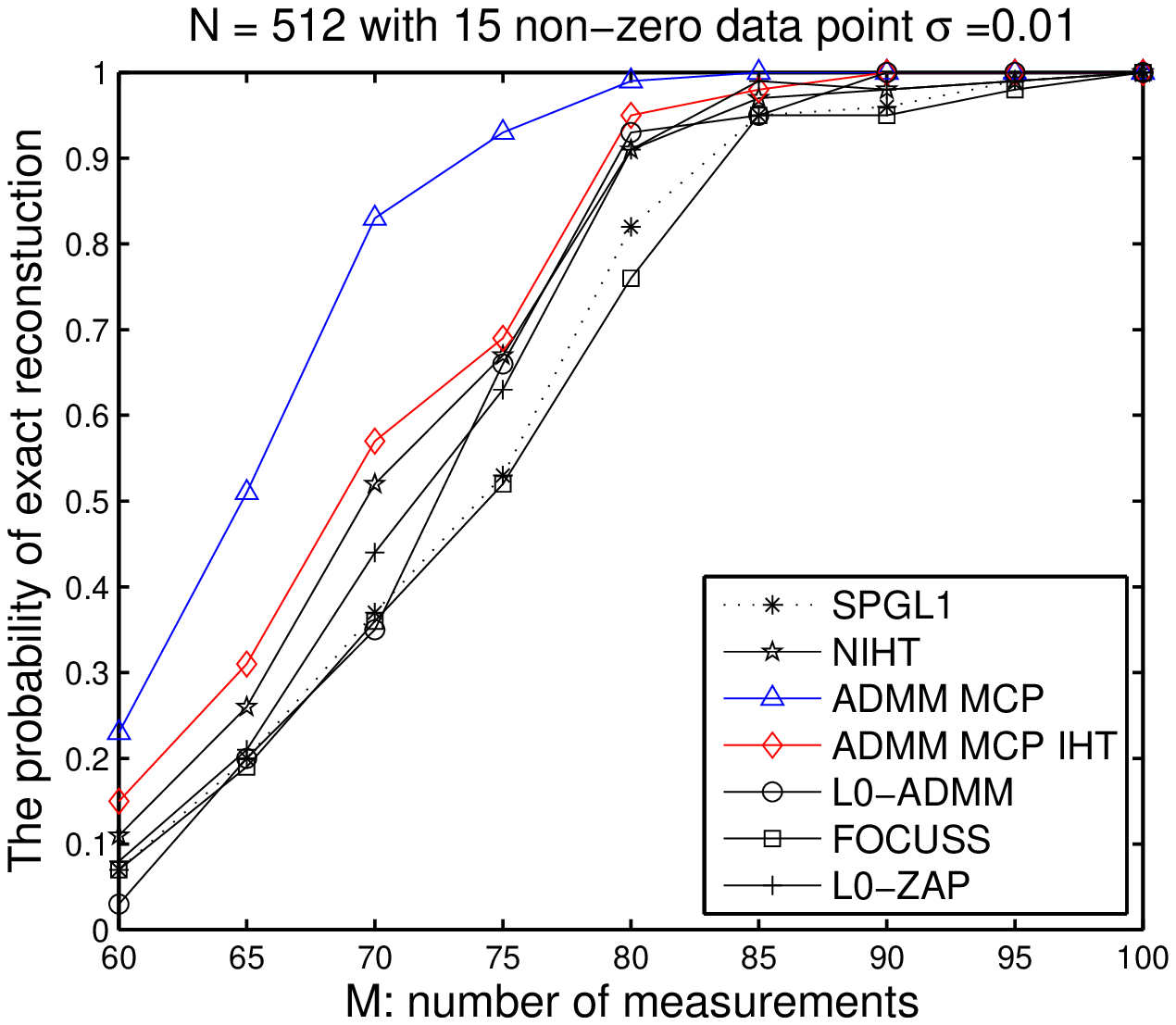,width=2.05in}} \\
\mbox{\epsfig{figure=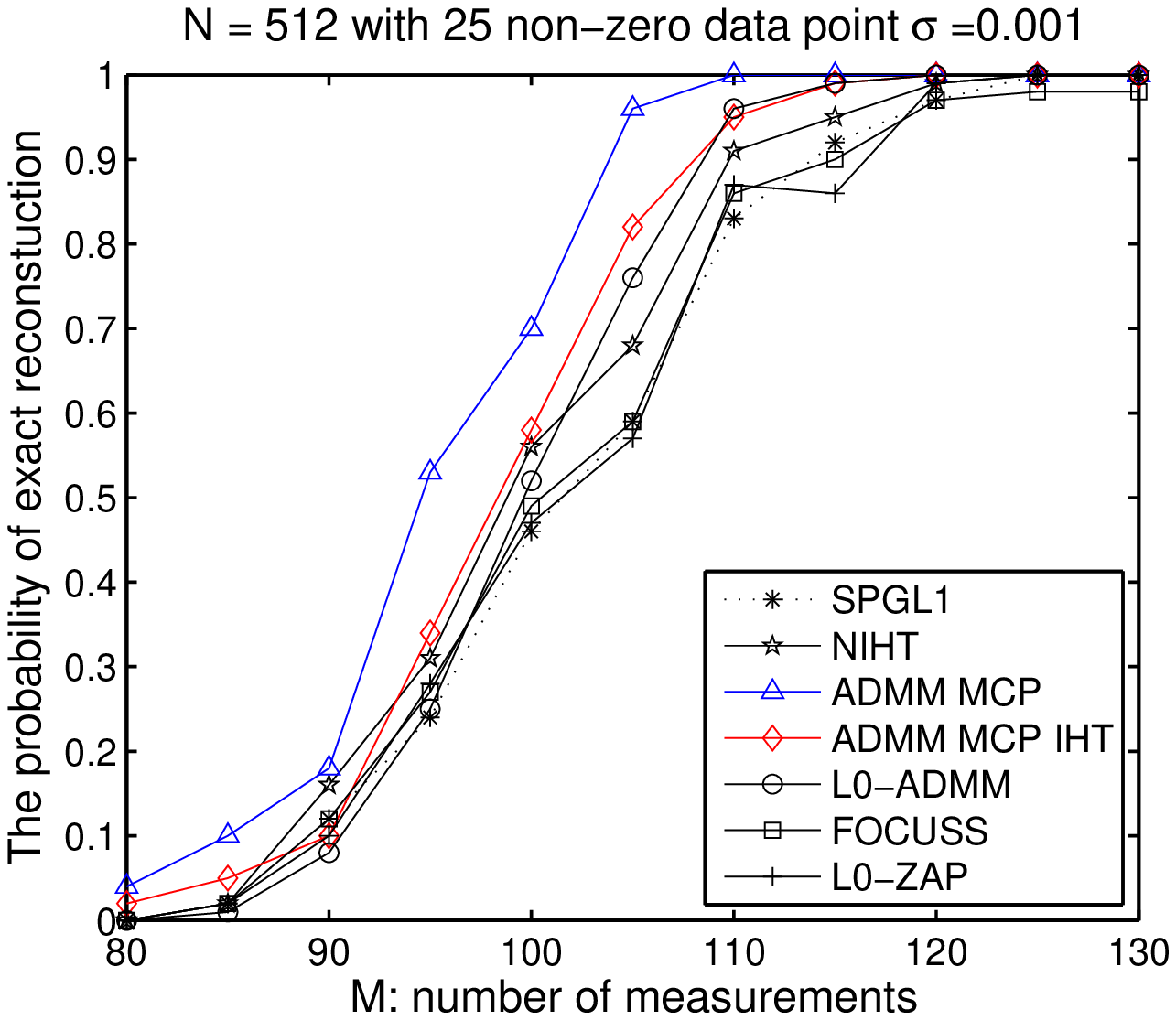,width=2.05in}} &
\mbox{\epsfig{figure=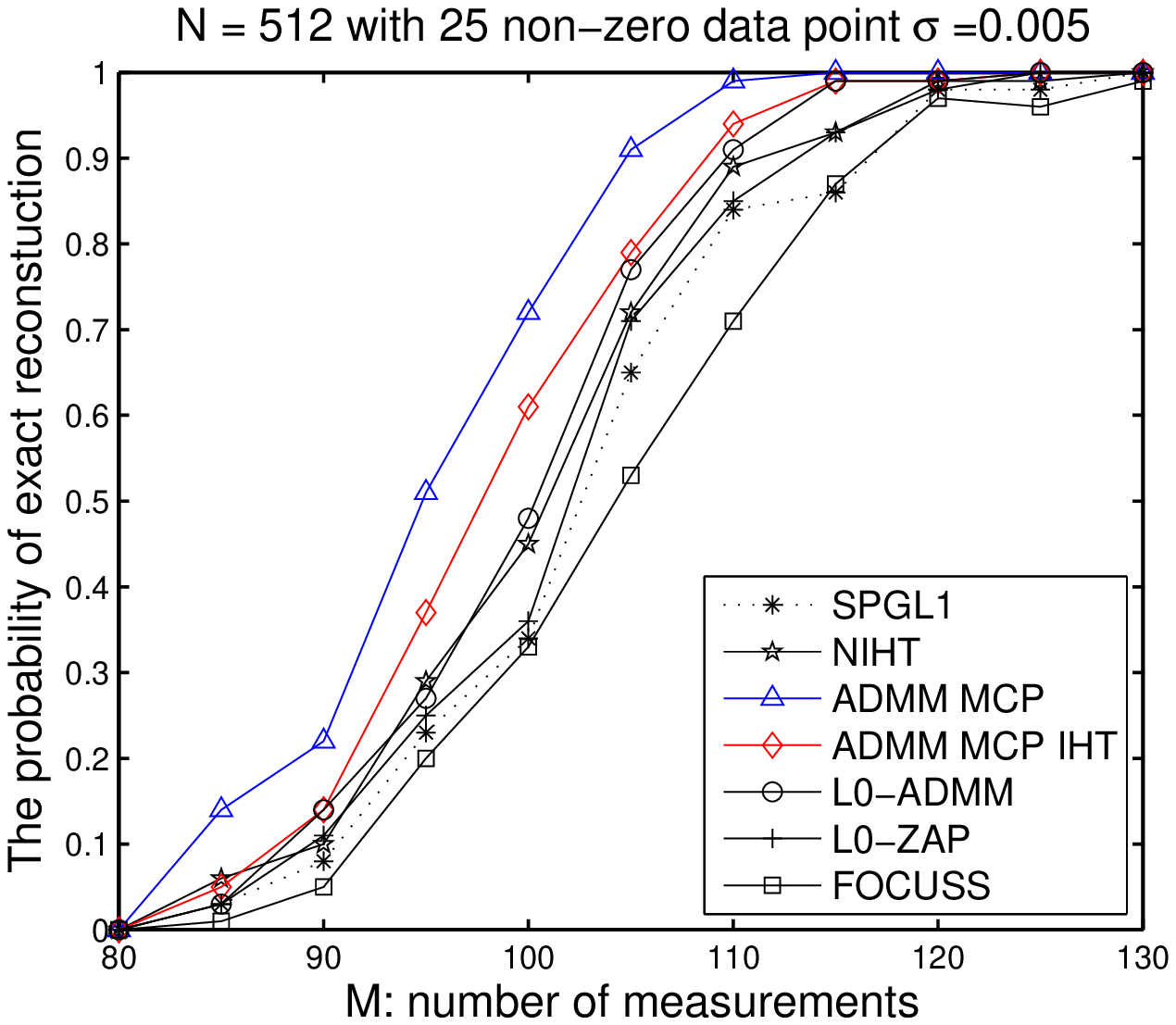,width=2.05in}} &
\mbox{\epsfig{figure=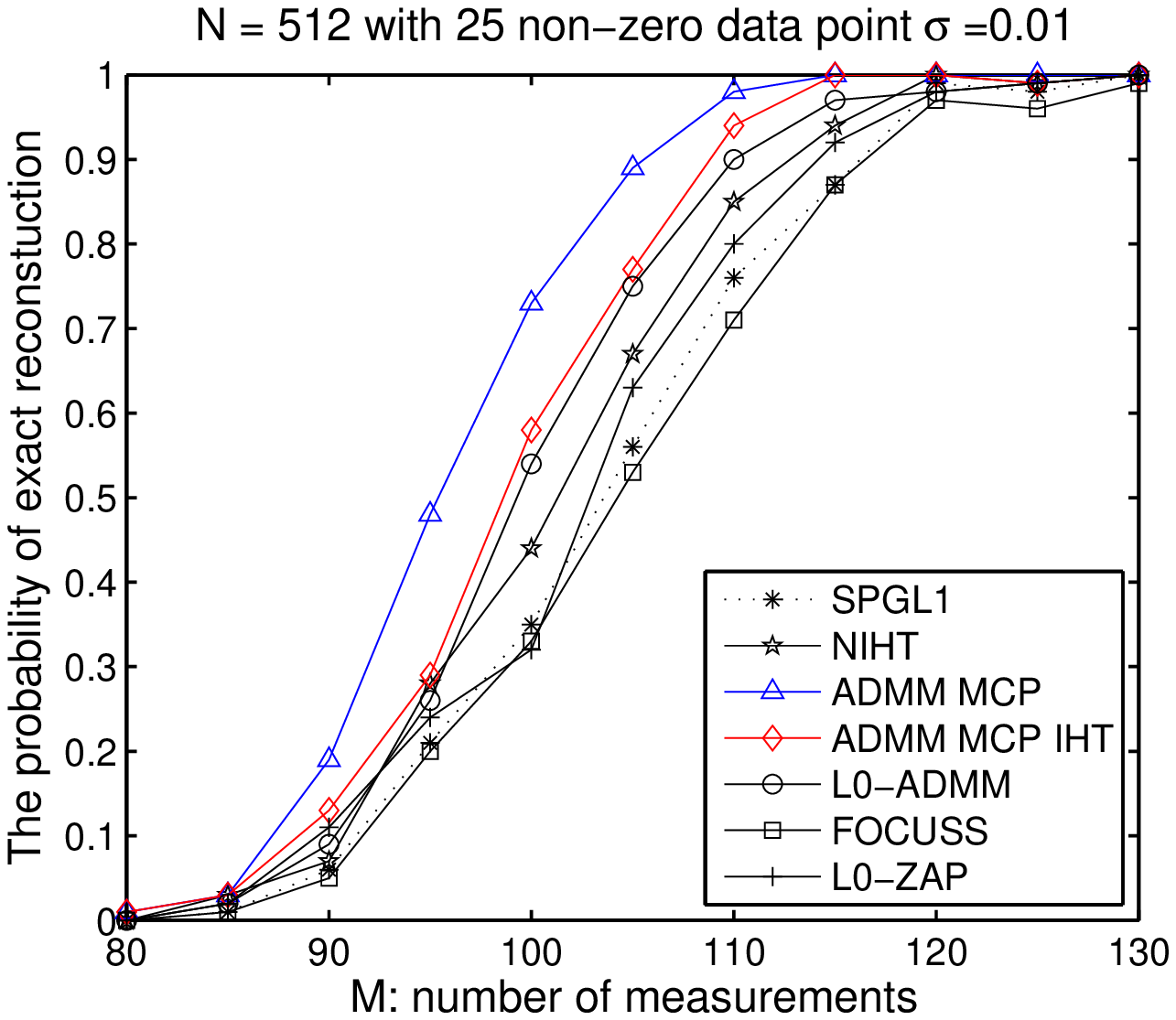,width=2.05in}} \\
\end{tabular}
\caption{Simulation results of different algorithms when $N=512$. For the first row, $\tau=15$. For the second row, $\tau=25$.  The three columns are based on three different noise levels with $\sigma=\{0.001,0.005,0.01\}$ respectively. The experiment is repeated 100 times at each number of measurements.}
\label{fig:512}
\end{figure*}

\begin{figure*}[!ht]
\centering
\begin{tabular}{c@{\extracolsep{2mm}}c@{\extracolsep{2mm}}c}
\mbox{\epsfig{figure=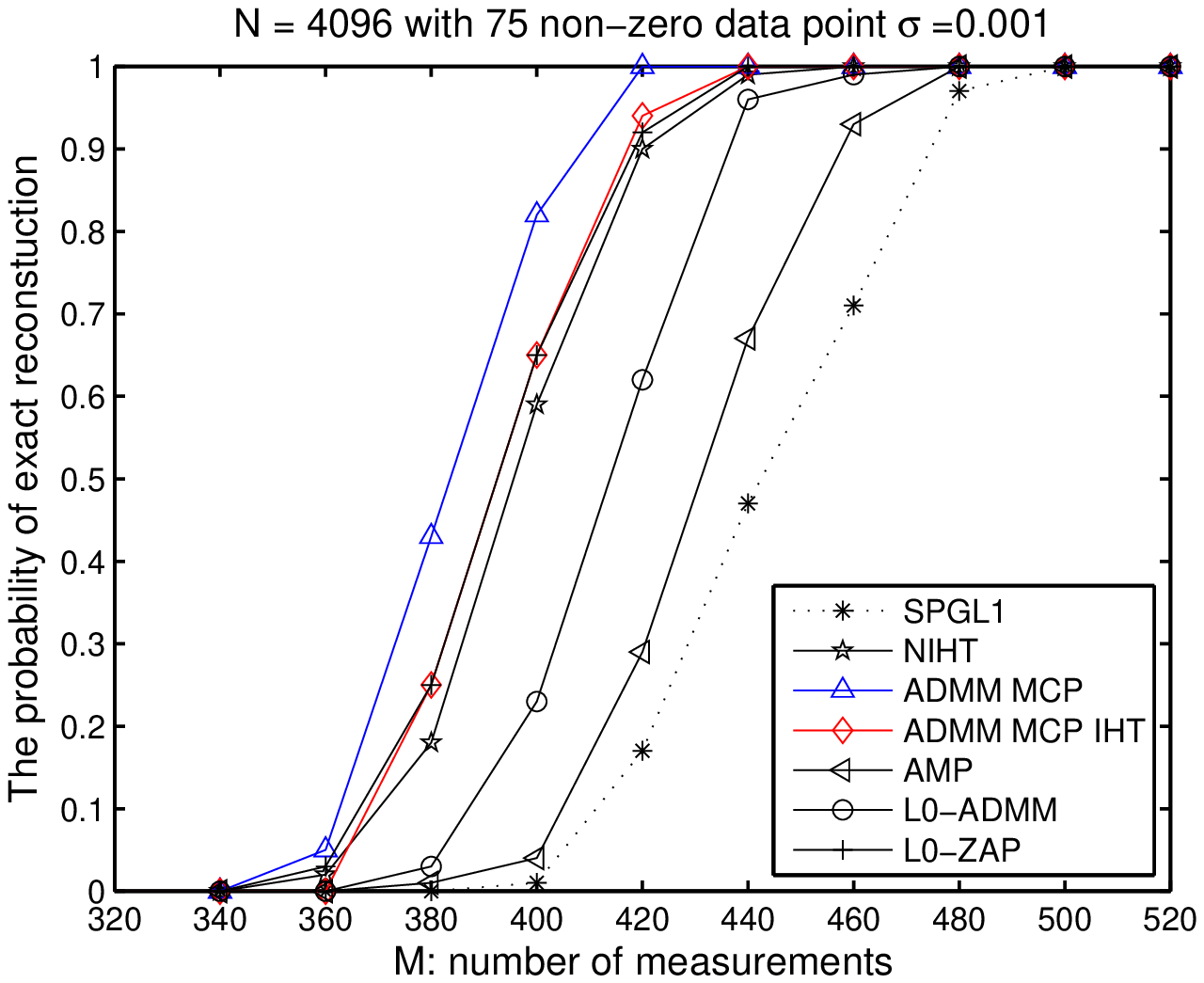,width=2.05in}} &
\mbox{\epsfig{figure=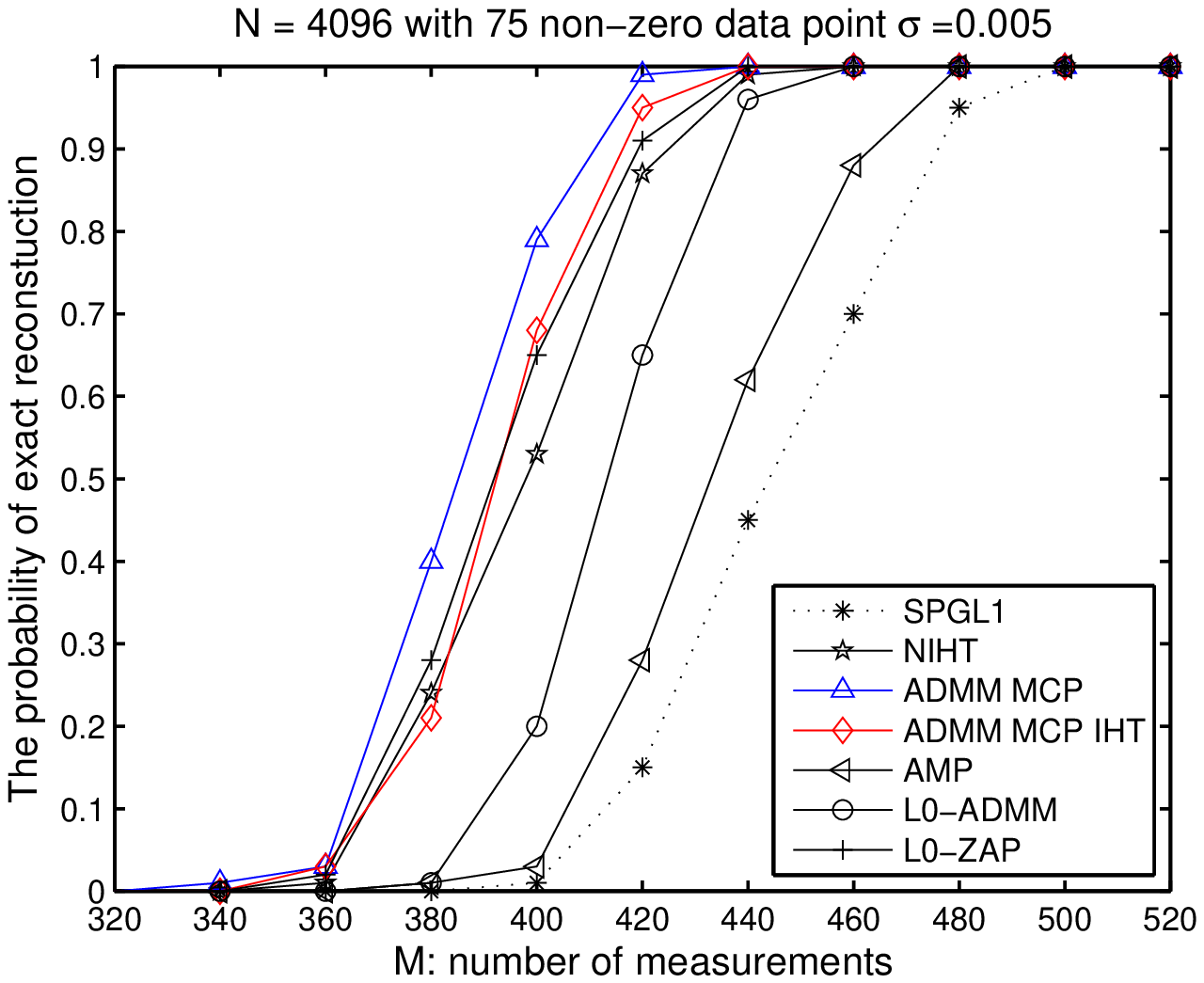,width=2.05in}} &
\mbox{\epsfig{figure=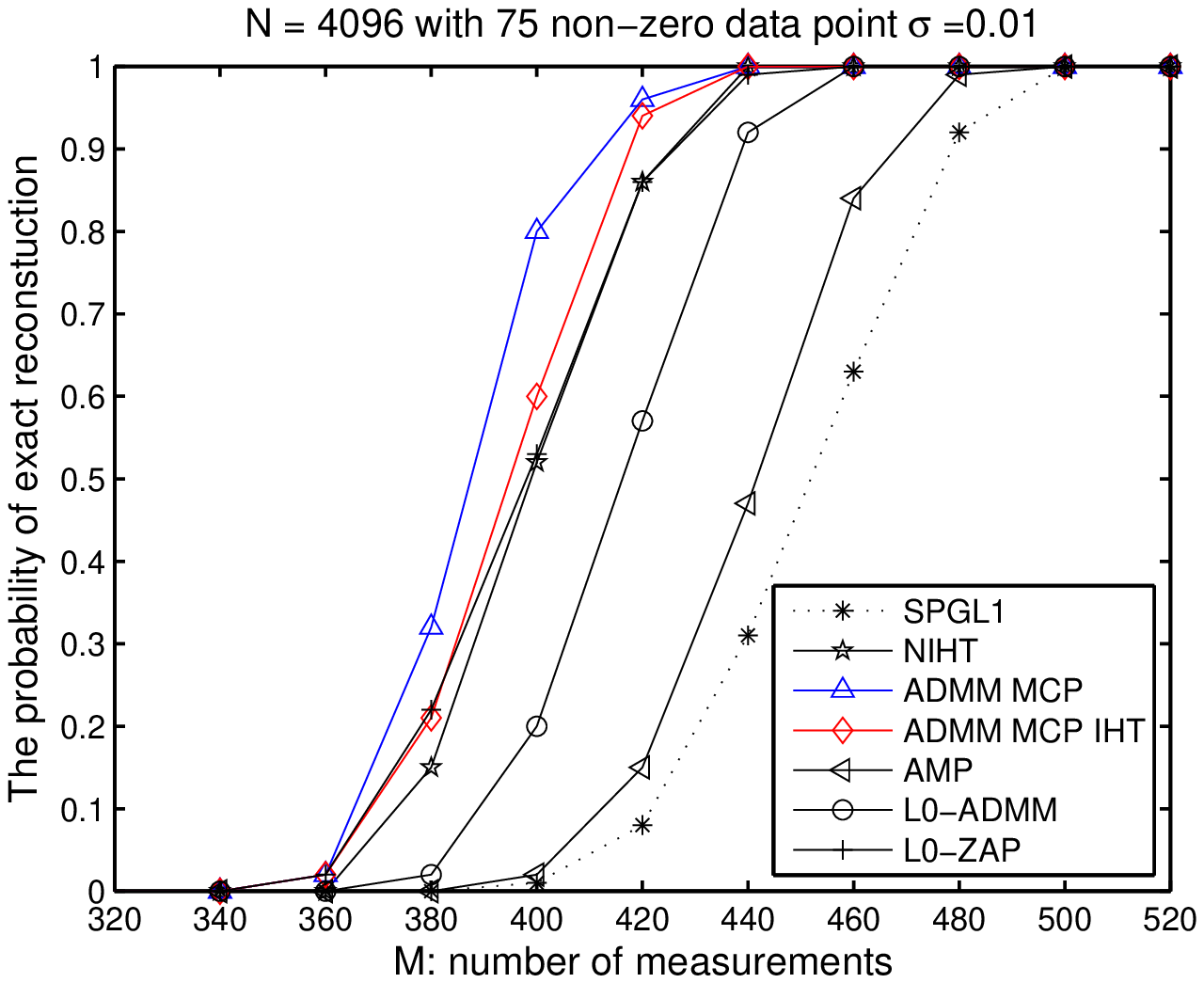,width=2.05in}} \\
\mbox{\epsfig{figure=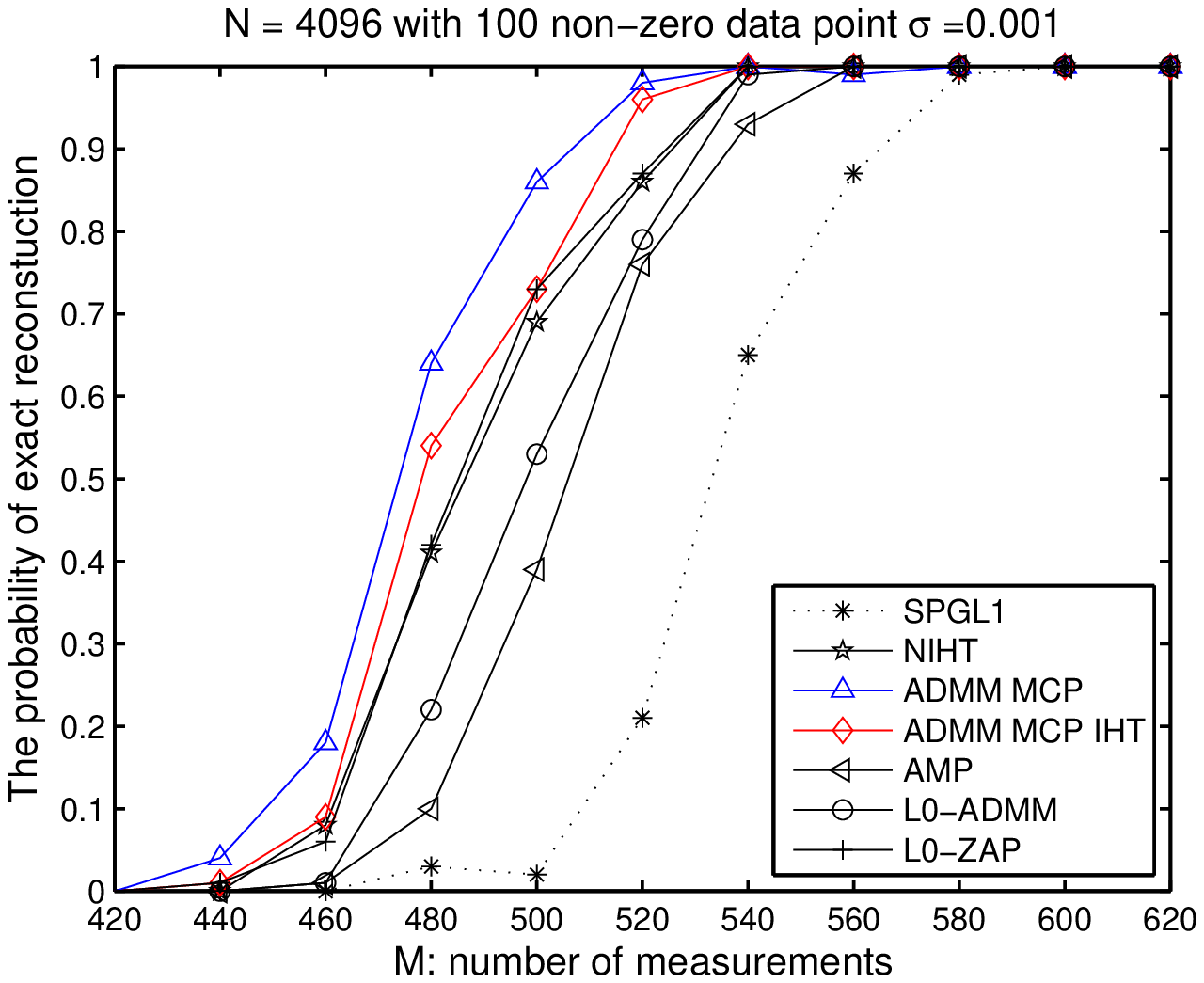,width=2.05in}} &
\mbox{\epsfig{figure=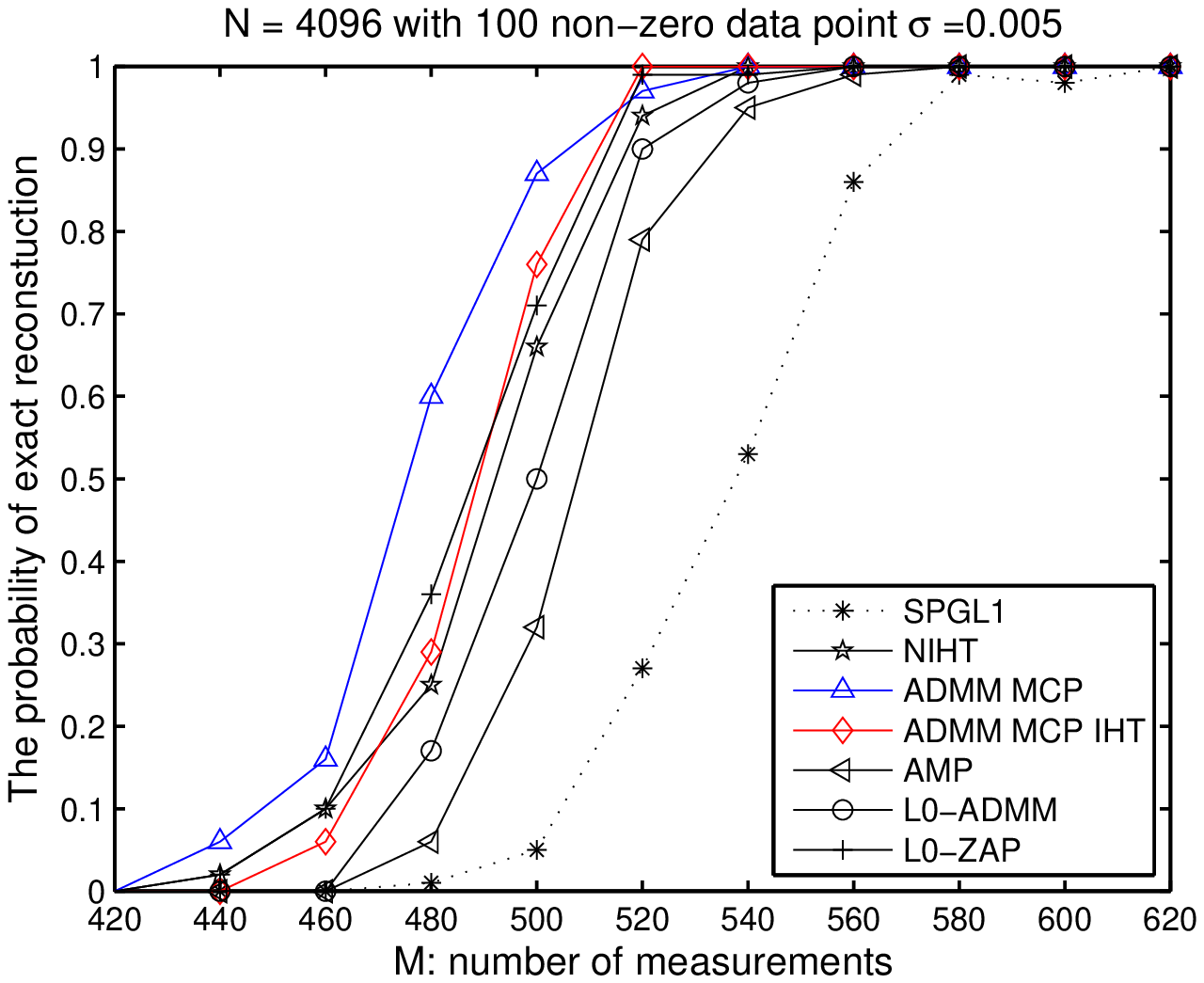,width=2.05in}} &
\mbox{\epsfig{figure=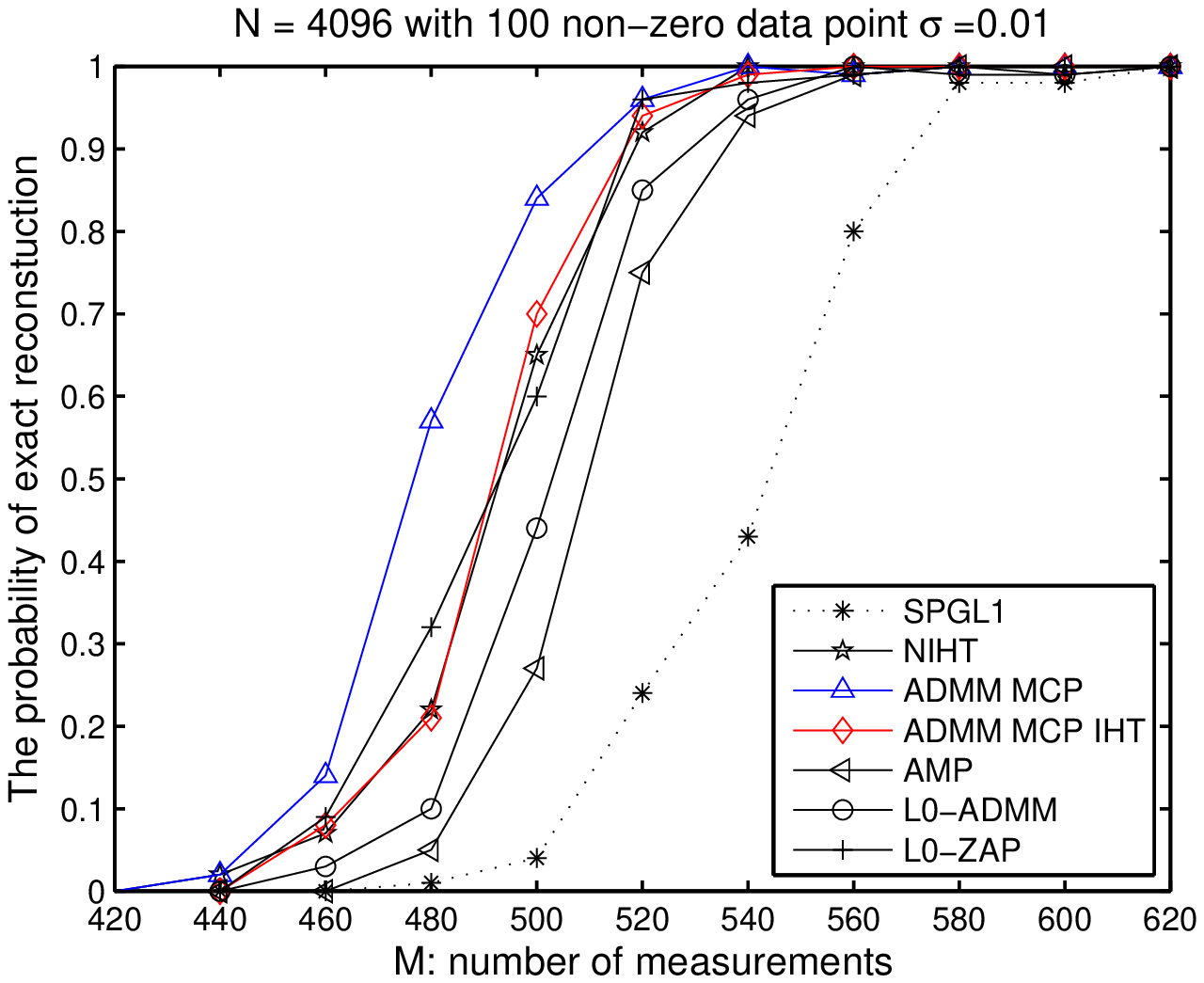,width=2.05in}} \\
\mbox{\epsfig{figure=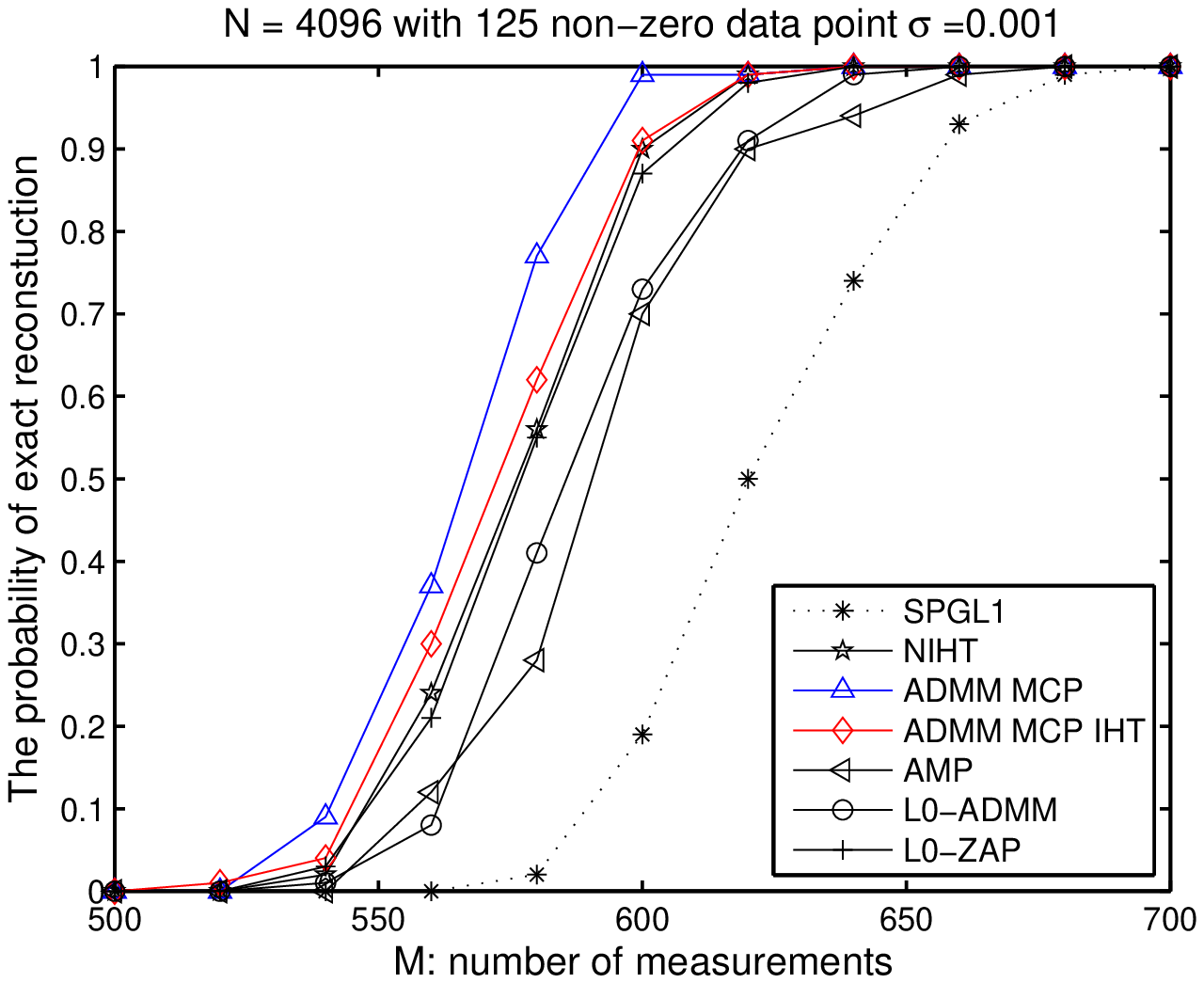,width=2.05in}} &
\mbox{\epsfig{figure=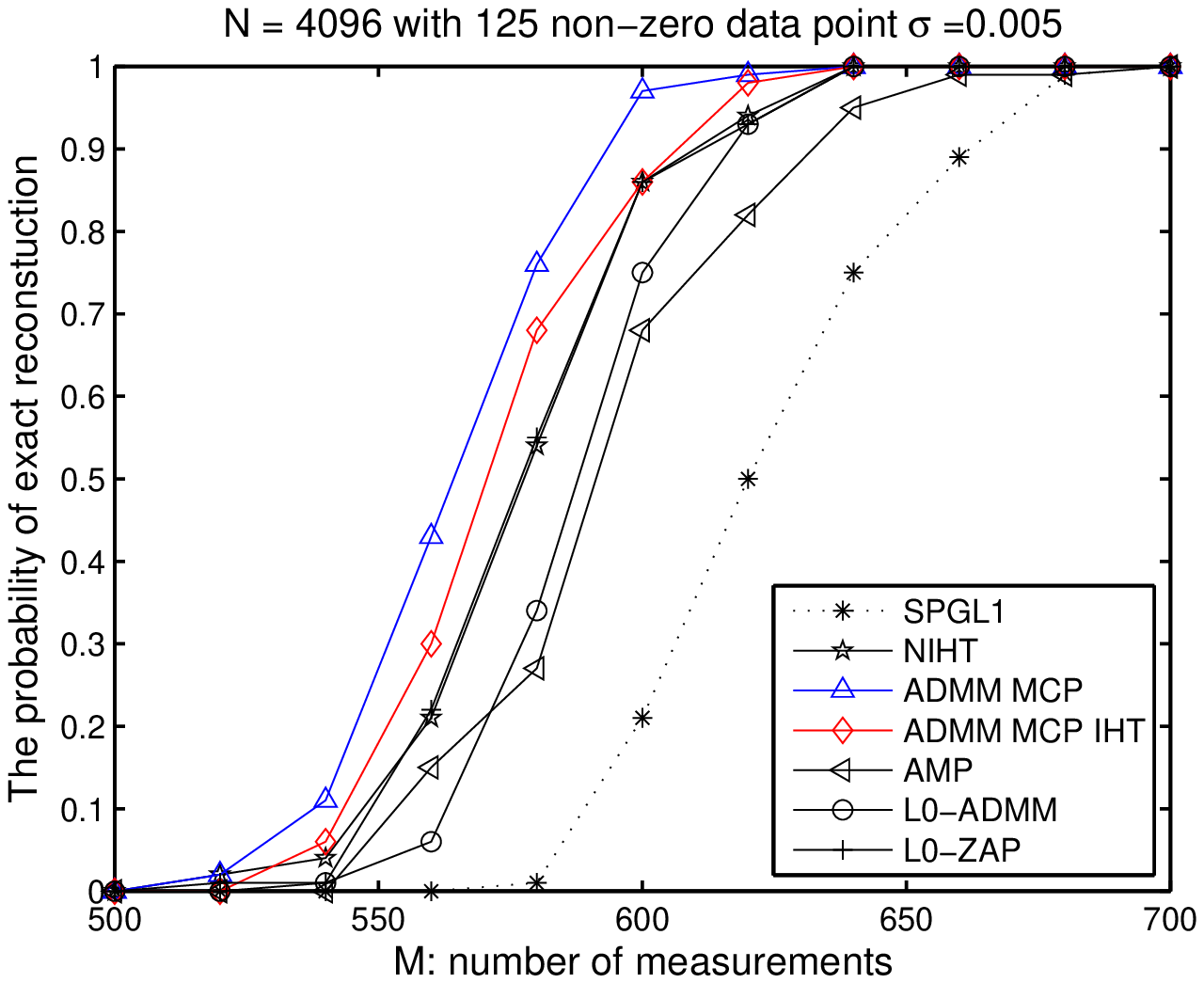,width=2.05in}} &
\mbox{\epsfig{figure=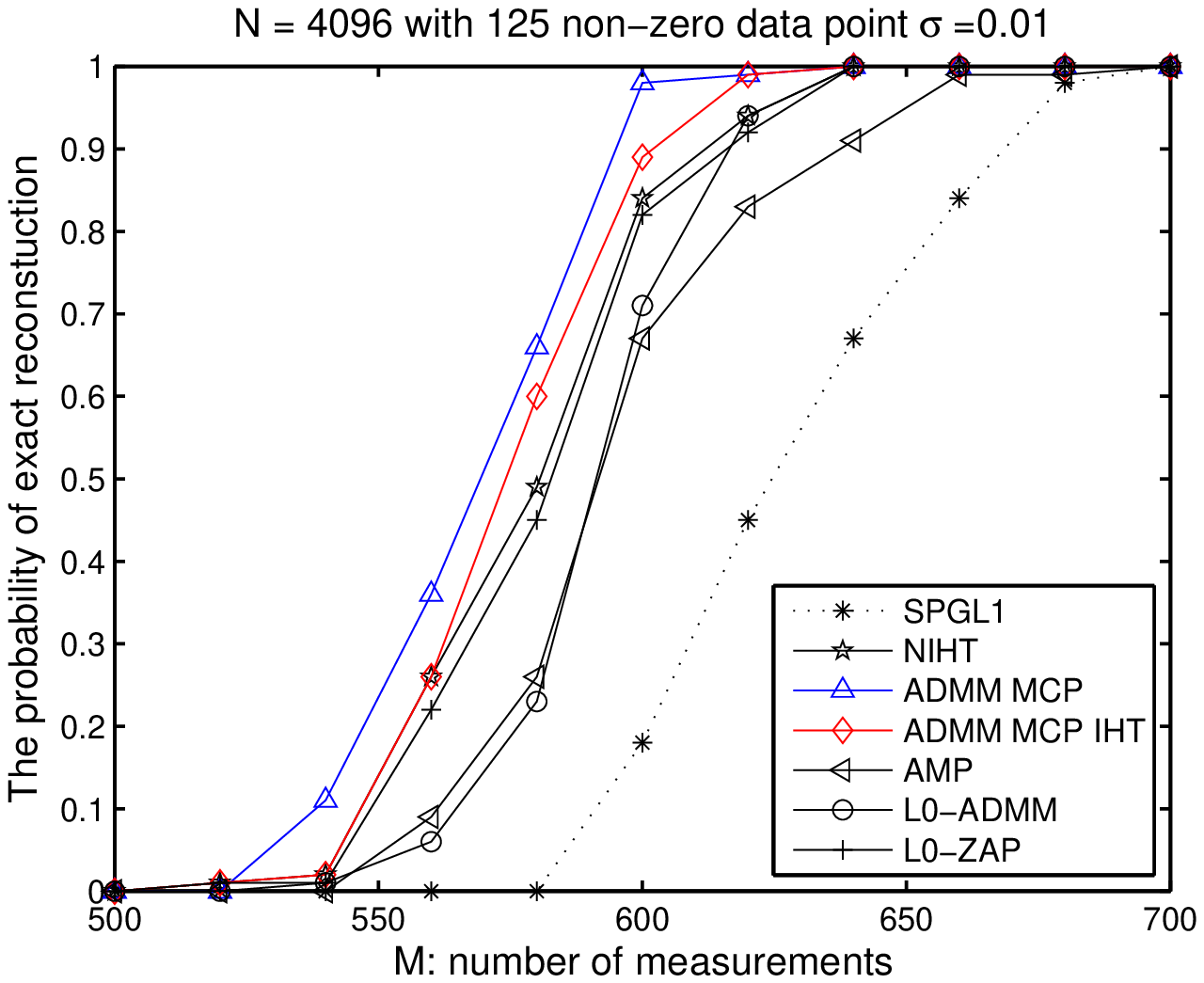,width=2.05in}} \\
\end{tabular}
\caption{Simulation results of different algorithms when $N=4096$. For the first row, $\tau=75$. For the second row, $\tau=100$. For the third row, $\tau=125$. The three columns are based on three different noise levels with $\sigma=\{0.001,0.005,0.01\}$ respectively. The experiment is repeated 100 times at each number of measurements.}
\label{fig:4096}
\end{figure*}

\section{Simulation and Experimental Result}\label{section6}

In this section, we conduct several experiments to verify the performance of our proposed method and compare it with several state-of-the-art sparse signal reconstruction approaches in compressed sensing including SPGL1~\cite{van2008probing,spgl12007}, FOCUSS~\cite{engan2000regularized,gorodnitsky1997sparse}, NIHT~\cite{blumensath2010normalized}, AMP~\cite{maleki2010approximate,metzler2016denoising},  L0-ZAP~\cite{jin2010stochastic}, and L0-ADMM~\cite{boyd2011distributed}. Except for SPGL1, all others are nonconvex $l_0$-norm approximate methods. FOCUSS uses a weighted norm to convert the nonconvex compressed sensing problem to a solvable convex one.
NIHT is a variant of IHT algorithm. Comparing with IHT, NIHT uses normalization which can further relax its convergence condition.
AMP combines the iterative thresholding algorithm with scalar threshold functions. This algorithm utilizes the statistical property of signal, hence it can only estimate the signal with enough length. For the above two algorithms, they all only have local convergence, hence their initializations are critical.
L0-ZAP is proposed by incorporating the adaptive filter and projection operation.
In our experiments, to make sure that above three methods can perform well, they are all initialized with the result of SPGL1. L0-ADMM uses the ADMM framework and the IHT. Its iterations are given in \eqref{meth1_u}-\eqref{meth1_w}. As initialization is not a key step in  this method, the initial values of all parameters are $\mathbf{0}$ in our experiments.

\subsection{Settings}
The basic settings of the experiments follow a standard model given by \cite{candes2005l1,ji2008bayesian}. We consider two signal lengths, $N=512$ and $N=4096$.
The measurement matrix $\bA$ is a random matrix with elements equaling $\pm 1/\sqrt{M}$ randomly.

For each $\ibx$, it includes $\tau$ nonzero elements, their locations are randomly chosen with uniform distribution. Their corresponding value is random $\pm 1$. When $N=512$ we let $\tau=\{15,25\}$, and when $N=4096$ we let $\tau=\{75,100,125\}$. The observation vector is generated by
\begin{equation*}
\ibb=\bA\ibx+\ibe,
\end{equation*}
where $\ibe$ is a zero mean Gaussian noise with standard deviation  $\sigma=\{0.001,0.005,0.01\}$. Each experiment repeats 100 times with different measurement matrix $\bA$, initial states and sparse signals. Following \cite{maleki2010approximate, jin2010stochastic}, for each run, we claim that it is successful if
\begin{equation*}
\frac{\|\ibx_0-\hat{\ibx}\|_2}{\|\ibx_0\|_2}\leq tol,
\end{equation*}
where $\ibx_0$ denotes the true signal, $\hat{\ibx}$ is the recovered signal, $tol$ is a reference value, which is chosen as $0.01$ in our experiment.

\subsection{Stability}
In Section~\ref{section4}, we theoretically prove that the proposed method has global convergence. Here we experimentally verify the convergence of our proposed method. Several typical results when $N=512$ are given in Fig.~\ref{fig:convergence512}. The first row denotes the case when $\tau = 15$, $M = 80$, $\sigma=0.01$. And the second row is the case with the same setting except $\sigma=0.001$. The third one is the case when $\tau = 25$, $M = 110$ and $\sigma=0.01$. The last row shows the result when $\tau = 25$, $M = 110$ and $\sigma=0.001$. The first three columns show the convergence of variables $\ibx$, $\ibu$, and $\ibw$. The final column displays the performance of our proposed algorithm in this trial. From Fig.~\ref{fig:convergence512}, it is observed that the proposed method can settle down within around 60-100 iterations and it can successfully recover the signal as long as we have enough observations. Comparing the first and second rows or the third and last rows in Fig.~\ref{fig:convergence512}, we see that the proposed algorithm is not very sensitive to the noise level.

When $N=4096$ and $\sigma=0.01$, the typical results are shown in Fig.~\ref{fig:convergence4096}. The first row shows the case when $\tau = 75$ and $M = 440$. The second and third row are the cases when $\tau = 100$, $M = 540$ and $\tau = 125$, $M = 640$ respectively. The meaning of each column is same as that in Fig.~\ref{fig:convergence512}. From Fig.~\ref{fig:convergence4096}, we see that the proposed method can settle down within around 100-150 iterations. Even $\sigma=0.01$, as long as we have a long enough observation vector, the performance of our proposed algorithm can be guaranteed.

\subsection{Comparison with other algorithms}
Fig.~\ref{fig:512} shows the simulation results of $N=512$. Because the signal is too short, it cannot offer sufficient statistical information to AMP method, i.e., the results of AMP in this setting is meaningless to some extent. Thus we have not implemented the AMP in this case.
For all the plots in Fig.~\ref{fig:512}, the x-axis denotes the number of the measurements $M$ (i.e. the length of the observation vector $\ibb$), the y-axis is the probability of exact reconstruction.
The blue line denotes our proposed method with parameter $\lambda$ selected by the first method in Section~\ref{section5}. While the red line is our proposed method with parameter $\lambda$ selected by the second method.
It is observed that for all the algorithms shown in Fig.~\ref{fig:512}, their performance is improved with the number of measurements. All nonconvex approximate $l_0$-norm solutions, except FOCUSS, are superior to the standard $l_1$-norm method SPGL1. Under high-level Gaussian noise, the performance of FOCUSS method is worse than SPGL1. Besides, tuning hyper-parameter for FOCUSS under noisy environment is cumbersome and time-consuming. Hence we will not implement this method in our following experiments.
From the blue line in Fig.~\ref{fig:512} we see that the first $\lambda$ selection method can offer an appropriate $\lambda$ for the proposed method, and its performance is excellent. Comparing with other algorithms, it needs fewer observations to obtain the same probability of exact reconstruction.
From the red line, we know that if we automatically regularize parameter $\lambda$ with our second parameter setting method, the performance of our proposed method is worse than that of the first $\lambda$ selection method but slightly better than others.

When $N=4096$, the AMP is used to replace the FOCUSS. The experimental results are shown in Fig.~\ref{fig:4096}. Under this case, we draw conclusions similar to the case when $N=512$.  
Besides, due to the global convergence, there is no need for us to worry about the initial value for the proposed method. So we let $\ibu^0=\mathbf{0}$, $\ibx^0=\mathbf{0}$, and $\ibw^0=\mathbf{0}$.

\section{Conclusion}\label{section7}
For solving the signal reconstruction problem in compressed sensing, we devise a novel approach based on ADMM and MCP. With reasonable assumptions, the global convergence of the proposed method is proved. Then, we discuss the parameter selection for our proposed algorithm. Finally, we show its convergence and performance experimentally. In future work, efficient parameter selection methods are still worth studying.

\appendices
\section{Proof of Lemma 1} \label{append:1}
Let $\iby^k=\ibu^k-\ibx^k$, $\alpha$ be a scalar, $h(\alpha)=\psi(\ibx^k + \alpha\iby^k)$. Then
\beq
&&\psi(\ibx^k+\iby^k)-\psi(\ibx^k)=h(1)-h(0) \nonumber\\
&=&\int_0^1\frac{dh(\alpha)}{d\alpha} d\alpha = \int_0^1 {\iby^k}^\mathrm{T}\nabla\psi(\ibx^k+\alpha\iby^k) d\alpha \nonumber\\
&\leq& \int_0^1 {\iby^k}^\mathrm{T}\nabla\psi(\ibx^k) d\alpha \nonumber\\ &&+\left|\int_0^1 {\iby^k}^\mathrm{T}(\nabla\psi(\ibx^k+\alpha\iby^k)-\nabla\psi(\ibx^k))d\alpha\right|
\nonumber\\
&\leq& \int_0^1{\iby^k}^\mathrm{T}\nabla\psi(\ibx^k) d\alpha \nonumber\\ &&+\int_0^1\|\iby^k\|_2\|\nabla\psi(\ibx^k+\alpha\iby^k)-\nabla\psi(\ibx^k)\|_2d\alpha
\nonumber\\
&\leq& {\iby^k}^\mathrm{T}\nabla\psi(\ibx^k)+\|\iby^k\|_2\int_0^1 l_{\psi}\alpha \|\iby^k\|_2 d\alpha
\nonumber\\
&=&{\iby^k}^\mathrm{T}\nabla\psi(\ibx^k)+\frac{l_{\psi}}{2}\|\iby^k\|_2^2 \nonumber \\
&=&\nabla\psi(\ibx^k)^\mathrm{T}(\ibu^k-\ibx^k)+\frac{l_{\psi}}{2}\|\ibu^k-\ibx^k\|_2^2
\eeq

\ifCLASSOPTIONcaptionsoff
  \newpage
\fi

\bibliographystyle{IEEEtran}
\bibliography{IEEEabrv,reference}

\end{document}